\documentclass[aps,prb,reprint,superscriptaddress]{revtex4-2}

\usepackage{tikz}
 \normalsize
\usepackage{scalerel}
\usepackage{amsthm}

\usepackage{amssymb}
\usepackage{suffix}
\usepackage{float}
\usepackage{mathtools}
\usepackage[utf8]{inputenc}
\usepackage{booktabs}
\usepackage{cases}
\usepackage[multiple]{footmisc}
\usepackage{dcolumn}
\usepackage{color,soul}
\usepackage{rotating}
\usepackage{perpage}
\usepackage{xcolor}
\usepackage{comment}
\usepackage{soul}
\usepackage{tikz}
\usepackage[T1]{fontenc}
\usepackage{etoolbox}
\usepackage{graphics}
\usepackage{siunitx}
\usepackage{hyperref}
\usepackage{physics}
\hypersetup{colorlinks}
\usepackage{float}	
\usepackage{collref}
\usepackage{multirow}
\usepackage{mathtools}
\usepackage{bm}
\usepackage{url}
\usepackage{wasysym}
\usepackage{orcidlink}

\begin{document}

\title{Entanglement and magic transitions in an all-to-all non-Hermitian spin model}
\author{Sasanka Dowarah\orcidlink{0000-0003-4072-8349}}
\email{sasanka.dowarah@utdallas.edu}
 \affiliation{Department of Physics, The University of Texas at Dallas, Richardson, Texas 75080, USA}
 \author{Michael Kolodrubetz\orcidlink{0000-0001-5628-3300}}
\email{mkolodru@utdallas.edu}
 \affiliation{Department of Physics, The University of Texas at Dallas, Richardson, Texas 75080, USA}
\date{\today}
\begin{abstract}
The long-time state of a non-Hermitian system is determined by the eigenvalue
with the largest imaginary part. In interacting many-body systems this eigenvalue usually cannot be tracked analytically, and the character of the state it selects is unknown. We construct a non-Hermitian spin ensemble of
$L$ spins with exactly $k$-local all-to-all interactions, in which this dominant
eigenvalue can be tracked analytically from the clean limit into the disordered
regime. Disorder produces a competition between an isolated spectral outlier
and the edge of a many-body spectral bulk. We study three cases: purely
anti-Hermitian disorder, purely Hermitian disorder, and mixed disorder of equal
strength. For purely anti-Hermitian and mixed disorder, we show that when the
bulk overtakes the outlier in imaginary part, the dominant eigenstate switches
from an outlier state with low entanglement and magic (nonstabilizerness) to a
bulk state with substantially larger entanglement and magic, with both changing
at the same threshold. \textcolor{red}{We show that for large-$k$ locality} $(k \gg \sqrt{L})$ the bulk \textcolor{red}{spectrum} has a sharp spectral edge
and the transition thresholds follow in closed form, while for \textcolor{red}{small-$k$ locality} $(k \ll \sqrt{L})$
\textcolor{red}{extended} spectral tails \textcolor{red}{change} the transition into a crossover. 
\textcolor{red}{In the purely Hermitian disorder case, we find no similar outlier-to-bulk switching of the dominant eigenvalue}. Finally,
we map the non-Hermitian evolution exactly onto postselected trajectories of a
monitored quantum system, connecting this spectral mechanism to
measurement-induced transitions.
\end{abstract}
\maketitle
{\allowdisplaybreaks}
 \parskip 0 pt
\section{Introduction}
Non-Hermitian Hamiltonians arise as effective descriptions of systems that involve gain and loss of particles, dissipation, and nonreciprocity \textcolor{red}{as well as quantum evolution conditioned on measurement outcomes}~\cite{Ashida02072020}. Hermitian Hamiltonians have real eigenvalues and an orthonormal eigenbasis. Non-Hermitian Hamiltonians generally have a complex spectrum and biorthogonal left and right eigenvectors~\cite{Ashida02072020, 10.5555/343374}. This complex nature of the eigenvalues makes their physical interpretation more subtle, but it can also give rise to new physics. In
particular, for a diagonalizable non-Hermitian Hamiltonian, normalized
time evolution generically selects the right eigenstate whose
eigenvalue has the largest imaginary part, provided that this state has nonzero overlap with the initial state.

This observation reduces a dynamical question to a spectral one. The state of these systems after long time is controlled not by the full complex spectrum, but by the eigenvalue with the largest imaginary part. Therefore, understanding when such an eigenvalue remains separated from the rest of the spectrum by a finite gap~\footnote{By gap, here we mean the usual Euclidean distance between two complex numbers in the complex plane.},  when such a gap exists, and how it changes as system parameters are varied is essential for predicting the long-time behavior of non-Hermitian many-body systems.

We use this observation to formulate a minimal problem: \textit{\textcolor{red}{Can one construct an interacting many-body model in which the eigenvalue governing long-time dynamics can be predicted analytically?}} The relevant problem in this case is to understand when a non-Hermitian many-body spectrum contains an isolated dominant eigenvalue with the largest imaginary part, how this eigenvalue is separated from the rest of the spectrum, and how this separation disappears as a function of system parameters. These are precisely the types of questions for which random-matrix theory provides a natural language. In deformed random-matrix models---in which a structured deterministic perturbation is added to a random-matrix---the spectrum is organized into a dense bulk together with possible isolated eigenvalues, or outliers, generated by the perturbation. The emergence and eventual absorption of such outliers into the bulk have been extensively studied in both Hermitian and non-Hermitian deformed random-matrix ensembles and are closely related to gap-closing phenomena~\cite{baik2004phasetransitionlargesteigenvalue, peche2005largesteigenvaluesmallrank, benaychgeorges2010eigenvalueseigenvectorsfinitelow, tao2014outliersspectrumiidmatrices, Rochet2017, han2025outliersboundedrankperturbation}.

Completely random matrices are not microscopic models of generic many-body Hamiltonians.  In such ensembles, the matrix elements are independent
up to global symmetry constraints, and correspond to all particles
interacting at once.  Physical Hamiltonians instead contain additional structure inherited from locality, few-body interactions, conservation
laws, and symmetry sectors~\cite{FRENCH1970449, BOHIGAS1971261, Torres_Herrera_2016, Benet_2003}. These constraints correlate the matrix elements in ways that unstructured random-matrix models do not capture. \textcolor{red}{We retain this structure by considering an all-to-all spin model whose interactions involve a prescribed number of spins. This tunable locality allows us to interpolate between physical models (small locality) to random matrices (large locality). We choose the interaction strength among the spins from Gaussian distributions with separately tunable means and variances. This allows us to separate the Hamiltonian into a deterministic part and a disordered part and further provides a natural mapping to deformed random matrices. We show that when the means are kept fixed, the variances of the interaction strength controls largest imaginary part of the Hamiltonian and therefore the steady state of the system.}

We examine two natural probes of the steady state: the entanglement entropy and magic, or nonstabilizerness. Entanglement entropy measures the amount of quantum correlation shared between a subsystem and the rest of the system~\cite{RevModPhys.81.865}. Physically, it distinguishes a product-like or weakly correlated selected eigenstate from a thermal state. Magic probes an independent notion of complexity; it measures how far the state lies from the set of stabilizer states generated by Clifford circuits~\cite{Veitch_2014, PhysRevA.71.022316}. As 
dissipation is tuned, one can ask how these two notions of complexity change.

\textcolor{red}{We choose our non-Hermitian Hamiltonian to be of the form $G = H_{R} + i H_{I}$}, where $H_{R}$ and $H_{I}$ are Hermitian matrices.  We impose structure on $G$ by taking these two components $H_{R}$ and $H_{I}$ to be independent \textcolor{red}{and} exactly $k_{R}$ and $k_{I}$--local spin Hamiltonians, with coupling coefficients drawn from two normal distributions with independently tunable means $\mu_{R}, \mu_{I}$ and variances $\sigma^{2}_{R}, \sigma^{2}_{I}$. The model is random enough to display non-Hermitian
spectral universality but structured enough to remain analytically
\textcolor{red}{tractable}. In the deterministic limit the model is exactly diagonalizable, with the spectrum given in closed form by Krawtchouk polynomials, and the eigenstates are
product states. In the fully disordered limit, the ensemble falls into
one of the standard non-Hermitian random-matrix classes, fixed by an
antiunitary symmetry tied to the parities of $L$, $k_{R}$, and $k_{I}$.
 
Within this model, we distinguish between \emph{outlier existence}, meaning spectral isolation
from the bulk, and \emph{dynamical dominance}, meaning possession of the largest imaginary part
and hence control of the long-time state. For purely anti-Hermitian disorder, these occur at
distinct thresholds, $\sigma_{I,c}^{\mathrm{outlier}}=\sqrt{2}$ and
$\sigma_{I,c}^{\mathrm{dom}}=2-\sqrt{2}$, leaving a finite regime in which the outlier remains
isolated but is no longer dynamically dominant. At the dominance transition, the selected right
eigenstate switches from the outlier branch to the bulk edge, producing simultaneous sharp
changes in entanglement and magic. Purely Hermitian disorder provides a contrasting case: the
bulk expands along the real direction and no analogous switching transition occurs. For balanced
mixed disorder, a Ginibre-like bulk produces a corresponding dominance transition at
$\sigma_{c,\mathrm{dom}}=1/\sqrt{2}$, with extensive post-transition entanglement. These results
show that the transition is controlled not by disorder alone, but by whether the random-matrix
bulk outgrows the deterministic outlier along the imaginary direction. All of these thresholds
are sharp in the regime $k \gg \sqrt{L}$, where the disordered bulk has a hard spectral edge; for
$k \ll \sqrt{L}$ the same competition survives but as a size-dependent crossover rather than a
sharp transition.

\textcolor{red}{These results have a direct interpretation in monitored quantum systems, where evolution conditioned on the absence of detection events is governed by a non-Hermitian effective Hamiltonian. We show that an imaginary identity shift maps our generator onto this form without changing its normalized dynamics, with explicit constructions of both spectral and $k_{I}$-body jump operators. The anti-Hermitian couplings therefore specify an engineered monitoring strength, and changing their disorder distribution can change the conditional long-time state and its entanglement and magic. This connects our spectral analysis to state selection in monitored many-body systems and, through the weak-measurement construction.}
 
The rest of the paper proceeds as follows. Section~\ref{sec:Models and methods}
defines the ensemble and the entanglement and magic probes.
Section~\ref{sec:Exactly solvable deterministic limit} solves the deterministic limit, classifies
the disordered ensemble by antiunitary symmetry, and derives the
outlier trajectory. Section~\ref{sec:Dynamical dominance and phase diagram} introduces the
dominance criterion and the resulting phase diagram separating outlier
existence from dynamical dominance. Section~\ref{sec:Outlier-to-bulk transitions of dominant eigenvalue}
analyzes the three disorder cases in turn---anti-Hermitian, Hermitian,
and mixed---and their consequences for entanglement and magic. Section~\ref{sec:postselected_monitored_dynamics} shows that the normalized dynamics along the anti-Hermitian ray is exactly a no-click trajectory of a monitored spin system, and constructs both a spectral and a $k_{I}$-local unraveling. We
conclude in Section~\ref{sec:discussion_and_outlook} with a synthesis of the results
and directions for future work.

\section{Model and diagnostics}\label{sec:Models and methods}
\subsection{The $k$--local non-Hermitian ensemble}
Any complex matrix $G$ acting on an $L$-qubit Hilbert space admits a unique
decomposition into Hermitian and anti-Hermitian parts,
\begin{equation}
    G
    =
    \frac{G+G^{\dagger}}{2}
    +
    i\;\frac{G-G^{\dagger}}{2i}
    \equiv H_{R}+iH_{I} ,
\end{equation}
where
\begin{equation}
    H_{R} \equiv \frac{G+G^{\dagger}}{2},
    \qquad
    H_{I} \equiv \frac{G-G^{\dagger}}{2i}
\end{equation}
are Hermitian matrices. This decomposition motivates a natural way to construct
structured non-Hermitian ensembles. Rather than drawing $G$ directly from a
random-matrix ensemble, one can independently specify the statistical
properties of $H_{R}$ and $H_{I}$ and assemble $G$ from them. 

Since $H_{R}$ and $H_{I}$ are Hermitian operators acting on an $L$-spin Hilbert space, each admits an expansion in the Pauli basis. We write
\begin{subequations}
\begin{align}
    P_{\boldsymbol{\alpha}}
    =
    \sigma_{\alpha_1}\otimes\sigma_{\alpha_2}\otimes\cdots\otimes\sigma_{\alpha_L},\label{eq:Pauli_string_definition}\\ 
    \boldsymbol{\alpha}=(\alpha_1,\ldots,\alpha_L)\in\{0,x,y,z\}^{L},
\end{align}
\end{subequations}
where $\sigma_{0}\equiv I, \sigma_{x}, \sigma_{y}, \sigma_{z}$ are the Pauli matrices. These operators form an orthogonal basis with respect to the Hilbert--Schmidt inner product,
\begin{equation}
    \mathrm{Tr} \left[P_{\boldsymbol{\alpha}}^{\dagger}P_{\boldsymbol{\beta}}\right]
    =
    2^{L}\delta_{\boldsymbol{\alpha},\boldsymbol{\beta}} .
\end{equation}
The weight of a Pauli string of the form in Eq.~\eqref{eq:Pauli_string_definition} is defined as
\begin{equation}
    |\boldsymbol{\alpha}|
    =
    \#\{j:\alpha_j\neq 0\},
\end{equation}
which counts the number of sites on which the string acts nontrivially. In a generic Pauli expansion, the Hermitian matrices $H_{R}$ and $H_{I}$ may contain components of all possible weights $0\leq k\leq L$. In this work, we restrict the Hermitian and anti-Hermitian components of $G$ to fixed Pauli-string weights. We call a Pauli string \emph{exactly $k$--local}, or equivalently $k$-body, if $|\boldsymbol{\alpha}|=k$~\cite{Brown2018}. We take $H_{R}$ to be an exactly $k_{R}$--local Hermitian spin operator, while $H_{I}$ to be an exactly $k_{I}$--local Hermitian spin operator. We write them as
\begin{widetext}
\begin{subequations}
\begin{eqnarray}
H_{R}(\mu_{R}, \sigma^{2}_{R}) &\equiv& \sum_{j_{1}< j_{2} < \cdots < j_{k_{R}}} \sum^{3}_{\alpha_{1}, \cdots, \alpha_{k_{R}} = 1}  J^{\alpha_{1} \alpha_{2} \cdots \alpha_{k_{R}}}_{j_{1}j_{2}\cdots j_{k_{R}}}
   \;\;\; \sigma^{(j_{1})}_{\alpha_{1}}\otimes\sigma^{(j_{2})}_{\alpha_{2}}\otimes
    \cdots \otimes\sigma^{(j_{k_{R}})}_{\alpha_{k_{R}}}, \\
H_{I}(\mu_{I}, \sigma^{2}_{I}) &\equiv& \sum_{j_{1}< j_{2} < \cdots < j_{k_{I}}} \sum^{3}_{\alpha_{1}, \cdots, \alpha_{k_{I}} = 1}  M^{\alpha_{1} \alpha_{2} \cdots \alpha_{k_{I}}}_{j_{1}j_{2}\cdots j_{k_{I}}}
   \;\;\; \sigma^{(j_{1})}_{\alpha_{1}}\otimes\sigma^{(j_{2})}_{\alpha_{2}}\otimes
    \cdots \otimes\sigma^{(j_{k_{I}})}_{\alpha_{k_{I}}}\;, \label{eq:H_Pauli_basis_model}
\end{eqnarray}
\end{subequations}
\end{widetext}
where $J^{\alpha_1\alpha_2\cdots\alpha_{k_{R}}}_
{j_1j_2\cdots j_{k_{R}}}$ and $ M^{\alpha_1\alpha_2\cdots\alpha_{k_{I}}}_{j_1j_2\cdots j_{k_{I}}}$ are independent normally distributed couplings. We choose their distributions as 
\begin{subequations}
\begin{eqnarray}
    \mathbb E
    \left[
    J^{\alpha_1\alpha_2\cdots\alpha_{k_{R}}}_
    {j_1j_2\cdots j_{k_{R}}}
    \right]
    &=&
    \frac{\mu_{R} L}
    {\sqrt{3^{k_{R}}}\binom{L}{k_{R}}}
    \equiv m_{R},
    \\[6pt]
    \mathbb E
    \left[
    M^{\alpha_1\alpha_2\cdots\alpha_{k_{I}}}_
    {j_1j_2\cdots j_{k_{I}}}
    \right]
    &=&
    \frac{\mu_I L}
    {\sqrt{3^{k_{I}}}\binom{L}{k_{I}}}
    \equiv m_{I} ,
    \\[6pt]
    \text{Var}
    \left[
    J^{\alpha_1\alpha_2\cdots\alpha_{k_{R}}}_
    {j_1j_2\cdots j_{k_{R}}}
    \right] &=& \frac{\sigma^{2}_{R}L^{2}}{3^{k_{R}}\binom{L}{k_{R}}}
    \equiv s_{R}^{2} , \\
    \text{Var}
    \left[
    M^{\alpha_1\alpha_2\cdots\alpha_{k_{I}}}_
    {j_1j_2\cdots j_{k_{I}}}
    \right] &=& \frac{\sigma^{2}_{I}L^{2}}{3^{k_{I}}\binom{L}{k_{I}}}
    \equiv s_{I}^{2} .
\end{eqnarray}\label{eq:Mean_J_M_couplings}
\end{subequations}
We fix the coupling statistics this way following the Kac prescription that is standard for mean-field spin models~\cite{Defenu_2023, Erd_s_2014, ng44-x1tv}. The variance is scaled inversely with the number of exactly $k$--local Pauli strings $3^{k} \binom{L}{k}$, so that the root-mean-square spectral width is extensive, $E_{\mathrm{RMS}} = \sqrt{\mathrm{Tr} H^{2}_{R/I}/2^{L}} = \sigma L$. The mean is normalized so that its contribution to the spectrum grows on the same scale, with the largest eigenvalue it produces scaling linearly with $L$. This choice puts the mean-value and variance contributions to the spectrum on a common $O(L)$ energy scale and gives a finite thermodynamic limit for the energy density.

We define our non-Hermitian model as
\begin{align}
    G(\mu_{R}, \mu_{I}, \sigma_{R}, \sigma_{I}) = \frac{1}{\sqrt{2}}[H_{R}(\mu_{R}, \sigma^{2}_{R}) + iH_{I}(\mu_{I}, \sigma^{2}_{I})]. \label{eq:Definition-of-G}
\end{align}
Because each coupling is Gaussian with the mean and variance fixed in
Eq.~\eqref{eq:Mean_J_M_couplings}, we write it in the location-scale form,
$J_{\boldsymbol{\alpha}} = m_R + s_R \xi_{\boldsymbol{\alpha}}$ and
$M_{\boldsymbol{\beta}} = m_I + s_I \zeta_{\boldsymbol{\beta}}$, with
$\xi_{\boldsymbol{\alpha}}, \zeta_{\boldsymbol{\beta}} \sim \mathcal{N}(0,1)$
independent of $\mu_{R},\sigma_{R},\mu_I,\sigma_{I}$. This separates each Hermitian
component into a fixed part and a zero-mean random part,
\begin{subequations}
\begin{align}
    H_{R}(\mu_{R}, \sigma_{R}^2) &= m_R \; H_R(1,0) + s_R \; H_R(0,1), \\
    H_{I}(\mu_I, \sigma_{I}^2) &= m_I \; H_I(1,0) + s_I \; H_I(0,1),
\end{align} \label{eq:Hermitian_deterministic_disordered_components}
\end{subequations}
where $H_R(1,0) \equiv \sum_{\boldsymbol{\alpha}} P_{\boldsymbol{\alpha}}$ is
obtained by setting every coupling to exactly $1$, and
$H_R(0,1) \equiv \sum_{\boldsymbol{\alpha}} \xi_{\boldsymbol{\alpha}} P_{\boldsymbol{\alpha}}$
is a realization of $H_R$ with couplings drawn from $\mathcal{N}(0,1)$;
$H_I(1,0)$ and $H_I(0,1)$ are defined analogously. Substituting
Eq.~\eqref{eq:Hermitian_deterministic_disordered_components} into
Eq.~\eqref{eq:Definition-of-G} gives
\begin{align}
G(\mu_{R},\mu_{I},\sigma_{R},\sigma_{I})
&= \frac{1}{\sqrt{2}}\bigg[
m_{R} H_{R}(1,0)
+ im_{I} H_{I}(1,0) \nonumber\\
&+ s_{R} H_{R}(0,1)
+ is_{I} H_{I}(0,1)
\bigg].
\label{eq:Mu-sigma-decomposition}
\end{align}
The first two terms depend only on the fixed operators $H_R(1,0)$ and
$H_I(1,0)$ and are independent of the disorder realization. We call this
the deterministic part of $G$, obtained by setting $\sigma_{R}=\sigma_{I}=0$.
The remaining two terms depend on the random couplings $\xi_{\boldsymbol{\alpha}}$,
$\zeta_{\boldsymbol{\beta}}$ and vanish when $\mu_{R}=\mu_I=0$; we call this
the disordered part. We refer to one particular draw of $\{\xi_{\boldsymbol{\alpha}}\}$
and $\{\zeta_{\boldsymbol{\beta}}\}$ as a disorder realization, and to
$\sigma_{R}, \sigma_{I}$ as disorder strengths.

Our construction is closely related in spirit to non-Hermitian SYK models, where complex random couplings generate many-body Hamiltonians with complex spectra and non-Hermitian random-matrix universality~\cite{PhysRevX.12.021040}. In contrast to those fermionic models, we work with an exactly $k$--local spin ensemble with tunable deterministic mean components.

\subsection{Dominant eigenvalue and long-time dynamics}\label{sec:Non-Hermitian spectral decomposition and long-time dynamics}
In this section, we briefly review some definitions concerning spectra of non-Hermitian Hamiltonians. Let $G$ be a diagonalizable non-Hermitian Hamiltonian with eigenvalues $z_{n} \in \mathbb{C}$. Then the right and left eigenstates of $G$ are defined as \cite{Ashida02072020}
\begin{eqnarray}
    G|R_{n}\rangle &=& z_{n}|R_{n}\rangle,\;\;\; \langle L_{n}|G = z_{n} \langle L_{n}|. \label{eq:Left-right-eigenvectors}
\end{eqnarray}
The right eigenvectors are eigenvectors of \(G\), while the left eigenvectors may equivalently be viewed as right eigenvectors of \(G^\dagger\):
\[
G^\dagger |L_n\rangle = z_n^* |L_n\rangle .
\]
Additionally, they generically obey the following properties:
\begin{align}
        \langle L_{m}|R_{n}\rangle &=& \delta_{mn},\;\; \langle L_{m}|L_{n}\rangle \neq \delta_{mn},\; \langle R_{m}|R_{n}\rangle \neq \delta_{mn}
\end{align}
The spectral decomposition of $G$ is
\begin{eqnarray}
    G &=& \sum_{n} z_{n} |R_{n}\rangle \langle L_{n}| .
\end{eqnarray}
\begin{figure}
    \centering
    \includegraphics[width=\linewidth]{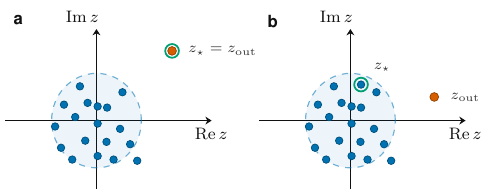}
\caption{Schematic definition of outlier, bulk, and dominant eigenvalue in a non-Hermitian spectrum. 
An outlier eigenvalue, which we denote by $z_{\rm out}$, is an eigenvalue that is isolated from the main spectral support by a finite distance in the complex plane. 
The remaining eigenvalues, which form the dense main spectral support, are referred to as the bulk. 
The dominant eigenvalue $z_{\star}$, which is shown as a green circle in the figure, is defined as the eigenvalue with the largest imaginary part. 
(a) The case when the outlier has the largest imaginary part, the outlier is the dominant eigenvalue, $z_{\star}=z_{\rm out}$. 
(b) The outlier $z_{\mathrm{out}}$ exists because it is still isolated from the rest of the spectrum, but it is no longer the dominant eigenvalue since it no longer has the largest imaginary part. Instead, $z_{\star}$ lies in the bulk. 
Thus, being an outlier and being dynamically dominant are distinct properties.
}
\label{fig:Bulk_outlier_dominant}
\end{figure}

Next, we define the notions of outlier, bulk, and dominant eigenvalue, which will be used throughout this work. 
We use the term \emph{outlier} in the random-matrix sense: an eigenvalue \(z_{\mathrm{out}}\) that remains separated from the main spectral support by a nonzero spectral distance in the large-system limit~\cite{baik2004phasetransitionlargesteigenvalue, tao2014outliersspectrumiidmatrices, Rochet2017, PhysRevLett.117.224101, han2025outliersboundedrankperturbation}. 
The remaining eigenvalues, which form the dense main spectral support, are referred to as the \emph{bulk}. 

To define the term \emph{dominant eigenvalue}, we first show that the long-time behavior of a non-Hermitian Hamiltonian is determined by the eigenvalue with the largest imaginary part \cite{Ashida02072020}. Suppose $\{z_{n} = x_{n} + i y_{n}, n = 1, 2, \cdots, N\}$ are the eigenvalues of a non-Hermitian Hamiltonian $G$ satisfying Eq.~\eqref{eq:Left-right-eigenvectors}. Then the time evolution of an initial state $|\psi(0)\rangle$ can be written as
\begin{eqnarray}
    |\psi(t)\rangle &=& \frac{\sum_{n} e^{-iz_{n}t}|R_{n}\rangle\langle L_{n}|\psi(0)\rangle}{||\sum_{n} e^{-iz_{n}t}|R_{n}\rangle\langle L_{n}|\psi(0)\rangle||} .
\end{eqnarray}
If $z_{\star}=x_{\star}+i y_{\star}$ has the largest  imaginary part, that is
\begin{eqnarray}
    y_{\star}>\max_{n\neq \star} y_{n},
\end{eqnarray}
and $\langle L_\star|\psi(0)\rangle\neq 0$, then the normalized state approaches the right eigenstate \(|R_\star\rangle\) at long times
\begin{eqnarray}
    |\psi(t)\rangle &\to& e^{-ix_{\star}t}|R_{\star}\rangle . \label{eq:Long-time-evolution_right_eigenstate}
\end{eqnarray}
We therefore define the \emph{dominant eigenvalue} \(z_{\star}\) as the eigenvalue with the largest imaginary part. This eigenvalue determines the long-time normalized non-Hermitian dynamics. An outlier eigenvalue and the dominant eigenvalue need not always coincide. An outlier may remain isolated from the bulk while a bulk eigenvalue has the largest imaginary part. We illustrate these definitions in Fig.~\ref{fig:Bulk_outlier_dominant}.
\subsection{Diagnostics of the long-time state}
\emph{Entanglement entropy}: We quantify the entanglement entropy (EE) of the right eigenstates $|R\rangle$
using the von Neumann entanglement entropy~\cite{j1r7-82ft}
\begin{eqnarray}
    S &=& - \mathrm{Tr}[\rho_{A} \log \rho_{A}] ,\label{eq:Entanglement_right_eigenstate}
\end{eqnarray}
where
\begin{eqnarray}
    \rho_{A} &=& \mathrm{Tr}_{A^{c}}\Bigg[\frac{|R\rangle\langle R|}{\langle R|R\rangle}\Bigg] . \label{eq:Reduced_density_right_eigenstate}
\end{eqnarray}
Here $A$ denotes a subsystem of the $L$ spins and $A^{c}$ its complement.
The normalization in Eq.~\eqref{eq:Reduced_density_right_eigenstate} is
important because right eigenvectors of a non-Hermitian matrix need not be
orthonormal under the usual Hilbert-space inner product.

Equations~\eqref{eq:Entanglement_right_eigenstate} and~\eqref{eq:Reduced_density_right_eigenstate} are different from the
biorthogonal entropy constructed from both left and right eigenvectors~\cite{j1r7-82ft, PhysRevLett.130.010401}. We use this definition because the density matrix in this case is Hermitian, positive semidefinite, has the usual interpretation as the density matrix of a normalized physical state, and it is easy to compute numerically.

\emph{Magic or nonstabilizerness}: The Gottesman--Knill theorem states that any quantum circuit composed entirely of 
Clifford gates---applied to computational basis states and measured in the Pauli basis---can be efficiently simulated on a classical computer~\cite{Nielsen_Chuang_2010, PhysRevA.70.052328}. Quantum 
states that can be prepared by such circuits are called stabilizer states. To achieve 
universal quantum computation, one must go beyond the Clifford group and access states 
that lie outside the stabilizer set. The resource that quantifies this non-Clifford 
content is called magic or nonstabilizerness~\cite{PhysRevA.71.022316, PRXQuantum.3.020333, 
xfp5-hhs4}. While entanglement measures correlations across a spatial bipartition, magic 
measures the non-Clifford structure of the many-body wavefunction; a state can be highly 
entangled yet have zero magic if it is a stabilizer state.

We quantify the magic of a right eigenstate $|R\rangle$ using the second 
stabilizer R\'enyi entropy. For a pure $L$-qubit normalized state $|R\rangle$, it is defined 
as~\cite{PhysRevLett.128.050402}
\begin{eqnarray}
    M_{2}(|R\rangle) &=& L \ln 2 - \ln \Bigg[ \sum_{P \in \mathcal{P}_{L}} 
    |\langle R | P | R\rangle|^{4}\Bigg], 
    \label{eq:Second_stabilizer_formula}
\end{eqnarray}
where $\mathcal{P}_{L} = \{ I, \sigma_{x}, \sigma_{y}, \sigma_{z}\}^{\otimes L}$ denotes 
the set of all $L$-qubit Pauli strings, modulo overall phases. With this convention, 
stabilizer states have $M_{2} = 0$, and a nonzero $M_{2}$ indicates that the state lies 
outside the stabilizer set. We use $M_{2}$ because it can be evaluated exactly by a fast Walsh–Hadamard transform at smaller sizes~\cite{Georges_2025}, estimated by importance sampling at larger sizes~\cite{PRXQuantum.4.040317, PhysRevLett.131.180401, PhysRevB.111.085144, pyzr-jmvw}, and measured experimentally~\cite{Oliviero2022, PhysRevLett.132.240602}.

Entanglement transitions have been identified in several non-Hermitian settings, including $\mathcal{PT}$-symmetric systems near Yang--Lee singularities~\cite{PhysRevB.104.L161107}, postselected free-fermion dynamics~\cite{SciPostPhys.14.5.138}, chaotic non-Hermitian spin chains~\cite{wy8r-7mr1}, and postselected non-Hermitian quantum mechanics near an
exceptional point~\cite{PhysRevLett.126.170503}, while the entanglement structure of Ginibre eigenstates has also been studied directly~\cite{PhysRevLett.130.010401}. 

Recent work has shown that magic-based diagnostics can display sharp
transitions and singular behavior in many-body systems, and can reveal
structure that is not captured by conventional thermodynamic or
entanglement observables~\cite{SciPostPhysCore.8.4.078, pyzr-jmvw, wang2025magictransitionmonitoredfree, SciPostPhysCore.9.1.012, 5c15-4g5n}. These results motivate magic as a
complementary probe of many-body structure beyond entanglement alone. A highly entangled
stabilizer state can have zero magic, while an unentangled product of
nonstabilizer states can have nonzero magic. Consequently, an entanglement
transition need not coincide with a transition in computational resource
content, which motivates studying the dynamics and transitions of magic
alongside entanglement.
\section{Spectral structure of the ensemble}\label{sec:Exactly solvable deterministic limit}
\subsection{Exactly solvable deterministic limit}
In this section, we determine the spectrum of the deterministic part of $G$. We follow Ref.~\cite{ng44-x1tv} closely. Setting $\sigma_{R} = \sigma_{I} = 0$ in Eq. \eqref{eq:Mu-sigma-decomposition} gives (with $G(\mu_{R}, \mu_{I}, 0, 0)  \equiv G_{0}(\mu_{R}, \mu_{I})$)
\begin{align}
   G_{0}(\mu_{R}, \mu_{I}) = \frac{1}{\sqrt{2}}\bigg[m_{R} H_{R}(1,0) + i m_{I} H_{I}(1,0)\bigg] .
\end{align}
For any $k_{R}$ and $k_{I}$, the deterministic Hermitian Hamiltonians commute: $[H_{R}(1, 0), H_{I}(1, 0)] = 0$. Therefore, $G_{0}(\mu_{R}, \mu_{I})$ is diagonal in their common eigenbasis. Let $U$ be the unitary that diagonalizes $G_{0}(\mu_{R}, \mu_{I})$. Then $U^{\dagger} G_{0}(\mu_{R}, \mu_{I}) U = D$~\cite{ng44-x1tv},
where
\begin{equation}
D =
\mathrm{diag}\!\Bigg(
\underbrace{d_{0},\ldots,d_{0}}_{\binom{L}{0}},
\underbrace{d_{1},\ldots,d_{1}}_{\binom{L}{1}},
\ldots,
\underbrace{d_{L},\ldots,d_{L}}_{\binom{L}{L}}
\Bigg), 
\label{eq:D-matrix-diagonal}
\end{equation}
with
\begin{eqnarray}
    d_{n} &=& \frac{L}{\sqrt{2}}\Bigg[ \mu_{R} \binom{L}{k_{R}}^{-1} K^{(L, 2)}_{k_{R}}(n) \nonumber\\
    && \hspace{1.2 cm} +\; i\mu_{I} \binom{L}{k_{I}}^{-1} K^{(L, 2)}_{k_{I}}(n)\Bigg] ,\label{eq:Exact_eigenvalues_of_G_deterministic}
\end{eqnarray}
where the Krawtchouk polynomials are defined as~\cite{Huffman_Pless_2003, kravchuk1929}
\begin{eqnarray}
    K^{(L, 2)}_{k}(n) = \sum^{k}_{j = 0} (-1)^{j} \binom{n}{j} \binom{L-n}{k-j},
\end{eqnarray}
and $n = 0, 1, \cdots, L$ and each eigenvalue $d_{n}$ has a degeneracy of $\binom{L}{n}$. For the rest of the work, we will set $\mu_{R} = \mu_{I} = 1$ unless stated otherwise, and denote $G_{0}(\mu_{R} = 1, \mu_{I} = 1) \equiv G_{0}$. 

In this deterministic limit, the dominant eigenvalue(s) originate from the two eigenvalues $d_{0}$ and $d_{L}$
\begin{eqnarray}
    d_{0} &=& \frac{L}{\sqrt{2}}(1+i), \; d_{L} = \frac{L}{\sqrt{2}}\bigg[ (-1)^{k_{R}} + i (-1)^{k_{I}}\bigg]. \label{eq:d_{0}_and_d_L_eigenvalues}
\end{eqnarray}
If $k_R$ and $k_I$ are both even, the dominant eigenvalue is exactly doubly degenerate, with $d_{0} = d_{L}$. For even $k_{R}$ and odd $k_{I}$, the dominant eigenvalue is unique. If $k_{R}$ is odd and $k_{I}$ is even, then there are two dominant eigenvalues related by $d_{L} = -d^{*}_{0}$. Finally, when $k_{R}$ and $k_{I}$ are both odd, the dominant eigenvalue is unique.

To determine the eigenstates of $G_{0}$, we first define the following single-site operator
\begin{eqnarray}
    P^{(j)} &\equiv & \frac{1}{\sqrt{3}} (\sigma^{(j)}_{x} + \sigma^{(j)}_{y} + \sigma^{(j)}_{z}).
\end{eqnarray}
Let $|u_{+}\rangle$ and $|u_{-}\rangle$ be the eigenstates of $P^{(j)}$. For a spin configuration
\begin{eqnarray}
    \Vec{s} &=& (s_{1}, s_{2}, \cdots, s_{L}), \hspace{1.2 cm} s_{j} = \pm,
\end{eqnarray}
we define the product states
\begin{eqnarray}
    |\Psi_{\Vec{s}}\rangle &=& \bigotimes^{L}_{j=1} |u_{s_{j}}\rangle. \label{eq:Psi_s}
\end{eqnarray}
Ref.~\cite{ng44-x1tv} shows that $|\Psi_{\Vec{s}}\rangle$ are the eigenstates of the Hermitian matrices $H_{R/I}(1,0)$.

Since $H_{R}(1, 0)$ and $H_{I}(1, 0)$ are Hermitian and commute with each other, the matrix $G_{0}$ is normal,
\begin{eqnarray}
    [G_{0}, G^{\dagger}_{0}] &=& 0. \label{eq:Normality-of-G_0}
\end{eqnarray}
Let $|R_{n}\rangle$ be a right eigenvector of $G_{0}$ with eigenvalue $z_{n}$:
\begin{eqnarray}
    G_{0}|R_{n}\rangle = z_{n}|R_{n}\rangle,\label{eq:Right-eigenvector-of-G_0}
\end{eqnarray}
and consider the norm
\begin{multline}
||(G^{\dagger}_{0} - z^{*}_{n} I)|R_{n}\rangle||^{2} =
    \langle R_{n} |(G_{0} - z_{n}I)(G^{\dagger}_{0} - z^{*}_{n} I) |R_{n}\rangle \\
    =\langle R_{n} |(G^{\dagger}_{0} - z^{*}_{n} I) (G_{0} - z_{n}I)|R_{n}\rangle = 0 , \label{eq:Norm-right-eigenvector}
\end{multline}
where in the last step we used Eqs.  \eqref{eq:Normality-of-G_0} and \eqref{eq:Right-eigenvector-of-G_0}. Hence
\begin{eqnarray}
    G^{\dagger}_{0} |R_{n} \rangle &=& z^{*}_{n}|R_{n}\rangle, \;\; \text{or}\;\; \langle R_{n}| G_{0} = z_{n} \langle R_{n}|.
\end{eqnarray}
In other words, the left eigenvectors can be chosen as the Hermitian conjugates of the right eigenvectors~\cite{Ashida02072020, 10.5555/343374}. Therefore, we conclude that
\begin{eqnarray}
    |R_{\Vec{s}}\rangle &=& |\Psi_{\Vec{s}}\rangle, \hspace{1.2 cm} \langle L_{\Vec{s}}| = \langle \Psi_{\Vec{s}}|. \label{eq:Right-left-eigenstate-of-G_0}
\end{eqnarray}
The spectral decomposition of $G_{0}$ is therefore
\begin{eqnarray}
    G_{0} &=& \sum_{\Vec{s}} d_{n(\Vec{s})} |\Psi_{\Vec{s}}\rangle \langle \Psi_{\Vec{s}}|,
\end{eqnarray}
where $n(\Vec{s})$ counts the number of minus signs in the vector $\Vec{s} = (s_{1}, s_{2}, \cdots, s_{L})$. The equality between left and right eigenvectors is special to the
deterministic limit $\sigma_{R} = \sigma_{I} = 0$, where $G_0$ is normal. Once $\sigma_{R}$ or $\sigma_{I}$
is nonzero, the full matrix $G$ is generically nonnormal and its left and
right eigenvectors no longer coincide.
\subsection{Symmetry classification of the disordered ensemble}\label{sec:Random-matrix structure of the zero-mean ensemble}
In this section, we consider the zero mean limit $\mu_{R} = \mu_{I} = 0$, in which the ensemble is fully disordered, and show that different combinations of $L, k_{R}, k_{I}$ completely determine its Ginibre random-matrix universality class. We define the time-reversal symmetry operator for the completely disordered Hermitian matrices $H_{R}(0,1)$ and $H_{I}(0,1)$ as
\begin{eqnarray}
    T &=& U \mathcal{K}, \quad U = \bigotimes^{L}_{j=1}(-i\sigma^{(j)}_{y}) ,\label{eq:Time-reversal-operator}
\end{eqnarray}
where $\mathcal{K}$ is the complex conjugation such that $\mathcal{K}i\mathcal{K}^{-1}= -i$, and $U$ is unitary~\cite{Sakurai_Napolitano_2020}.
The operator $T$ acts on an exactly $k$--local term as
\begin{eqnarray}
    &&T(\sigma^{(j_{1})}_{\alpha_{1}}\otimes\sigma^{(j_{2})}_{\alpha_{2}}\otimes
    \cdots \otimes\sigma^{(j_{k})}_{\alpha_{k}})T^{-1}\nonumber\\
    &&\hspace{1.2 cm}=(-1)^{k}\;(\sigma^{(j_{1})}_{\alpha_{1}}\otimes\sigma^{(j_{2})}_{\alpha_{2}}\otimes
    \cdots \otimes\sigma^{(j_{k})}_{\alpha_{k}}).
\end{eqnarray}
From this, we conclude the action of $T$ on the exactly $k$--local Hermitian matrix $Q \equiv H_{R}, H_{I}$:
\begin{eqnarray}
    T Q T^{-1} &=& (-1)^{k}Q. \label{eq:Action-of-sigma_y-K}
\end{eqnarray}
From the definition of $T$ [Eq.~\eqref{eq:Time-reversal-operator}], we get
\begin{eqnarray}
    T^{2} &=& (-1)^{L}\mathbf{1}.\label{eq:Sigma_y_K^2}
\end{eqnarray}
Using Eq.~\eqref{eq:Action-of-sigma_y-K}, we write the action of the time-reversal operator on $G$ as
\begin{eqnarray}
    T G T^{-1} &=& \frac{1}{\sqrt{2}}\bigg[(-1)^{k_{R}} H_{R} - i (-1)^{k_{I}} H_{I}\bigg].\label{eq:T_acting_on_matrix_G}
\end{eqnarray}
Note that this is equivalent to
\begin{eqnarray}
    T G^{\dagger} T^{-1} &=& \frac{1}{\sqrt{2}}\bigg[(-1)^{k_{R}} H_{R} + i (-1)^{k_{I}} H_{I}\bigg] .\label{eq:T_acting_on_matrix_G_dagger}
\end{eqnarray}
Using the results of Ref.~\cite{ng44-x1tv}, we conclude that in the completely disordered limit of $\mu_{R} = \mu_{I} = 0$, the parity combinations of $L$, $k_{R}$ and $k_{I}$ completely determine the random-matrix ensembles of $H_{R}$ and $H_{I}$, which we summarize in Table~\ref{tab:nonHermitian_symmetry_classification}. We further show that the random-matrix ensembles of $H_{R}$ and $H_{I}$ put spectral constraints on $G$ (proofs are in Appendix~\ref{sec:Antiunitary spectral constraints}), and each of the parity combinations falls into one of the known Ginibre random-matrix ensembles (proofs are in Appendix~\ref{sec:Relation to existing Non-Hermitian classification}).

\begin{table*}
\centering
\caption{
Antiunitary symmetry classification of the zero mean non-Hermitian ensemble. The GOE/GUE/GSE labels refer to the zero-mean random components $H_{R}(0,1)$ and $H_{I}(0,1)$. The antiunitary spectral constraints on $G$ hold also for the full nonzero-mean ensemble. The final column gives the corresponding non-Hermitian symmetry class, and the label non-Ginibre means the distribution is different from Ginibre's ensemble~\cite{PhysRevResearch.2.023286}.
}
\label{tab:nonHermitian_symmetry_classification}

\begin{ruledtabular}
\begin{tabular}{ccccccccc}
$L$ & $k_{R}$ & $k_{I}$ & $T^2$ &
$H_{R}(0,1)$ &
$H_{I}(0,1)$ &
$TGT^{-1}$ &
Spectral constraint &
NH class/statistics \\
\colrule
Even & Even & Even &
$+1$ &
GOE & GOE &
$G^\dagger$ &
$U(|L\rangle)^{*} \propto |R\rangle$&
AI$^\dagger$/non-Ginibre\\

Even & Even & Odd &
$+1$ &
GOE & GUE &
$G$ &
$(z,z^\ast)$ &
AI/GinUE\\

Even & Odd & Even &
$+1$ &
GUE & GOE &
$-G$ &
$(z,-z^\ast)$ &
D$^\dagger$/GinUE\\

Even & Odd & Odd &
$+1$ &
GUE & GUE &
$-G^\dagger$ &
$(z,-z)$ &
D/GinUE\\
\hline
Odd & Even & Even &
$-1$ &
GSE & GSE &
$G^\dagger$ &
Every $z$ is doubly degenerate &
AII$^\dagger$/non-Ginibre\\

Odd & Even & Odd &
$-1$ &
GSE & GUE &
$G$ &
$(z,z^\ast)$; real $z$ can be doubly degenerate &
AII/GinUE\\

Odd & Odd & Even &
$-1$ &
GUE & GSE &
$-G$ &
$(z,-z^\ast)$; imaginary $z$ can be doubly degenerate &
C$^\dagger$/GinUE\\

Odd & Odd & Odd &
$-1$ &
GUE & GUE &
$-G^\dagger$ &
$(z,-z)$; $z=0$ can be doubly degenerate &
C/GinUE\\
\end{tabular}
\end{ruledtabular}
\end{table*}

The deterministic spectrum obeys the same parity constraints as the full
ensemble. To see this explicitly, let us use the Krawtchouk identity
\begin{eqnarray}
K_k^{(L,2)}(L-n)=(-1)^k K_k^{(L,2)}(n),
\end{eqnarray}
in Eq.~\eqref{eq:Exact_eigenvalues_of_G_deterministic}, which gives
\begin{align}
d_{L-n}
=
\frac{L}{\sqrt{2}}
\Bigg[
\mu_{R}(-1)^{k_{R}}\binom{L}{k_{R}}^{-1}K_{k_{R}}^{(L,2)}(n)\nonumber\\
+
i\mu_I(-1)^{k_{I}}\binom{L}{k_{I}}^{-1}K_{k_{I}}^{(L,2)}(n)
\Bigg]. \label{eq:Deterministic_eigenvalues_G0_matrix}
\end{align}
Therefore the deterministic eigenvalues are paired by the same maps:
$(z, z^\ast)$, $(z, -z^\ast)$, or $(z, -z)$ dictated by
the parities of $k_{R}$ and $k_{I}$. The nonzero mean part therefore does not break the antiunitary spectral
constraints in Table~\ref{tab:nonHermitian_symmetry_classification}.

\subsection{Spectral bulk and locality regimes}\label{sec:Spectrum of the disordered Hamiltonian}

\subsubsection{Elliptic law}
\begin{figure}
    \centering
    \includegraphics[width=\linewidth]{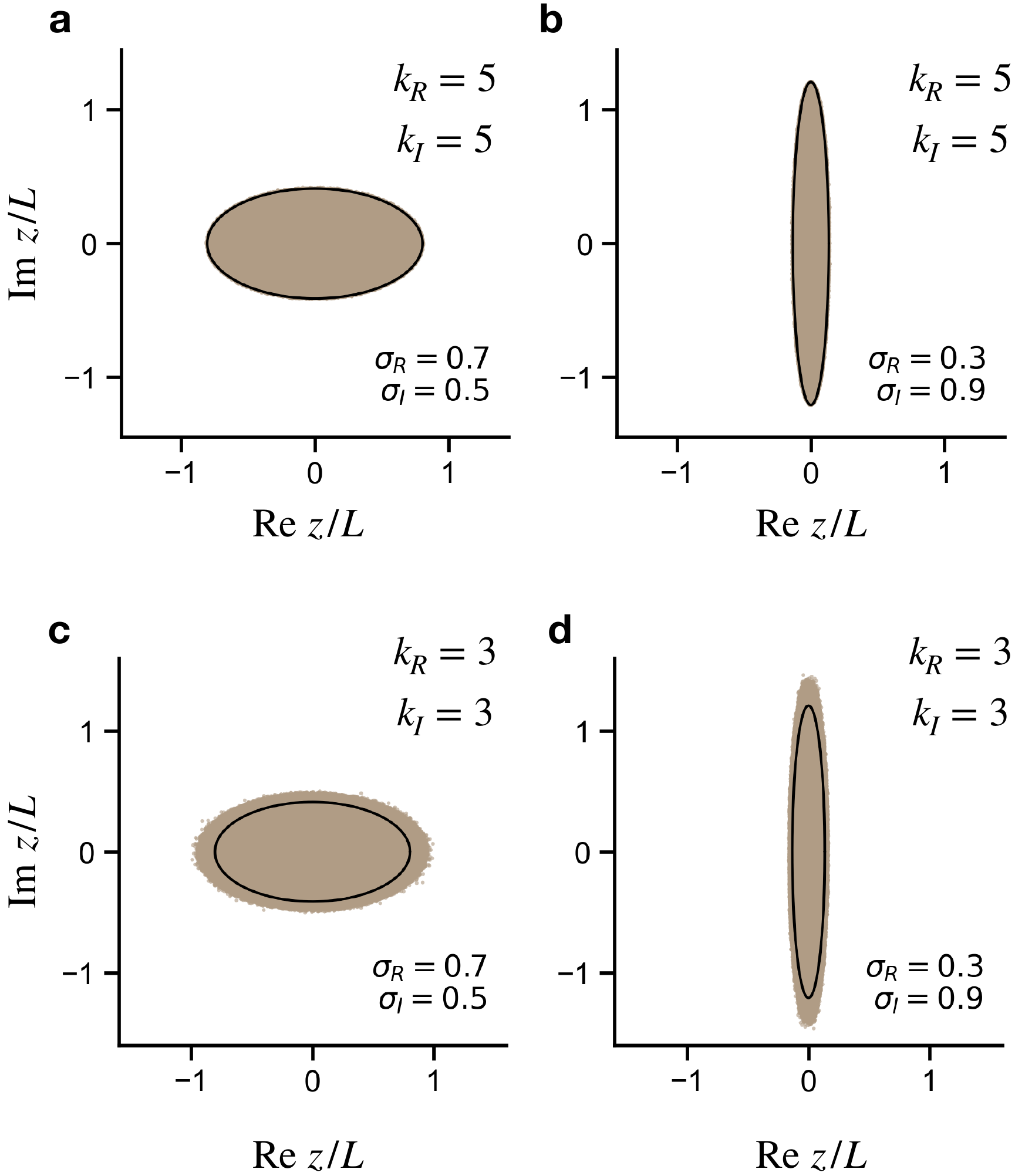}
\caption{
Validity of the elliptic-law description of the spectral bulk as a function of locality $k$ for $L=12$.
(a),(b) $k_{R}=k_{I}=5$, averaged over $64$ disorder realizations; 
(c),(d) $k_{R}=k_{I}=3$, averaged over $128$ disorder realizations.
The left column corresponds to $\sigma_{R}=0.7$ and $\sigma_{I}=0.5$ 
($\tau\simeq 0.32$), while the right column corresponds to 
$\sigma_{R}=0.3$ and $\sigma_{I}=0.9$ ($\tau=-0.8$).
The black solid curves denote the elliptic-law boundary predicted by
Eq.~\eqref{eq:elliptic_law_main}, with the eigenvalues rescaled by $L$.
For $k_{R}=k_{I}=5$ [panels (a),(b)], the numerical spectral support is
well described by the predicted ellipse, with only approximately
$0.4\%$ and $0.95\%$ of the eigenvalues lying outside the predicted
boundary, respectively.
In contrast, for $k_{R}=k_{I}=3$ [panels (c),(d)], pronounced spectral
tails extend beyond the elliptic boundary, with approximately
$14.3\%$ and $16.3\%$ of the eigenvalues lying outside it,
respectively. This breakdown of a hard elliptic edge is consistent
with the small-locality regime $k\ll\sqrt{L}$.
}\label{fig:elliptic_law_numerical}
\end{figure}

Having classified the random-matrix ensembles of the zero-mean part of the Hamiltonian, we now investigate the distribution of its eigenvalues. We start by computing the disorder-averaged positive and signed second trace moments of $G$,
\begin{equation}
    m_{11} = \frac{1}{N}\left\langle \mathrm{Tr}\left(GG^{\dagger}\right)\right\rangle, \qquad
    m_{20} = \frac{1}{N}\left\langle \mathrm{Tr}\left(G^{2}\right)\right\rangle ,
    \label{eq:trace_moments_definition}
\end{equation}
where $N = 2^{L}$ and $\langle \cdot \rangle$ denotes the average over disorder realizations. A full derivation using the Pauli-string expansion of $H_{R}$ and $H_{I}$ is given in Appendix~\ref{app:trace_moments}; the result, for $k_{R}=k_{I}=k$, is
\begin{equation}
    m_{11} = \frac{L^{2}}{2}\left[\sigma_{R}^{2} + \sigma_{I}^{2} + (\mu_{R}^{2}+\mu_{I}^{2})/\binom{L}{k}\right],
    \label{eq:m11_main}
\end{equation}
\begin{equation}
    m_{20} = \frac{L^{2}}{2}\left[\sigma_{R}^{2} - \sigma_{I}^{2} + (\mu_{R}+i\mu_{I})^{2}/\binom{L}{k}\right].
    \label{eq:m20_main}
\end{equation}
In a standard elliptic ensemble, the entries of an $N\times N$ random-matrix $X$ satisfy $\langle X_{ij}\rangle = \langle X_{ji}\rangle = 0$, $\langle X_{ij}^{2}\rangle = \langle X_{ji}^{2}\rangle = 1$, and $\langle X_{ij}X_{ji}\rangle = \rho$ for $i\neq j$ and $|\rho|<1$~\cite{Girko1986EllipticLaw, naumov2012ellipticlawrealrandom, nguyen2014ellipticlaw, alexeev2016singularvaluesdistributionsquares}. The eigenvalues of such matrices fill an ellipse in the complex plane, hence the name. Our exactly $k$--local ensemble does not have i.i.d. entries, so it is not an elliptic ensemble at the level of individual matrix elements. The Pauli-string structure correlates the matrix elements of $G$. Nevertheless, we show that we can often use the elliptic ensemble as an effective description of the zero-mean spectral bulk, with effective scale and ellipticity set by the two trace moments,
\begin{equation}
    s = \sqrt{m_{11}} = L\sqrt{\frac{\sigma_{R}^{2}+\sigma_{I}^{2}}{2}}, \qquad
    \tau = \frac{m_{20}}{m_{11}} = \frac{\sigma_{R}^{2}-\sigma_{I}^{2}}{\sigma_{R}^{2}+\sigma_{I}^{2}} .
    \label{eq:elliptic_scale_ellipticity}
\end{equation}
For a standard elliptic ensemble, the spectral support is bounded by the ellipse
\begin{equation}
    \left(\frac{\mathrm{Re}\;z}{a_R}\right)^2
    +
    \left(\frac{\mathrm{Im}\;z}{a_I}\right)^2
    = 1,
    \;
    a_R=s(1+\tau),\;
    a_I=s(1-\tau),
    \label{eq:elliptic_law_main}
\end{equation}
with the eigenvalues filling its interior in the large-matrix limit. We test this effective elliptic-law description numerically in Fig.~\ref{fig:elliptic_law_numerical} for two representative disorder anisotropies, using $L=12$ and $k_{R}=k_{I}=k$ with $k=3$ and $k=5$. For $k\gtrsim \sqrt{L}$, the numerical spectral support is well captured by the elliptic boundary in Eq.~\eqref{eq:elliptic_law_main}. By contrast, for $k\lesssim \sqrt{L}$, the spectrum develops appreciable tails beyond the nominal ellipse, indicating the absence of a hard spectral edge. This behavior is consistent with the known Gaussian-to-semicircle crossover in exactly $k$--local quantum spin-glass Hamiltonians: the density of states is approximately Gaussian for $k\ll\sqrt{L}$ and approaches the Wigner semicircle law for $k\gg\sqrt{L}$~\cite{Erd_s_2014}. Since the random part of our non-Hermitian ensemble is constructed from two independent exactly $k$--local Hermitian matrices, the same locality-controlled crossover naturally manifests itself in the shape and sharpness of the complex spectral boundary.
\subsubsection{Ginibre universality of the bulk}\label{sec:Ginibre universality of the bulk}
For $\sigma_{R}=\sigma_{I}=\sigma$ and $\mu_{R}=\mu_{I}=0$, the ellipticity vanishes, $\tau = 0$, and Eq.~\eqref{eq:elliptic_law_main} reduces to the circular-law disk of radius $L\sigma$, with $m_{11} = L^{2}\sigma^{2}$. Girko's circular law strictly applies to i.i.d. Ginibre matrices, so we again use it as a random-matrix benchmark rather than an exact prediction. We test it against two independent diagnostics: the cumulative radial eigenvalue distribution, compared with the circular law, and the complex level-spacing ratio, compared with Ginibre-ensemble statistics.

\textit{Girko's circular law:} The eigenvalues of an $N\times N$ random-matrix with independent and identically distributed entries of variance $1/N$ lie in the unit disk $|z|=1$ as $N\to\infty$~\cite{tao2014outliersspectrumiidmatrices, Girko1986EllipticLaw, tao2008randommatricescircularlaw, tao2009randommatricesuniversalityesds, meckes2021eigenvaluesrandommatrices}, with cumulative radial probability
\begin{equation}
    P(|z|<r) = \begin{cases} r^{2}, & 0\leq r\leq 1, \\ 1, & r\geq 1. \end{cases}
    \label{eq:circular_law_cdf_main}
\end{equation}

\textit{Complex level spacing ratios:} For a set of eigenvalues $\{z_{n}, n=1,\ldots,2^{L}\}$, the complex level spacing ratio is defined as~\cite{PhysRevX.10.021019}
\begin{equation}
    u_{n} = \frac{z_{n}^{\mathrm{NN}}-z_{n}}{z_{n}^{\mathrm{NNN}}-z_{n}} ,
    \label{eq:complex_spacing_ratio_main}
\end{equation}
where $z_{n}^{\mathrm{NN}}$ and $z_{n}^{\mathrm{NNN}}$ denote the nearest and next-nearest neighbors of $z_{n}$ under the Euclidean distance. By construction $|u_{n}|\leq 1$, so the $u_{n}$ fill the unit disk. For an uncorrelated Poisson spectrum, the $u_{n}$ are uniformly distributed over this disk, giving a radial density $P_{\mathrm{Poi}}(r) = 2r$, $r=|u_{n}|$, and a uniform angular density $P_{\mathrm{Poi}}(\theta) = 1/(2\pi)$, $\theta = \arg u_{n}$. Level repulsion in a correlated, Ginibre-type spectrum instead suppresses $P(r)$ at small $r$ and $P(\theta)$ near $\theta=0$, distinguishing Ginibre statistics from an uncorrelated Poisson spectrum~\cite{PhysRevX.10.021019}.

\begin{figure}
    \includegraphics[width=\linewidth]{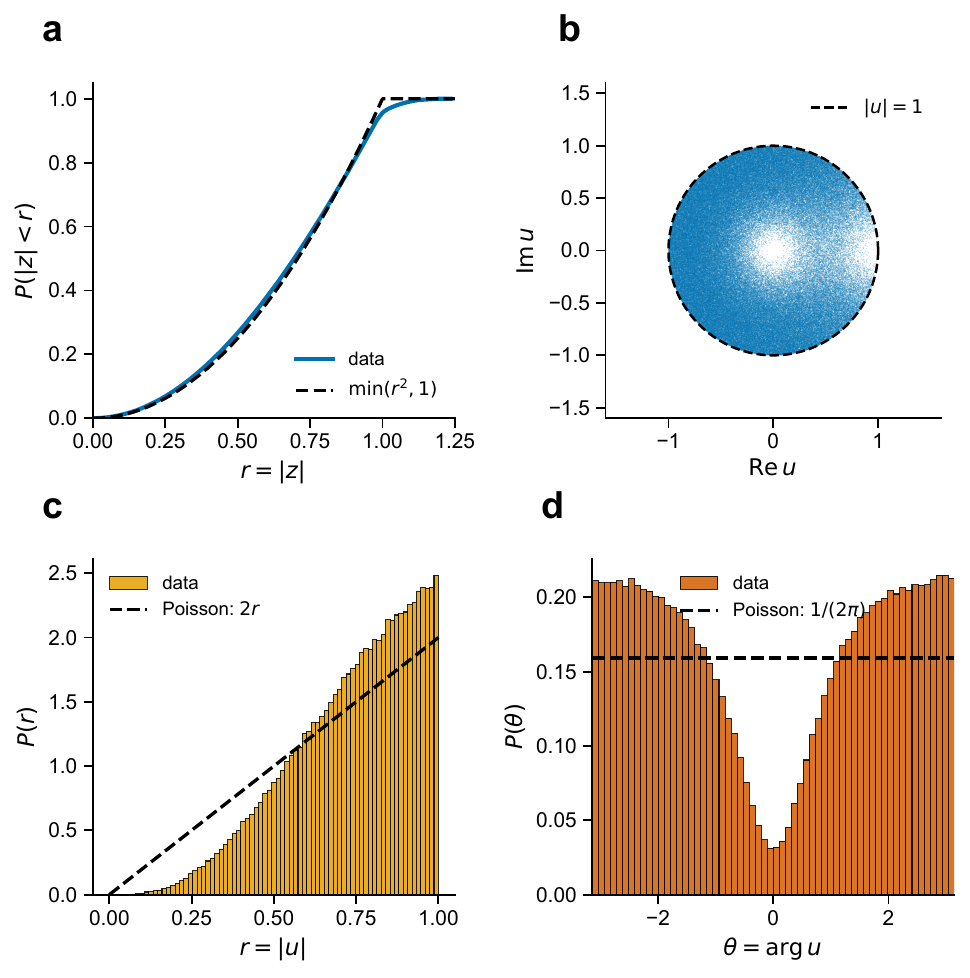}
    \caption{Numerical verification of Ginibre universality of the completely disordered part $G(\mu_{R}=\mu_{I}=0, \sigma_{R}=\sigma_{I}=1)$ for $L=12$, $k_{R}=k_{I}=3$, averaged over $128$ random disorder realizations. (a) Cumulative radial distribution $P(|z|<r)$ of the eigenvalues, compared with the flat-disk prediction $\min(r^{2},1)$ from Girko's circular law. (b) Complex spacing ratios $u_{n}$ [Eq.~\eqref{eq:complex_spacing_ratio_main}] in the complex plane. (c) Radial density $P(r)$, $r=|u|$, compared with the Poisson prediction $P_{\mathrm{Poi}}(r)=2r$. (d) Angular distribution $P(\theta)$, $\theta=\arg u$, compared with the uniform Poisson value $1/(2\pi)$.}
    \label{fig:complex_spacing_ratio_ginibre}
\end{figure}

We verify Ginibre universality for $L=12$, $k_{R}=k_{I}=3$ in Fig.~\ref{fig:complex_spacing_ratio_ginibre}, finding clear agreement with Ginibre predictions across all four diagnostics. This agreement is a nontrivial check of the symmetry classification in Table~\ref{tab:nonHermitian_symmetry_classification}, which fixes only the universality class, not whether the correlated, non-i.i.d.\ entries of our $k$--local construction reproduce the corresponding statistics quantitatively.
\subsection{Outlier trajectory}\label{sec:Outlier and dominant eigenvalues}
In this section we derive the analytical expressions for the outlier eigenvalues when the deterministic and the disordered parts are added together. We start by writing Eq.~\eqref{eq:Mu-sigma-decomposition} in the eigenbasis of $G_{0}$,
\begin{align}
    G = D + \frac{s_{R}}{\sqrt{2}} U^{\dagger} H_{R}(0, 1) U + i \frac{s_{I}}{\sqrt{2}} U^{\dagger} H_{I}(0,1) U . \label{eq:Complete_G_in_eigenbasis_of_G_0}
\end{align}
For notational simplicity, we define
\begin{subequations}
\begin{eqnarray}
    A_{k_{R}} &\equiv& \frac{1}{\sqrt{3^{k_{R}}\binom{L}{k_{R}}}} U^{\dagger} H_{R}(0,1)\; U, \\
    B_{k_{I}} &\equiv& \frac{1}{\sqrt{3^{k_{I}}\binom{L}{k_{I}}}} U^{\dagger} H_{I}(0,1)\; U.
\end{eqnarray} \label{eq:A_B_under_unitary_conjugations}
\end{subequations}
The matrices $A_{k_{R}}$ and $B_{k_{I}}$ are the disordered Hermitian
matrices written in the eigenbasis of $G_0$. Since $U$ is fixed by the deterministic Hamiltonian and is independent of the random couplings, this change of basis does not alter the underlying antiunitary symmetry class (see the proof in Appendix~\ref{sec:Invariance of the random-matrix ensembles under unitary conjugation}). Therefore, from now on we treat $A_{k_{R}}$ and $B_{k_{I}}$ as belonging to the same random-matrix universality class as $H_{R}(0,1)$ and $H_{I}(0,1)$, respectively, in Table~\ref{tab:nonHermitian_symmetry_classification}.

In the eigenbasis of $G_{0}$, we write the non-Hermitian ensemble in Eq.~\ref{eq:Mu-sigma-decomposition} as $(\mu_{R} = \mu_{I} = 1)$
\begin{eqnarray}
    G(\sigma_{R}, \sigma_{I}) &=& D + \frac{L}{\sqrt{2}} \bigg[\sigma_{R} A_{k_{R}} + i\sigma_{I} B_{k_{I}}\bigg].\label{eq:G_in_eigenbasis_of_deterministic}
\end{eqnarray}
Random matrices of this form, in which a Hermitian or a non-Hermitian random-matrix is perturbed by a diagonal matrix, can be analyzed using deformed random-matrix theory~\cite{Doussal_2025, tao2014outliersspectrumiidmatrices, Rochet2017, han2025outliersboundedrankperturbation, PhysRevLett.117.224101, orourke2014lowrankperturbationslarge}. In the absence of $D$, the eigenvalues of $\sigma_{R} A_{k_{R}} + i \sigma_{I} B_{k_{I}}$ lie densely together. For nonzero $D$, the matrix $D$ acts as a perturbation to the random-matrix and gives rise to outlier eigenvalues that are detached from the bulk of the spectrum. The exact locations of these outliers depend on the specific form and relative strength of the perturbation $D$ compared to $A_{k_{R}}, B_{k_{I}}$.

We now apply results on low-rank deformations of elliptic random matrices to Eq.~\eqref{eq:G_in_eigenbasis_of_deterministic}, specifically those of Ref.~\cite{orourke2014lowrankperturbationslarge}: for a bounded-rank deterministic perturbation $D$ added to an elliptic bulk, every eigenvalue $d_{a}$ of $D$ satisfying $|d_{a}| > s$ produces an isolated outlier at
\begin{equation}
    z_{\mathrm{out}} = d_{a} + \frac{\tau s^{2}}{d_{a}} ,
    \label{eq:General_outlier_formula}
\end{equation}
where $s$ and $\tau$ are the scale and ellipticity of the bulk, whose limiting spectral distribution is unchanged by the perturbation. In Eq.~\eqref{eq:G_in_eigenbasis_of_deterministic}, however, the perturbation $D$ has full rank $N = 2^{L}$, so these results do not apply directly. We use Eq.~\eqref{eq:General_outlier_formula} for the location of the outlier generated by $d_{0}$ as an effective description rather than a rigorous result and verify it numerically in Fig.~\ref{fig:Sigma_I_outlier_numerical_verification}.

\begin{figure}
    \centering
    \includegraphics[width=\linewidth]{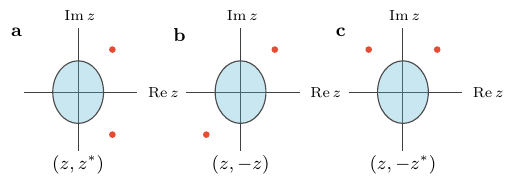}
    \caption{Schematic illustration of the three possible symmetry-related configurations of a pair of isolated outlier eigenvalues in the complex spectrum: (a) $(z,z^{*})$, (b) $(z,-z)$, and (c) $(z,-z^{*})$. The shaded region denotes the bulk spectral support, while the red points indicate the isolated outlier eigenvalues.
}
    \label{fig:Three_symmetry_case}
\end{figure}

The deterministic deformation can generate outlier eigenvalues, including the symmetry-related partners (see Table~\ref{tab:nonHermitian_symmetry_classification}). We illustrate these three cases in Fig.~\ref{fig:Three_symmetry_case}. This gives us two possible scenarios---a single outlier with the largest imaginary part (Figs.~\ref{fig:Three_symmetry_case}(a) and~\ref{fig:Three_symmetry_case}(b)), and two outliers with the same largest imaginary part ( Fig.~\ref{fig:Three_symmetry_case} (c)).
In the single outlier case the eigenvalue $d_{0} = L(1+i)/\sqrt{2}$ produces the outlier eigenvalue with the largest imaginary part. Substituting $d_{0}$ and the ellipticity parameters from Eq.~\eqref{eq:elliptic_scale_ellipticity} in Eq.~\eqref{eq:General_outlier_formula}, we get the outlier trajectory
\begin{align}
    z_{\mathrm{out}}(\sigma_{R}, \sigma_{I}) = \frac{L}{\sqrt{2}}\Bigg[ 1 + \frac{\sigma^{2}_{R} - \sigma^{2}_{I}}{2} + i\bigg( 1 + \frac{\sigma^{2}_{I} - \sigma^{2}_{R}}{2}\bigg)\Bigg]\label{eq:Outlier_trajectory_general}
\end{align}
In the case of two outliers with degenerate imaginary parts, the other outlier comes from $d_{L}$ and is given by $-z^{*}_{\mathrm{out}}(\sigma_{R}, \sigma_{I})$. The outlier exists (remains separated from the bulk eigenvalues) for $|d_{0}| > s$, equivalently~\cite{orourke2014lowrankperturbationslarge}
\begin{eqnarray}
    \sigma^{2}_{R} + \sigma^{2}_{I} < 2 \label{eq:General_outlier_existence_threshold}
\end{eqnarray}
Other deterministic eigenvalues in $D$ can also generate secondary
outliers. In particular, within the same deformed-random-matrix
estimate, the \textcolor{red}{eigenvalue} $d_{1}$ produces an outlier provided $\sqrt{\sigma_{R}^2+\sigma_{I}^2}<
   \sqrt{2}\,|d_{1}|/{L}
    \label{eq:d1_outlier_existence}$. Appendix~\ref{sec:Secondary outliers never become dynamically dominant}
shows that, whenever a secondary outlier exists, its imaginary part
remains below that of the leading $d_{0}$ outlier. Once the upper bulk
edge overtakes the $d_{0}$ branch, it therefore also lies above every
secondary branch. Consequently, the dominant eigenvalue always comes
from either the leading $d_{0}$ outlier branch---including any
symmetry-related degenerate partner---or the upper bulk edge.
\begin{figure}
    \centering
    \includegraphics[width=\linewidth]{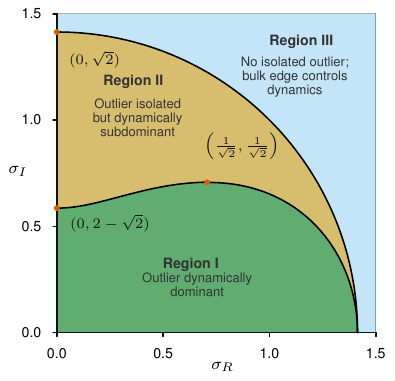}
        \caption{
Phase diagram in the $(\sigma_{R},\sigma_{I})$ plane showing the distinction
between outlier existence and dynamical dominance. The outer circular
boundary, $\sigma_{R}^2+\sigma_{I}^2=2$, marks the disappearance of the isolated
outlier. The inner curve marks the dynamical-dominance transition, where the
upper edge of the bulk overtakes the outlier in imaginary part. Region I has
an isolated outlier that is also dynamically dominant. Region II has an
isolated outlier that is dynamically subdominant. Region III has no isolated
outlier. The marked points indicate the pure anti-Hermitian dominance
transition $(0,2-\sqrt{2})$, the balanced-disorder transition
$(1/\sqrt{2},1/\sqrt{2})$, and the pure anti-Hermitian outlier-absorption
point $(0,\sqrt{2})$.
    }
    \label{fig:phase_diagram_sigmaR_sigmaI}
\end{figure}

\section{Phase diagram: existence versus dominance}\label{sec:Dynamical dominance and phase diagram}
As mentioned earlier, the outlier $z_{\rm out}$ and the eigenvalue with the largest imaginary part, $z_{\star}$, need not be the same eigenvalue. For weak disorder, the outlier sits well above the bulk in imaginary part, so $z_{\star} = z_{\rm out}$ and the outlier alone controls the long-time state. As $\sigma_{R}$ and $\sigma_{I}$ grow, the bulk edge rises while the outlier's position \textcolor{red}{grows slower than the bulk along the imaginary direction}, so at some critical combination of disorder strengths the bulk edge overtakes the outlier along the imaginary direction. Beyond that point $z_{\star}$ switches to an eigenvalue at the upper bulk edge, even though $z_{\rm out}$ can remain spectrally isolated. The physical consequences of this switch are not yet visible from the spectrum alone, but they become clear in Sec.~\ref{sec:Outlier-to-bulk transitions of dominant eigenvalue} where we show that the outlier and bulk eigenstates carry sharply different entanglement and magic, so the point at which $z_{\star}$ switches branch is exactly the point at which these two quantities undergo an abrupt change.

By the elliptic-law description of Sec.~\ref{sec:Spectrum of the disordered Hamiltonian}, in the $k_{R}, k_{I} \gg \sqrt{L}$ limit, the disordered part of $G$ fills an ellipse with semi-axes $a_R=s(1+\tau)$, $a_I=s(1-\tau)$ [Eq.~\eqref{eq:elliptic_scale_ellipticity}], giving the edge of the bulk along the imaginary direction
\begin{eqnarray}
    \mathrm{Im}\; z^{\rm max}_{\rm bulk} &=& \frac{\sqrt{2}\;L\;\sigma_{I}^2}{\sqrt{\sigma_{R}^2+\sigma_{I}^2}} ,
    \label{eq:Imaginary_semiaxis_general}
\end{eqnarray}
which is the largest imaginary part attained anywhere in the bulk. To determine the disorder strengths at which the dominant eigenvalue switches from the outlier to the bulk or vice versa, we compare Eq.~\eqref{eq:Outlier_trajectory_general} with Eq.~\eqref{eq:Imaginary_semiaxis_general}, which gives
\begin{eqnarray}
    1 + \frac{\sigma^{2}_{I} - \sigma^{2}_{R}}{2} &=& \frac{2 \sigma^{2}_{I}}{\sqrt{\sigma^{2}_{R} + \sigma^{2}_{I}}} .
\end{eqnarray}
This implicit equation can be solved in terms of polar coordinates: $\sigma_{R} = r \cos\theta, \sigma_{I} = r \sin\theta$, which gives the solution
\begin{eqnarray}
    \sigma_{R}(\theta) = r(\theta) \cos\theta,\;\;
    \sigma_{I}(\theta) = r(\theta) \sin\theta, \label{eq:Dominance_curve_equation}
\end{eqnarray}
where $0 \leq \theta \leq \pi/2$ and
\begin{eqnarray}
    r(\theta) &=& \frac{2}{1 - \cos(2\theta) + \sqrt{1 + \cos^{2}(2\theta)}} .
\end{eqnarray}
Eq.~\eqref{eq:General_outlier_existence_threshold} and Eq.~\eqref{eq:Dominance_curve_equation} split the $(\sigma_{R},\sigma_{I})$ plane into three regions as shown in Fig.~\ref{fig:phase_diagram_sigmaR_sigmaI}. Region~I, outlier isolated and dominant ($z_{\star}=z_{\rm out}$); Region~II, outlier isolated but $z_{\star}$ belongs to the bulk; Region~III, no isolated outlier.

\begin{figure}
    \centering
    \includegraphics[width=\linewidth]{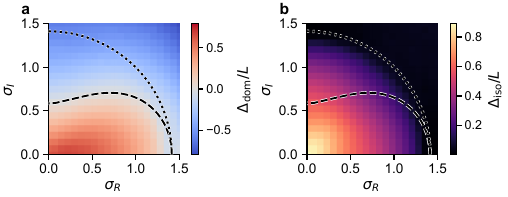}
    \caption{Numerical diagnostics of the outlier-to-bulk transition for $L=10$, $k_{R}=k_{I}=5$, $\mu_{R}=\mu_I=1$ and averaged over $30$ disorder realizations per point. (a) Dominance gap $\overline{\Delta_{\rm dom}}/L$. (b) Isolation gap $\overline{\Delta_{\rm iso}}/L$. The dashed curve is the analytic dominance boundary and the dotted curve is the analytic existence boundary from Eq.~\eqref{eq:General_outlier_existence_threshold}.}
    \label{fig:combined_phase_diagram}
\end{figure}

Next we test these three regions numerically. We define the dominance gap 
\begin{eqnarray}
\Delta_{\rm dom} &=& \mathrm{Im}\;z_{\rm out}-\mathrm{Im}\;z^{\rm max}_{\rm bulk},\label{eq:Dominance_gap_definition}
\end{eqnarray}
which measures the gap between the outlier and the bulk along the imaginary direction, and the isolation gap
\begin{eqnarray}
\Delta_{\rm iso}=|z_{\rm out}-z_{\rm bulk}|, \label{eq:Isolation_gap_definition}    
\end{eqnarray} 
the distance in the complex-plane from $z_{\rm out}$ to the nearest bulk eigenvalue. Figure~\ref{fig:combined_phase_diagram} shows the disorder averaged quantities $\overline{\Delta}_{\rm dom}/L$ and $\overline{\Delta}_{\rm iso}/L$ against the analytical boundaries.
\section{Outlier-to-bulk transitions of dominant eigenvalue}\label{sec:Outlier-to-bulk transitions of dominant eigenvalue}
In Sec.~\ref{sec:Dynamical dominance and phase diagram}, we derived the analytical boundaries for dynamical dominance and existence of the outlier eigenvalue for general $(\sigma_{R},\sigma_{I})$. We now solve it along three rays through the phase diagram---pure anti-Hermitian disorder $(\sigma_{R} = 0)$, pure Hermitian disorder $(\sigma_{I} = 0)$ and mixed disorder $(\sigma_{R} = \sigma_{I} = \sigma)$, chosen to isolate the role of the bulk's growth direction relative to the outlier.
\subsection{Purely anti-Hermitian disorder}
In this section, we consider the case when $\sigma_{R} = 0$, for which Eq.~\eqref{eq:G_in_eigenbasis_of_deterministic} reduces to
\begin{eqnarray}
    G(\sigma_{I}) &=& D + i\frac{L\sigma_{I}}{\sqrt{2}} B_{k_{I}}. \label{eq:G_sigma_I_equation}
\end{eqnarray}
Let us first consider the case when the outlier is unique, then
from Eq.~\eqref{eq:Outlier_trajectory_general}, we get the outlier trajectory
\begin{eqnarray}
    z_{\mathrm{out}}(\sigma_{I}) &=& \frac{L}{\sqrt{2}}\Bigg[ 1 - \frac{\sigma^{2}_{I}}{2} + i\bigg(1 + \frac{\sigma^{2}_{I}}{2} \bigg)\Bigg] ,\label{eq:General_outlier_sigma_I}
\end{eqnarray}
which is valid for (Eq.~\eqref{eq:General_outlier_existence_threshold})
\begin{eqnarray}
    \sigma_{I} < \sqrt{2} \equiv \sigma^{\mathrm{outlier}}_{I, c}. \label{eq:Critical_disorder_outlier_bulk_merge}
\end{eqnarray}
This is the disorder threshold below which the isolated outlier branch associated with $d_{0}$ exists. For $k_{I} \ll \sqrt{L}$ the outlier threshold is less sharp due to the long tails of the eigenvalues beyond the elliptic law boundary.

\begin{figure}
    \centering
    \includegraphics[width=\linewidth]{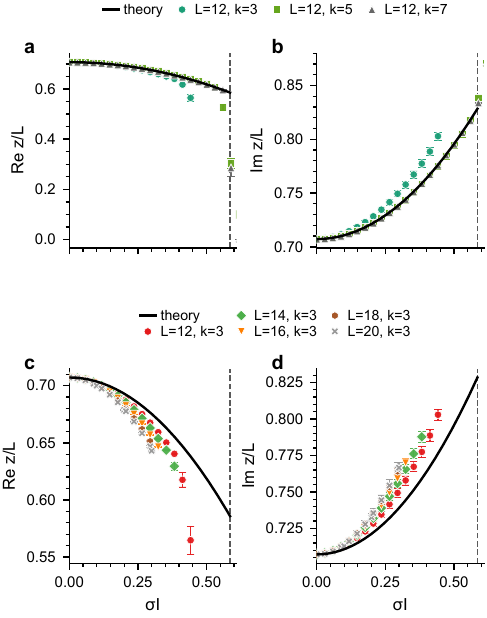}
    \caption{
Comparison between the analytical outlier formula (Eq.~\eqref{eq:General_outlier_sigma_I}) and numerically tracked eigenvalues for purely anti-Hermitian disorder ($\sigma_{R} = 0$). Panels (a) and (b) show $\mathrm{Re}[z]/L$ and $\mathrm{Im}[z]/L$ for fixed system size $L = 12$ and varying locality $k = 3, 5, 7$. Panels (c) and (d) show the same quantities for fixed locality $k = 3$ and varying system size $L = 12$ to $20$. The solid black line is the analytical prediction. The markers show the eigenvalue with the largest imaginary part, averaged over  120 random disorder realizations. Error bars show the standard error of the mean. The dashed vertical line marks $\sigma^{\mathrm{dom}}_{I, c} = 2 - \sqrt{2}$, the dynamical dominance threshold. Agreement with theory is strong at small $\sigma_{I}$. Deviations grow near the dominance threshold, and they grow faster for smaller $k$ and larger $L$. This happens because the theory line follows the isolated outlier branch, while the markers show the largest imaginary eigenvalue in the full spectrum. As the bulk edge approaches the outlier, rare disorder realizations push the largest eigenvalue above the outlier before the average threshold is reached. This effect is strongest when $k$ is small compared to $\sqrt{L}$, where the bulk has long Gaussian tails instead of a sharp semicircle edge. Larger $L$ increases the number of bulk eigenvalues, which increases the chance of an early extreme value, so the deviation grows with system size.}
    \label{fig:Sigma_I_outlier_numerical_verification}
\end{figure}

There is an alternative way to derive the outlier trajectory and the critical disorder threshold by writing Eq.~\eqref{eq:G_sigma_I_equation} as
\begin{eqnarray}
    -i G(\sigma_{I}) &=& -i D + \frac{L \sigma_{I}}{\sqrt{2}} B_{k_{I}}.
\end{eqnarray}
This equation can be viewed as a Hermitian random-matrix ensemble deformed by a non-Hermitian perturbation $-iD$, a setting studied in Ref.~\cite{Rochet2017}. We find that the outlier equation obtained using the results of Ref.~\cite{Rochet2017} agrees with the trajectory derived above. In Fig.~\ref{fig:Sigma_I_outlier_numerical_verification}, we compare our analytical outlier formula against numerical results for different cases of $L$, $k_{R}$ and $k_{I}$.

An interesting scenario arises when there are two distinct outliers with the same imaginary part---the case when $L, k_{I}$ are even and $k_{R}$ is odd.  The outliers are distributed according to  $(z(\sigma_{I}), -z^{*}(\sigma_{I}))$. This degeneracy in the imaginary part of the outlier gives rise to persistent oscillations inside this manifold at late times. To see this explicitly, suppose the corresponding eigenstates of $d_{0}$ and $d_{L}$ are $|R_{0}\rangle$ and $|R_{L}\rangle$. Then for an initial state of the form
\begin{eqnarray}
    |\psi(0)\rangle &\sim& c |R_{0}\rangle + (c+\delta)|R_{L}\rangle,
\end{eqnarray}
for $\delta \ll 1$ and $|c|\sim 1$, the long-time state of the system becomes
\begin{eqnarray}
    |\psi(t\to\infty)\rangle &\sim& c_{+} e^{-iL(1-\sigma^{2}_{I}/2)t/\sqrt{2}}|R_{0}\rangle +\nonumber\\
    && c_{-} e^{iL(1-\sigma^{2}_{I}/2)t/\sqrt{2}}|R_{L}\rangle.
\end{eqnarray}
This state shows persistent oscillation with oscillation frequency
\begin{eqnarray}
    \Omega (\sigma_{I}) &=& \sqrt{2} L \bigg(1 - \frac{\sigma^{2}_{I}}{2}\bigg).
\end{eqnarray}

\subsubsection{Dynamical dominance threshold}
We now determine the disorder threshold at which the outlier
eigenvalue ceases to be dominant. Because the elliptic law behaves
differently in the $k \gg \sqrt{L}$ and $k \ll \sqrt{L}$ regimes, we treat the
two cases separately.
\paragraph{Hard-edge regime, $k_{I}\gg\sqrt{L}$.}
In this regime the elliptic-law description of Sec.~\ref{sec:Spectrum of the disordered Hamiltonian} is expected to hold, and Eq.~\eqref{eq:Imaginary_semiaxis_general} evaluated at $\sigma_{R} = 0$ predicts the imaginary bulk edge,
\begin{equation}
    \mathrm{Im}\;z^{\rm max}_{\rm bulk} \approx \sqrt{2}\;L\;\sigma_{I} .
    \label{eq:Imaginary_part_eigenvalue_sigma_I}
\end{equation}
The same edge can be derived by noticing that $B_{k_{I}}$ is itself an exactly $k_{I}$--local Hermitian random-matrix, so by Ref.~\cite{Erd_s_2014} its density of states approaches the Wigner semicircle law for $k_{I}\gg\sqrt{L}$,
\begin{equation}
       \rho_{\mathrm{sc}}(E_{B}) = \begin{cases} \dfrac{1}{2\pi} \sqrt{4 - E^{2}_{B}} , & |E_{B}| < 2  \\[4pt]
       0, & \text{otherwise,}
\end{cases}
\label{eq:Wigner-semicircle_law_H_1_0}
\end{equation}
with a sharp cutoff at $\lambda_B\in[-2,2]$ in the thermodynamic limit~\footnote{At finite $L$, the eigenvalues near this edge show Tracy--Widom-type fluctuations~\cite{tracy1997distributionlargesteigenvaluegaussian}.}, so $\lambda_{\mathrm{max}}(B_{k_{I}})=2$ and
\begin{equation}
    \mathrm{Im}\; z^{\rm max}_{\mathrm{bulk}} \approx \frac{L \sigma_{I}}{\sqrt{2}}\;\lambda_{\mathrm{max}}(B_{k_{I}}) = \sqrt{2}\; L \sigma_{I} ,
\end{equation}
which is the same as Eq.~\eqref{eq:Imaginary_part_eigenvalue_sigma_I}.

Equating this bulk edge to the outlier trajectory Eq.~\eqref{eq:General_outlier_sigma_I}, gives the dominance threshold,
\begin{eqnarray}
    \frac{L}{\sqrt{2}}\left(1 + \frac{(\sigma^{\mathrm{dom}}_{I, c})^{2}}{2}\right) &=& \sqrt{2}\; L\; \sigma^{\mathrm{dom}}_{I,c}.
\end{eqnarray}
Solving for $\sigma^{\rm dom}_{I, c}$ gives
\begin{eqnarray}
    \sigma^{\mathrm{dom}}_{I,c} &=& 2 - \sqrt{2} \approx 0.586.    \label{eq:Critical_disorder_B_kI}
\end{eqnarray}
This is the disorder strength below which the outlier remains dynamically dominant. Above this disorder level, the dominant eigenvalue comes from the upper imaginary edge of the bulk.
\paragraph{Long-tail regime, $k_{I}\ll\sqrt{L}$.}
Here $B_{k_{I}}$ has an approximately Gaussian density of states with long tails rather than a hard semicircle edge, so the location of the largest eigenvalue is instead governed by extreme-value statistics. In Appendix~\ref{sec:max_eigenvalue_bound}, we derive the probabilistic upper bound
\begin{equation}
    \mathrm{Im}\;z^{\rm max}_{\rm bulk}
    \leq
    \sigma_I L \sqrt{L\ln 2+\ln(1/\delta)},
    \label{eq:Gaussian_edge}
\end{equation}
which holds with probability at least $1-\delta$. For any fixed failure probability $\delta$, the bound scales as $O(L^{3/2})$, in contrast to the $O(L)$ edge of the hard-edge regime. This scaling is unchanged even if the failure probability is chosen to be exponentially small in system size. In particular, taking $\delta=e^{-cL}$ gives
\begin{equation}
    \mathrm{Im}\;z^{\rm max}_{\rm bulk}
    \leq
    \sigma_I L^{3/2}\sqrt{\ln 2+c},
\end{equation}
so that only the prefactor changes, while the $L^{3/2}$ exponent remains the same. Comparing this scale with the $O(L)$ outlier trajectory therefore gives the characteristic crossover scale
\begin{equation}
    \sigma_{I,c}\sim L^{-1/2}.
    \label{eq:Sigma_c_scaling_approximate}
\end{equation}
\subsubsection{Dynamical dominance transition and long-time overlap}
We now look at the dynamical signature for the dominant eigenvalue switching from the outlier to the bulk at the critical disorder value. For each chosen \(\sigma_{I}\) value and each disorder
realization, we numerically track the outlier eigenvalue and the upper bulk edge eigenvalue. This gives us two right eigenstates:
\(|R_{\rm out}\rangle\) for the outlier and \(|R_{\rm bulk}\rangle\) for the upper edge of the bulk. We then time-evolve the system using
\begin{equation}
    |\psi(t)\rangle
    =
    \frac{
    e^{-i G t}|\psi(0)\rangle
    }{
    \|e^{-i G t}|\psi(0)\rangle\|
    } ,
    \label{eq:normalized_time_evolved_state_overlap_section}
\end{equation}
and then compute the final-time overlaps
\begin{subequations}
\begin{align}
    O_{\rm out}(t_f)
    &=
    \left|
    \langle R_{\rm out}|\psi(t_f)\rangle
    \right|^2, \\
    O_{\rm bulk}(t_f)
    &=
    \left|
    \langle R_{\rm bulk}|\psi(t_f)\rangle
    \right|^2 ,
\end{align}
\label{eq:outlier_bulk_final_overlap_definitions}
\end{subequations}
These overlaps provide a direct diagnostic of which right eigenstate the normalized dynamics approaches.

\begin{figure}
    \centering
    \includegraphics[width= \linewidth]{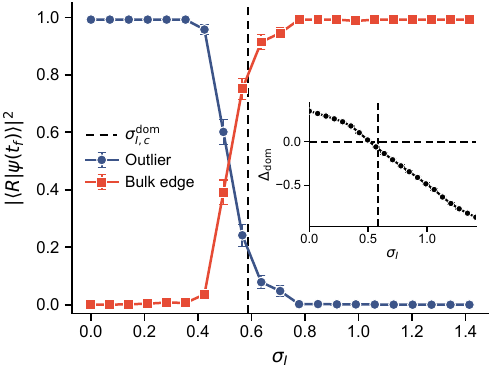}
\caption{
Disorder-averaged overlap of the normalized time-evolved state after $t = 3000$ with the outlier right eigenstate of $G(\sigma_{I})$ for \(L=12\), \(k_{R}=k_{I}=3\), and \(\mu_{R}=\mu_{I}=1\). The vertical dashed line marks
\(\sigma^{\mathrm{dom}}_{I,c}=2-\sqrt{2}\).  The long-time state switches from outlier
dominated to bulk-edge dominated near \(\sigma^{\mathrm{dom}}_{I,c}\). The inset shows the dominance gap $\Delta_{\mathrm{dom}} = \mathrm{Im}\;z_{\mathrm{out}} - \mathrm{Im}\;z^{\rm max}_{\rm bulk}$ which changes its sign at $\sigma_{I} = \sigma^{\mathrm{dom}}_{I,c}$ indicating the switch in the largest imaginary part from the outlier to the bulk edge. The error bars indicate the standard error of the mean over $128$ disorder realizations.}
\label{fig:final_overlap_transition}
\end{figure}

Figure~\ref{fig:final_overlap_transition} shows the disorder-averaged
final-time overlaps as a function of \(\sigma_{I}\).  For weak
anti-Hermitian disorder, the evolved state has nearly unit overlap with
the outlier and negligible overlap with the bulk edge.  Thus the
long-time state is outlier dominated.  As \(\sigma_{I}\) is increased, the
outlier overlap decreases while the bulk-edge overlap increases.  The
two curves cross near $\sigma^{\mathrm{dom}}_{I,c}$
shown by the vertical dotted line.  For \(\sigma_{I}>\sigma^{\mathrm{dom}}_{I,c}\), the
final state has nearly unit overlap with the bulk-edge eigenstate and
negligible overlap with the outlier.

The mechanism behind this switch is made explicit by the dominance gap defined in Eq.~\eqref{eq:Dominance_gap_definition}.
In the inset of Fig.~\ref{fig:final_overlap_transition}, we plot this dominance gap which shows
\(\Delta_{\rm dom}\) changes sign near the same value
\(\sigma^{\mathrm{dom}}_{I,c}\) signaling the loss of dynamical dominance of the outlier eigenvalue.
In the next section, we discuss the effect of this outlier-to-bulk switch on the entanglement entropy and magic of the state of the system at long times.
\subsubsection{Entanglement and magic signatures of dynamical dominance}
\label{sec:pure_anti_Hermitian_entanglement}

\begin{figure*}
\includegraphics[width=\linewidth]{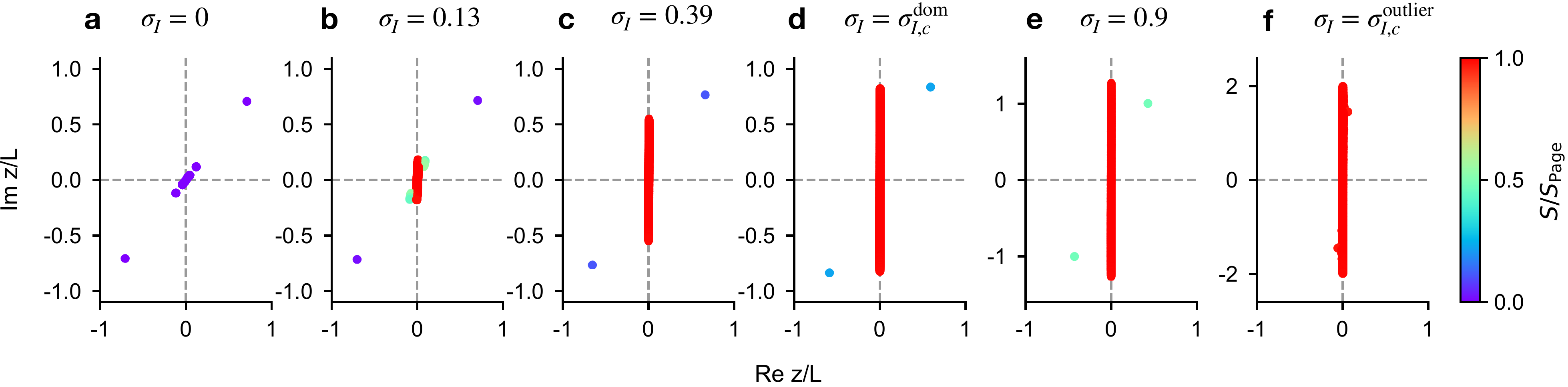}
    \caption{
Eigenvalue spectrum of the scaled non-Hermitian matrix \(G/L\) in the complex plane for
\(L=12\), \(k_{R}=k_{I}=5\), and \(\mu_{R}=\mu_I=1\), with increasing anti-Hermitian disorder
strength \(\sigma_{I}\), shown for a single set of random couplings.  We set \(\sigma_{R}=0\).  The horizontal and vertical axes show
\(\mathrm{Re}\;z/L\) and \(\mathrm{Im}\;z/L\), respectively.  At \(\sigma_{I}=0\), the spectrum
consists of the discrete deterministic eigenvalues of \(G_0/L\).  For nonzero \(\sigma_{I}\),
the random anti-Hermitian component broadens the spectrum into a bulk while the deterministic
outlier remains separated.  As \(\sigma_{I}\) is increased, the bulk extends along the imaginary
direction and eventually overtakes the tracked outlier in imaginary part. The colorbar represents the entanglement entropy for each right eigenstate calculated using Eq.~\eqref{eq:Entanglement_right_eigenstate}.
}
\label{fig:eigenvalues_G_sigma_I}
\end{figure*}
Having established when the dominant eigenvalue switches from the outlier to the bulk edge, we now discuss the physical consequences of that switch on the state of the system at long times. In Fig.~\ref{fig:eigenvalues_G_sigma_I}, we plot the spectrum of $G(\sigma_{I})$ for $L=12$, $k_{R} = k_{I} = 5$ for a single disorder realization, with each eigenvalue colored by the half-system entanglement entropy of the corresponding right eigenstate, normalized by the Page value $S_{\rm Page}=(L\ln2-1)/2$~\cite{Page_1993}. We observe that the bulk eigenstates carry  high entanglement typical of random-matrix eigenstates, while the outliers, tied to the deterministic eigenvalue $d_{0}$, remain weakly entangled throughout. Consequently, while the outlier is
dominant, the selected long-time state is weakly entangled. Once the
upper bulk edge overtakes the outlier, a highly entangled bulk
eigenstate becomes dominant and replaces it as the long-time state.
Thus, for an individual disorder realization, the asymptotic
entanglement changes abruptly at the outlier-to-bulk dominance
crossing.

To investigate this effect on the entanglement of the long-time state, we now turn to a systematic, disorder-averaged study. For numerics, we focus on even $L$ with $k_R=k_I=3$, where there is only one outlier with the largest imaginary part. For each disorder realization we identify $z_\star(\sigma_I)$, the eigenvalue with the largest imaginary part, and compute the half-system von Neumann entanglement entropy $S_{\star}$ of the corresponding normalized right eigenstate $|R_\star\rangle$ [Eq.~\eqref{eq:Entanglement_right_eigenstate}], averaged over independent cuts.

\begin{figure*}
    \centering
    \includegraphics[width=\linewidth]{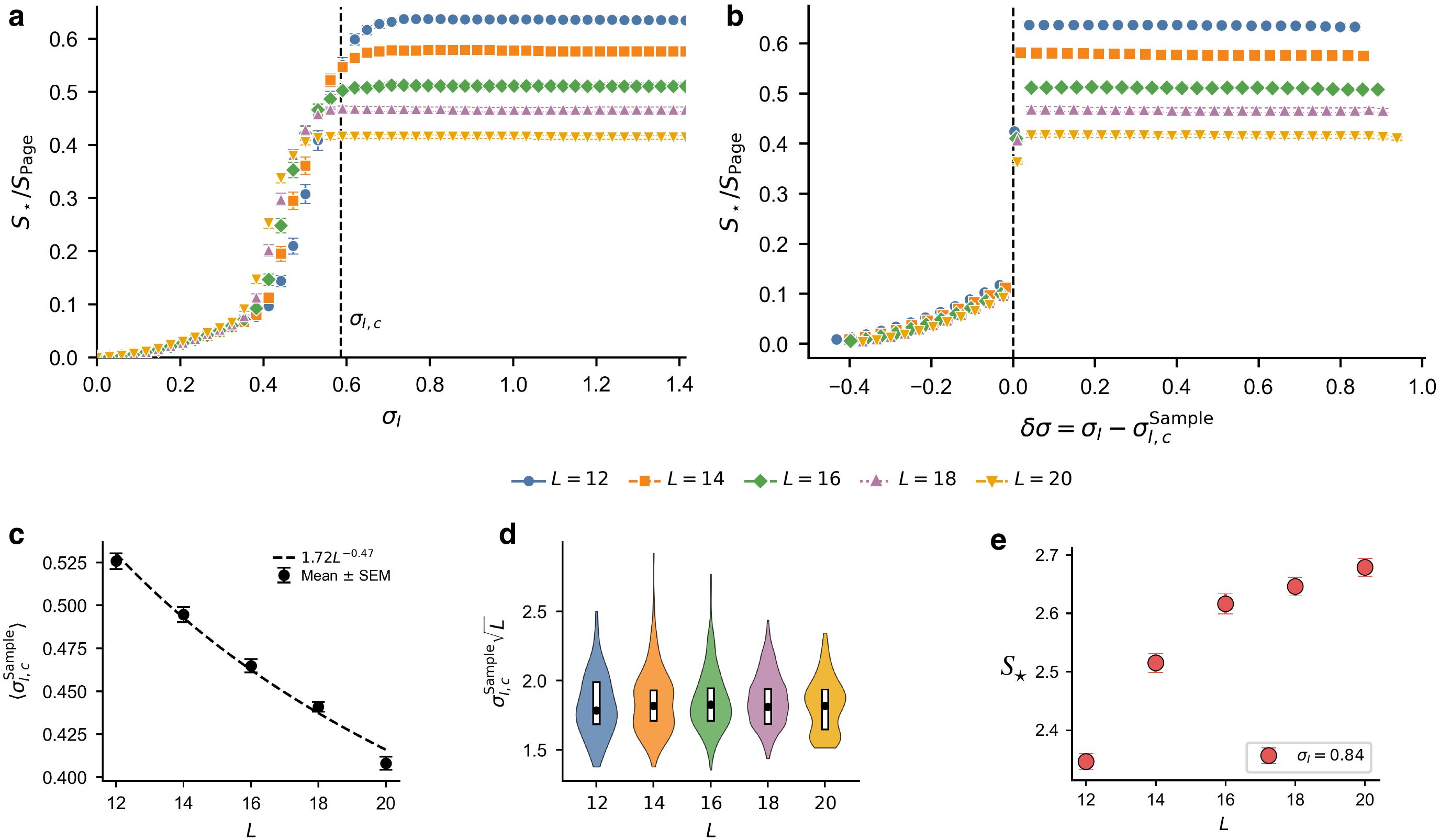}
\caption{
Entanglement of the long-time state for purely anti-Hermitian disorder with
$k_{R}=k_{I}=3$.
(a) Normalized entanglement entropy $S_{\star}/S_{\rm Page}$ as a function of the disorder strength $\sigma_{I}$ for $L=12$--$20$, averaged over $130$ disorder realizations. The vertical dashed line marks the analytical prediction
$\sigma_{I,c}^{\rm dom}=2-\sqrt{2}$ .
(b) The same data after realigning each disorder realization by its sample-dependent pseudocritical point,
$\delta\sigma_{I} =\sigma_{I} -\sigma_{I,c}^{\rm Sample}$.
(c) The pseudocritical point
$\langle\sigma_{I,c}^{\rm Sample}\rangle$ as a function of system size averaged over different disorder realizations. Error bars denote the standard error of the mean over $130$ disorder realizations. The dashed curve is a fit to
$aL^{-b}$, yielding $a=1.72$ and $b=0.47$, consistent with the asymptotic $L^{-1/2}$ scaling predicted by Eq.~\eqref{eq:Sigma_c_scaling_approximate}.
(d) Distributions of the sample-dependent pseudocritical points after rescaling by $\sqrt{L}$.
(e) Saturated post-transition entanglement entropy $S_{\star}$ as a function of $L$ at $\sigma_{I} = 0.84$. The monotonic growth is consistent with extensive entanglement. However, the accessible system sizes are insufficient to distinguish reliably between competing finite-size forms and no fit is therefore shown. Error bars in this panel denote the standard error of the mean over $600$ disorder realizations.}
\label{fig:sigma-c-sample-distribution}
\end{figure*}
Figure~\ref{fig:sigma-c-sample-distribution}(a) shows the disorder-averaged $S_{\star}/S_{\rm Page}$ against $\sigma_{I}$ for $L=12 , 14, 16, 18, 20$. The entanglement stays low below $\sigma^{\rm dom}_{I, c}=2-\sqrt2$ and climbs sharply above it. We observe that instead of a sharp jump, the disorder averaged entanglement varies continuously as a function of $\sigma_{I}$. This is because the bulk eigenvalues have long Gaussian tails in this locality regime $k_{I} \ll \sqrt{L}$. As a result the disorder strength at which a particular realization's bulk overtakes its outlier varies widely from sample to sample, and averaging over realizations makes the data continuous. To recover the intrinsic sharpness of the transition in each realization, we define a realization-dependent pseudocritical disorder $\sigma_{I,c}^{(\rm Sample)}$, the value of $\sigma_{I}$ at which $z_{\star}$ switches branch in that particular realization, and replot the entanglement against the shifted variable $\delta\sigma_{I} = \sigma_{I} -\sigma_{I,c}^{(\rm Sample)}$. This isolates the transition that each realization undergoes individually. Fig.~\ref{fig:sigma-c-sample-distribution}(b) shows the result: a sharp jump at $\delta\sigma=0$, in place of the smooth curve in Fig.~\ref{fig:sigma-c-sample-distribution}(a).

This entanglement transition has a different origin from other
entanglement transitions reported in non-Hermitian systems. Jian et al.
found a first-order entanglement transition tied to the Yang–Lee edge
singularity in $\mathcal{PT}$-symmetric Hamiltonians~\cite{PhysRevB.104.L161107}. Here there is no $\mathcal{PT}$
symmetry and no critical point in the usual sense, only a spectral
outlier losing a competition with the random-matrix bulk. The transition
is likewise distinct from that of Gopalakrishnan and Gullans~\cite{PhysRevLett.126.170503}, where
a postselected non-Hermitian Hamiltonian undergoes a spectral transition
at a single exceptional point at which $\mathcal{PT}$ symmetry is spontaneously broken.

The outlier-to-bulk switching of the dominant eigenvalue leaves an equally sharp signature in the magic of the long-time state. Magic transitions have so far been studied mainly as consequences of measurement and feedback---in monitored and error-corrected dynamics~\cite{Niroula2024}, measurement-only circuits~\cite{Tarabunga2025}, and related monitored settings~\cite{PRXQuantum.5.030332, wang2025magictransitionmonitoredfree, heinrich2026criticalbehaviorsmagicparticipation}, as well as in ground-state and equilibrium contexts~\cite{PhysRevA.106.042426, PhysRevA.106.062405}. A related
diagnostic role for magic has also been identified at exceptional points in
non-Hermitian spin chains~\cite{moca2025nonstabilizernessdiagnosticcriticalityexceptional} where the relevant transition is tied to a
single exceptional point. The mechanism here requires none of these ingredients. It requires only that an isolated spectral outlier lose dynamical dominance to a random matrix bulk eigenvalue.

\begin{figure}
    \centering
    \includegraphics[width=\linewidth]{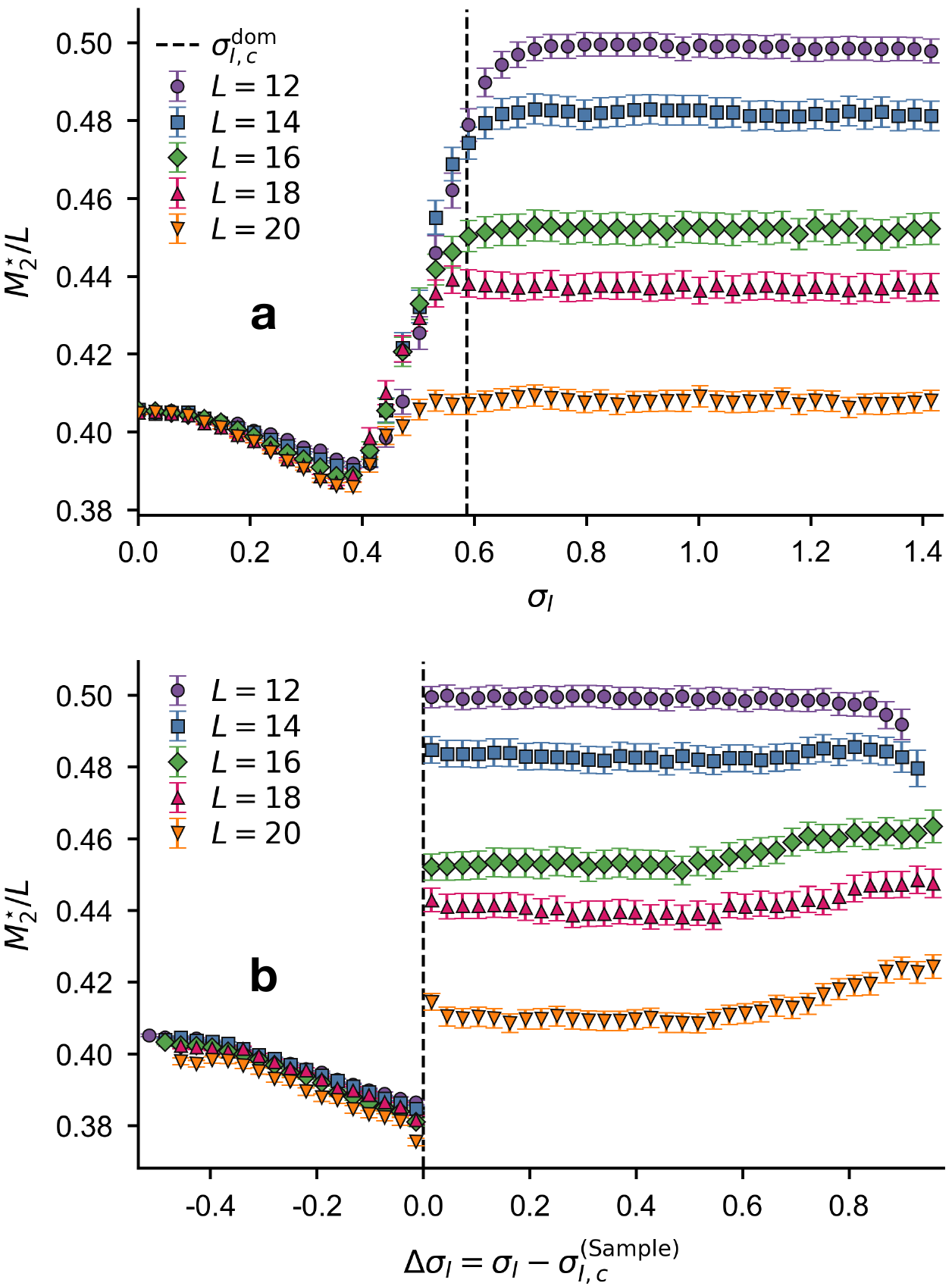}
    \caption{(a) Disorder-averaged second stabilizer 
    R\'{e}nyi entropy per site, $M_{2}(|R_{\star}\rangle)/L$, 
    as a function of $\sigma_{I}$ for $L = 12, 14, 16, 18, 20$, 
    $k_{R} = k_{I} = 3$, and $\mu_{R} = \mu_{I} = 1$. 
    The dashed vertical line marks $\sigma^{\mathrm{dom}}_{I,c} = 2 - \sqrt{2}$ 
    and the dashed horizontal line marks the exact 
    product-state value $\ln(3/2) \approx 0.405$ nats. 
    Error bars denote the standard error of the mean over 
    $197$ disorder realizations. (b) Realigned magic density 
    plotted against $\delta\sigma_{I} = \sigma_{I} - 
    \sigma_{I, c}^{(\mathrm{Sample})}$, where 
    $\sigma_{I, c}^{(\mathrm{Sample})}$ is the sample-resolved 
    jump location extracted from the largest nearest-neighbor 
    increase in $M_{2}/L$. The outer edges of the realigned curves should not be interpreted as physical features. Different disorder realizations cover different disorder ranges after shifting by their sample-dependent transition points, the number of contributing samples decreases toward the boundaries. The physically relevant feature is therefore the sharp change near $\delta\sigma_{I} = 0$.}
    \label{fig:Magic_vs_sigma_I}
\end{figure}

For $L \le 16$ we evaluate the second stabilizer R\'enyi entropy $M_{2}$ exactly via the fast Walsh--Hadamard
transform~\cite{Georges_2025}. For larger system sizes, where enumeration of all $4^L$ Pauli
strings is no longer tractable, we estimate $M_2$ using a
Metropolis--Hastings importance-sampling scheme~\cite{PhysRevB.111.054301}. Figure~\ref{fig:Magic_vs_sigma_I}(a) shows the disorder-averaged $M^{\star}_{2}/L$ for $L = 12, 14, 16, 18, 20$ for $k_{R} = k_{I} = 3$. The magic density is low and drifts slightly downward as $\sigma_{I}\to\sigma^{\rm dom}_{I, c}$, then increases to a higher value at $\sigma^{\rm dom}_{I, c}$ and saturates beyond it.

The magic of the dominant right eigenstate at $\sigma_{I} = 0$ can be computed exactly. Using Eq.~\ref{eq:Psi_s}, we get $\langle u_+|\sigma_\alpha|u_+\rangle=1/\sqrt3$ for each $\alpha\in\{x,y,z\}$, a Pauli string of weight $k$ contributes $(1/3)^{2k}$ to $|\langle\psi|P|\psi\rangle|^4$, and summing over all Pauli strings gives
\begin{equation}
    \sum_{P \in \mathcal{P}_{L}} 
    \left|\langle u_{+}^{\otimes L}|P|u_{+}^{\otimes L}
    \rangle\right|^{4} = \sum_{k=0}^{L} \binom{L}{k} 
    3^{k} \cdot \frac{1}{3^{2k}} = \left(\frac{4}{3}
    \right)^{L}.
\end{equation}
Substituting into Eq.~\eqref{eq:Second_stabilizer_formula} gives the magic density,
\begin{eqnarray}
    M^{\star}_{2} (\sigma_{I} = 0) \textcolor{red}{/ L} &=& \frac{M_{2}(|u_{+}\rangle^{\otimes L})}{L} = 
    \ln\frac{3}{2} \approx 0.405 .
\end{eqnarray}
This is exactly the value of magic at zero disorder in Fig.~\ref{fig:Magic_vs_sigma_I}(a). Above $\sigma^{\rm dom}_{I, c}$, the saturated magic density sits well above $\ln(3/2)$ but still short of the Haar value $\ln2\approx0.693$.

Realigning each disorder realization by its own $\sigma_{I,c}^{(\rm Sample)}$ [Fig.~\ref{fig:Magic_vs_sigma_I}(b)] produces a similar sharp jump in magic seen for entanglement entropy. Below the dominant eigenvalue switch, the long-time state is poor in both entanglement and magic, signaling its product-state origin. Above it, being a random-matrix eigenstate, the state is rich in both.
\subsubsection{Finite-size character of the transition}
\label{sec:Critical disorder scaling}
In Figs.~\ref{fig:sigma-c-sample-distribution}(a) and~\ref{fig:sigma-c-sample-distribution}(b) we observe that the Page value normalized post-transition entanglement entropy decreases with the system size. To quantify the finite size drift of the transition in this locality regime, we look at the distribution of the critical disorders $\sigma^{\rm (Sample)}_{I, c}$ for $L = 12, 14, 16, 18, 20$ at fixed $k_{R}=k_{I}=3$. Since Eq.~\eqref{eq:Sigma_c_scaling_approximate} provides a probabilistic upper bound, we test numerically the finite-size scaling
\begin{equation}
    \sigma_{I,c}(L) =  a L^{-b},
    \label{eq:sigma_c_scaling}
\end{equation} Figures~\ref{fig:sigma-c-sample-distribution}(c) and~\ref{fig:sigma-c-sample-distribution}(d) test this prediction directly. For each disorder realization, we extract a sample-dependent pseudocritical point $\sigma_{I,c}^{\rm Sample}$ from the location of the entanglement jump in $S_{\star}/S_{\rm Page}$ [panel (b)], and then average over realizations at fixed $L$. Panel (c) shows $\langle \sigma_{I,c}^{\rm Sample}\rangle$ as a function of $L$. The dashed curve is a two-parameter fit to Eq.~\eqref{eq:sigma_c_scaling}, with $a = 1.72, b = 0.47$.

The disorder average alone in panel (c) does not show whether the observed scaling is representative of individual disorder realizations. To probe the sample-to-sample distribution, we look at the distribution of the rescaled variable $ \sigma_{I,c}^{\rm Sample}\sqrt{L}$ in panel (d) for each system size. If the sample dependent critical disorder scales approximately as $L^{-1/2}$ then the distribution of $ \sigma_{I,c}^{\rm Sample}\sqrt{L}$ should be approximately independent of system size. The violin plots in panel (d) show the full distribution for each $L$, while the white boxes indicate the interquartile range and the black markers indicate the median. The weak variation of the rescaled distribution across $L$ shows the approximate collapse across system sizes.

We therefore conclude that, for $k_{R} = k_{I} = 3$, the observed transition is a finite-size crossover whose characteristic location drifts toward smaller disorder with increasing $L$. This behavior is consistent with the extreme-value statistics of the soft spectral edge, rather than with a transition at an $L$-independent critical disorder.
\subsubsection{Convergence to analytically predicted disorder for $k_{I} \gg \sqrt{L}$}
\begin{figure}
    \centering
    \includegraphics[width=\linewidth]{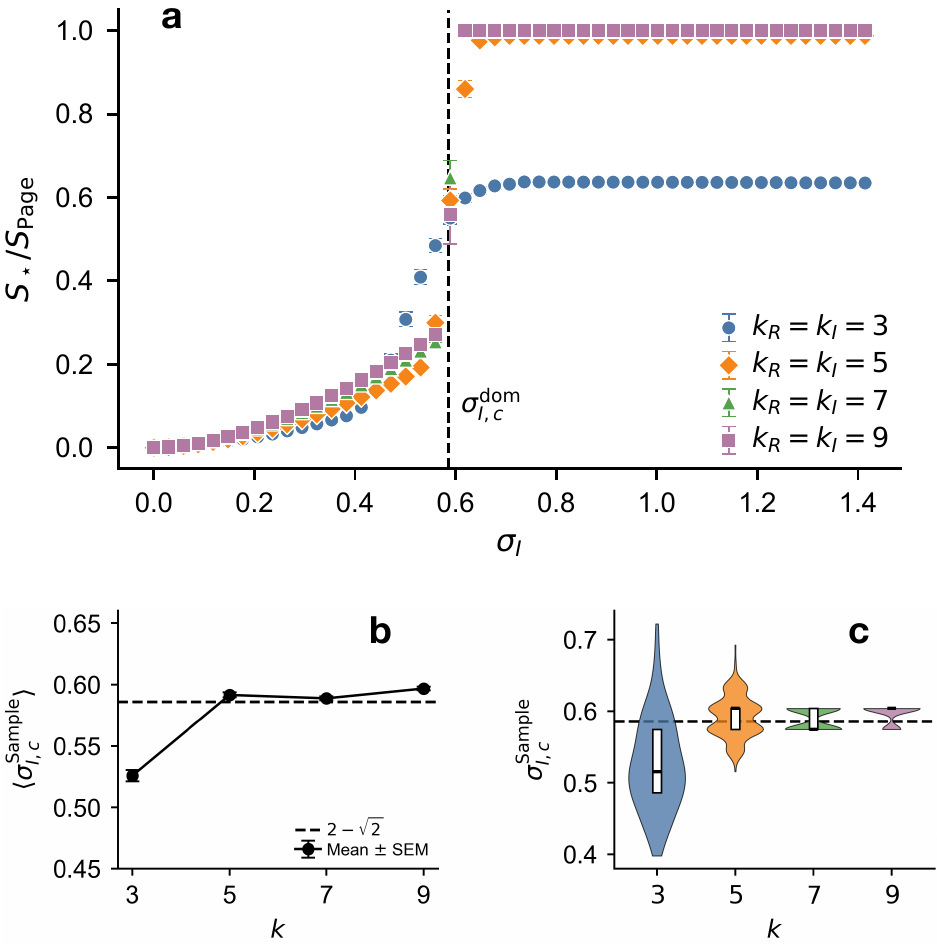}
\caption{Entanglement entropy of the dominant eigenstate at fixed system size $L=12$, $k_{R}=k_{I}=k$, for $k=3,5,7,9$ (all odd, so $k_{R},k_{I}$ stay in the same antiunitary symmetry class throughout the scan). (a) Disorder-averaged normalized entanglement entropy $S_{\star}/S_{\mathrm{Page}}$ versus $\sigma_{I}$ for each $k$. (b) Mean sample-resolved pseudocritical disorder $\langle \sigma_{I,c}^{\mathrm{Sample}} \rangle$ versus $k$, with error bars denoting the standard error of the mean. (c) Violin plots of $\sigma_{I,c}^{\mathrm{Sample}}$ versus $k$, showing the sample-to-sample spread directly. In (b) and (c), the dashed line marks the large-locality analytic threshold $\sigma^{\mathrm{dom}}_{I,c} = 2-\sqrt{2}$. As $k$ crosses $\sqrt{L}\approx 3.46$, the mean and the spread of $\sigma_{I,c}^{\mathrm{Sample}}$ converge to $\sigma^{\mathrm{dom}}_{I,c}$. Statistics were computed over $200, 200, 74$, and $66$ disorder realizations for $k=3,5,7,9$, respectively.}
\label{fig:k_scan_threshold_recovery}
\end{figure}

While the critical disorder in the $k_{I} \ll \sqrt{L}$ regime does not converge to the analytical prediction $\sigma^{\mathrm{dom}}_{I,c}$, we now show that  the transition point does converge to the analytical prediction for the $k \gg \sqrt{L}$ case. Because entanglement and magic originate from the same outlier-to-bulk switching event, we use only entanglement to demonstrate the convergence toward the analytical prediction.
In Fig.~\ref{fig:k_scan_threshold_recovery}(a) we plot the bipartite entanglement entropy as a function of disorder $\sigma_{I}$ for $k_{R} = k_{I} = k = 3, 5, 7, 9$ at fixed $L = 12$ ($\sqrt{L} \approx 3.46$). The transition sharpens visibly with increasing $k$. At $k=3$ the entanglement rises smoothly through $\sigma^{\mathrm{dom}}_{I,c}$, while by $k=9$ it is nearly a step. Panels (b) and (c) quantify this directly through the sample-resolved pseudocritical disorder $\sigma_{I,c}^{\mathrm{Sample}}$, extracted independently for each disorder realization. As $k$ increases past $\sqrt{L}$, the distribution of $\sigma_{I,c}^{\mathrm{Sample}}$ narrows sharply and its mean converges onto the analytic threshold $\sigma^{\mathrm{dom}}_{I,c}$. The sample count decreases at larger $k$ (200, 200, 74, and 66 realizations for $k=3,5,7,9$, respectively).
\subsection{Purely Hermitian disorder}
We now consider the case when $\sigma_{I} = 0$. Eq.~\eqref{eq:G_in_eigenbasis_of_deterministic} gives
\begin{eqnarray}
    G(\sigma_{R}) &=& D + \frac{ L \sigma_{R}}{\sqrt{2}} A_{k_{R}}. \label{eq:Sigma_h=0}
\end{eqnarray}
Using Eq.~\eqref{eq:General_outlier_formula}, we get the outlier
\begin{eqnarray}
    z_{\mathrm{out}}(\sigma_{R}) &=& \frac{L}{\sqrt{2}}\Bigg[ 1 + \frac{\sigma^{2}_{R}}{2} + i\bigg(1 - \frac{\sigma^{2}_{R}}{2} \bigg)\Bigg],
\end{eqnarray}
in the single outlier case, and $(z_{\mathrm{out}}(\sigma_{R}), -z^{*}_{\mathrm{out}}(\sigma_{R}))$ in the two outliers case. Similar to the anti-Hermitian disorder case, Eq.~\eqref{eq:Sigma_h=0} can also be solved as a non-Hermitian perturbation to a Hermitian matrix $A_{k_{R}}$. The outlier exists for $\sigma_{R} < \sqrt{2} \equiv \sigma^{\mathrm{outlier}}_{R, c}$.

\begin{figure}
    \includegraphics[width=\linewidth]{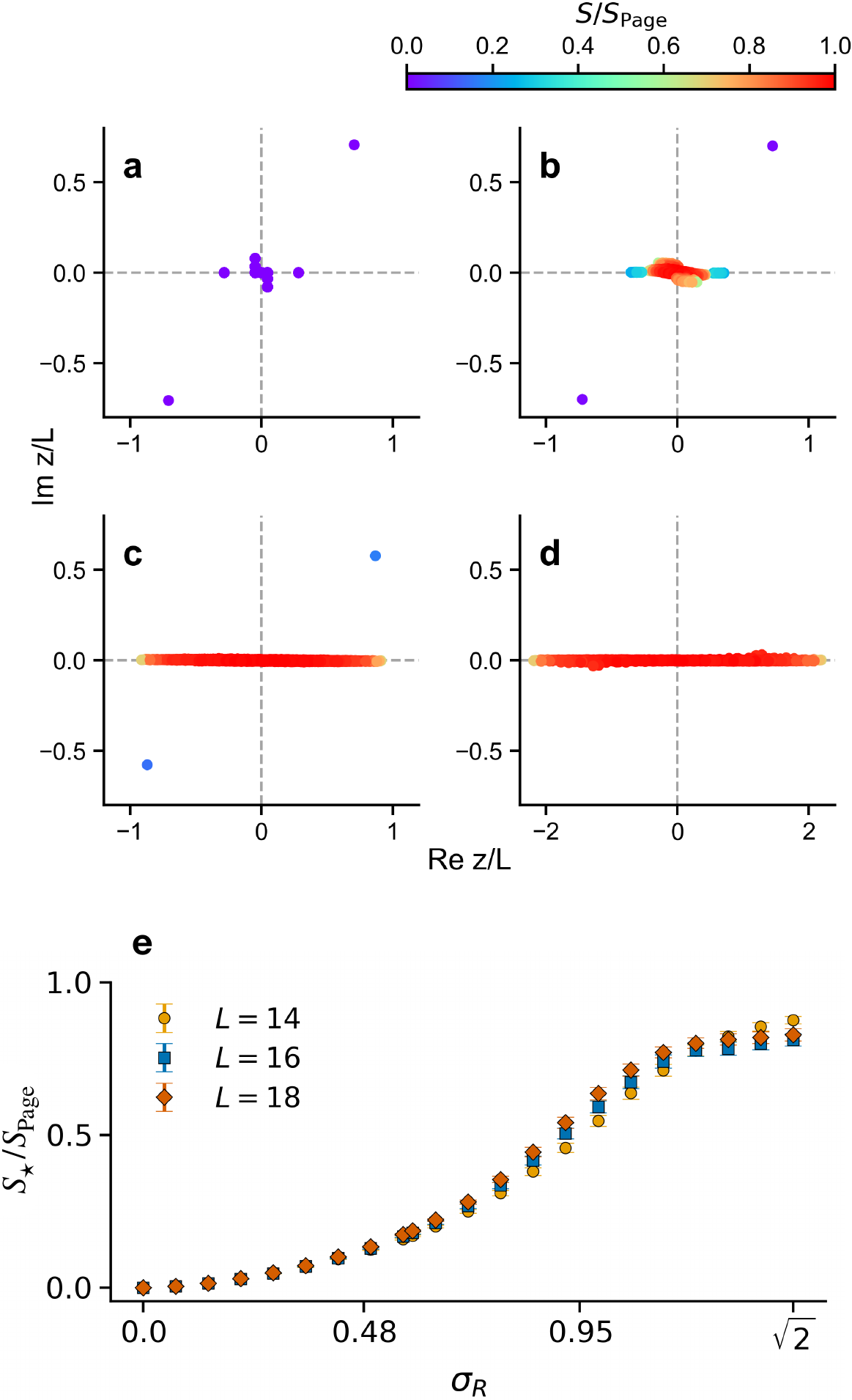}
    \caption{(a) - (d) Complex eigenvalue spectra of $G(\sigma_{R})/L$ 
    for $L = 10$, $k_{R} = 3$, $k_{I} = 5$, and 
    $\mu_{R} = \mu_{I} = 1$, shown for representative values of 
    $\sigma_{R}$ for a single realization. Each eigenvalue is colored by the 
    bipartite von Neumann entanglement entropy of its 
    corresponding right eigenstate. The bulk spreads 
    along the real axis as $\sigma_{R}$ increases, 
    while the dominant outlier $z_{\star}$ retains the 
    largest imaginary part throughout, with no 
    competing bulk edge in the imaginary direction and therefore no sharp entanglement transition. (e) Disorder-averaged normalized entanglement 
    entropy $S_{\star}/S_{\mathrm{Page}}$ of the dominant 
    right eigenstate as a function of $\sigma_{R}$ for 
    $L = 14, 16, 18$, $k_{R} = k_{I} = 3$, and 
    $\mu_{R} = \mu_{I} = 1$. The data show a smooth crossover without 
    any sharp jump, in contrast to the anti-Hermitian 
    and mixed disorder cases. Error bars denote one 
    standard error of the mean over $60$ disorder 
    realizations.}
    \label{fig:Eigenvalues_entanglement_sigma_R}
\end{figure}

In Figs.~\ref{fig:Eigenvalues_entanglement_sigma_R}(a) -~\ref{fig:Eigenvalues_entanglement_sigma_R}(d), we plot the spectrum of $G(\sigma_{R})$ for a set of representative $\sigma_{R}$ values, with each eigenvalue colored by the half-system entanglement entropy of the corresponding right eigenstate. 
We observe that unlike anti-Hermitian disorder, Hermitian disorder does not produce a
random-matrix bulk edge that grows upward along the imaginary direction. Instead the bulk grows along the real axis.

Figure~\ref{fig:Eigenvalues_entanglement_sigma_R}(e) shows the average
normalized entanglement entropy of the dominant right 
eigenstate as a function of $\sigma_{R}$ for $L = 14, 
16, 18$. In contrast to the anti-Hermitian and mixed 
disorder cases, entanglement evolves as smooth 
crossovers from a product-like state at small $\sigma_{R}$ 
to a more highly entangled state at larger $\sigma_{R}$, even at a single disorder realization level. Thus there is no analog of the outlier-to-bulk transition threshold
$\sigma^{\mathrm{dom}}_{I,c}$.
This comparison shows that the sharp transitions 
in entanglement and magic observed in the other two 
cases are not caused by adding randomness in general, but whether the random 
perturbation introduces a bulk edge that grows faster 
in the imaginary direction than the deterministic 
outlier---a condition that is absent when the disorder 
is purely Hermitian.

\subsection{Equal mixed-disorder}
In this section, we consider the equal mixed-disorder case of $\sigma_{R} = \sigma_{I} = \sigma$, for which Eq.~\eqref{eq:G_in_eigenbasis_of_deterministic} becomes
\begin{eqnarray}
    G(\sigma) &=& D + \frac{L \sigma}{\sqrt{2}} \bigg[ A_{k_{R}} + iB_{k_{I}}\bigg].\label{eq:Mixed_equal_disorder_case}
\end{eqnarray}
Next we derive the dominance and outlier transition and show that the long-time state undergoes a change in entanglement and magic as a function of the disorder strength.

From Eq.~\eqref{eq:Outlier_trajectory_general}, we find that the outlier 
position is independent of the disorder strength $\sigma$. 
In the case of a single outlier, it is given by
\begin{eqnarray}
    z_{\mathrm{out}} &=& \frac{L}{\sqrt{2}}(1+i),\label{eq:Outlier_sigma_R_sigma_I_equal}
\end{eqnarray}
and in the case of two outliers, the other outlier is given by $-z^{*}_{\mathrm{out}}$.

\subsubsection{Outlier existence threshold}
For locality $k_{R}, k_{I} \gg \sqrt{L}$, using the elliptic law from Eq.~\eqref{eq:elliptic_law_main}, we find that the bulk expands radially outward with radius $|z|_{\mathrm{bulk}} \simeq L\sigma$.
Since the outlier position is fixed at a distance $|d_{0}| = L$ from the origin,
the outlier is absorbed into the bulk when $L\sigma = |d_{0}|$, which gives us the outlier existence threshold
\begin{equation}
    \sigma^{\rm outlier}_{c} = 1. \label{eq:Outlier_equation_equal_sigma}
\end{equation}
For locality $k_{R}, k_{I} \ll \sqrt{L}$, the existence threshold is less sharp due to the long tails of the distribution of the eigenvalues beyond the elliptic law boundary.
\subsubsection{Dynamical dominance threshold}
\begin{figure*}
    \includegraphics[width=\linewidth]{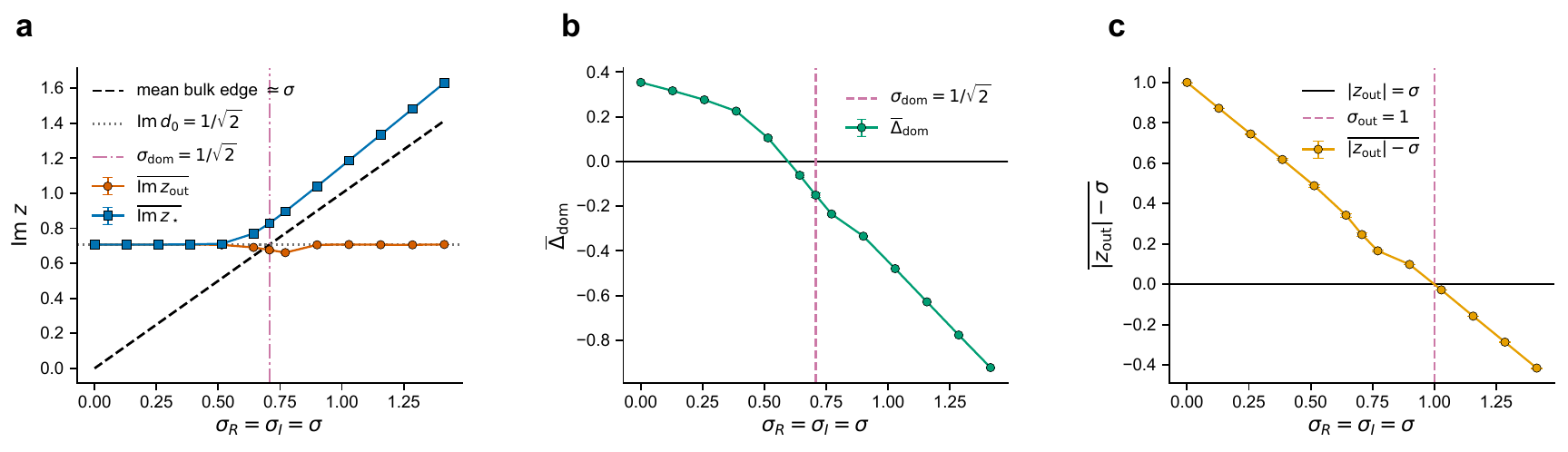}
    \caption{
Disorder dependence of the dominant and outlier eigenvalues of $G$, for $\sigma_{R} = \sigma_{I} = \sigma$, $L = 12$, $k_{R} = k_{I} = 3$. Throughout, $z \equiv \lambda(G)/L$ denotes an eigenvalue of $G$ rescaled by the system size. Data are averaged over $128$ random disorder realizations, and error bars denote the standard error of the mean.
(a) Disorder-averaged imaginary parts of the tracked outlier, $\overline{\mathrm{Im}\,z_{\rm out}}$, and of the dominant eigenvalue, $\overline{\mathrm{Im}\,z_{\star}}$, compared with the mean bulk edge $\mathrm{Im}\,z_{\rm bulk}^{\max} \simeq \sigma$.
(b) Dominance gap $\overline{\Delta}_{\rm dom} = \overline{\mathrm{Im}\,z_{\rm out} - \mathrm{Im}\,z_{\rm bulk}^{\max}}$, which changes sign near $\sigma_{\rm dom} = 1/\sqrt{2}$.
(c) Radial gap $\overline{|z_{\rm out}| - \sigma}$, showing that the outlier merges with the bulk near $\sigma_{\rm out} = 1$.
}
    \label{fig:Sigma_R_sigma_I_numerics}
\end{figure*}
To determine the dominance threshold, we note that for $k_{R}, k_{I} \gg \sqrt{L}$, the upper edge of the bulk grows as
\begin{eqnarray}
    \mathrm{Im} \; z^{\mathrm{max}}_{\mathrm{bulk}} \simeq L \sigma . \label{eq:Upper_edge_bulk}
\end{eqnarray}
At some critical disorder value $\sigma_{c, \mathrm{dom}}$, Eq.~\eqref{eq:Upper_edge_bulk} overtakes Eq.~\eqref{eq:Outlier_sigma_R_sigma_I_equal} along the imaginary direction. This happens when
\begin{eqnarray}
    L\sigma_{c, \mathrm{dom}} &=& \frac{L}{\sqrt{2}} \implies \sigma_{c, \mathrm{dom}} = \frac{1}{\sqrt{2}} \label{eq:Mixed_equal_disorder_dominance_threshold}.
\end{eqnarray}
Thus there is an intermediate regime,
\[
    \frac{1}{\sqrt{2}}<\sigma<1,
\]
in which the outlier still exists as an isolated eigenvalue, but it is no
longer the dominant eigenvalue controlling the long-time non-Hermitian
dynamics. For $k_{R}, k_{I} \ll \sqrt{L}$, the dominance threshold predicted by Eq.~\eqref{eq:Mixed_equal_disorder_dominance_threshold} becomes less accurate due to the long tails of the distribution of the bulk eigenvalues.  In
Fig.~\ref{fig:Sigma_R_sigma_I_numerics}, we test these predictions against numerical data for $L = 12, k_{R} = k_{I} = 3$. Figure~\ref{fig:Sigma_R_sigma_I_numerics}(a) shows that the disorder-averaged imaginary part of the outlier remains pinned near \(\mathrm{Im}\;d_{0}/L=1/\sqrt{2}\), while the imaginary
part of the dominant eigenvalue \(z_{\star}\) follows the outlier eigenvalue for
\(\sigma<1/\sqrt{2}\), and then grows approximately as
\(\mathrm{Im}\;z_{\star}/L\simeq \sigma\) for $\sigma > \sigma_{\mathrm{dom},c}$, signaling that it originates from the bulk.  Figure~\ref{fig:Sigma_R_sigma_I_numerics}(b) shows the
corresponding dominance gap, which changes sign at the same scale, confirming
that the outlier loses dynamical dominance near
\(\sigma_{\rm dom}=1/\sqrt{2}\). The radial separation of the outlier from the bulk is shown in Fig.~\ref{fig:Sigma_R_sigma_I_numerics}(c).  This
quantity vanishes only near \(\sigma_{c, \rm outlier}=1\), demonstrating the outlier threshold $\sigma^{\rm outlier}_{c}$.

\subsubsection{Entanglement and magic transitions in mixed disorder case}
\begin{figure*}[t]
    \includegraphics[width=\linewidth]{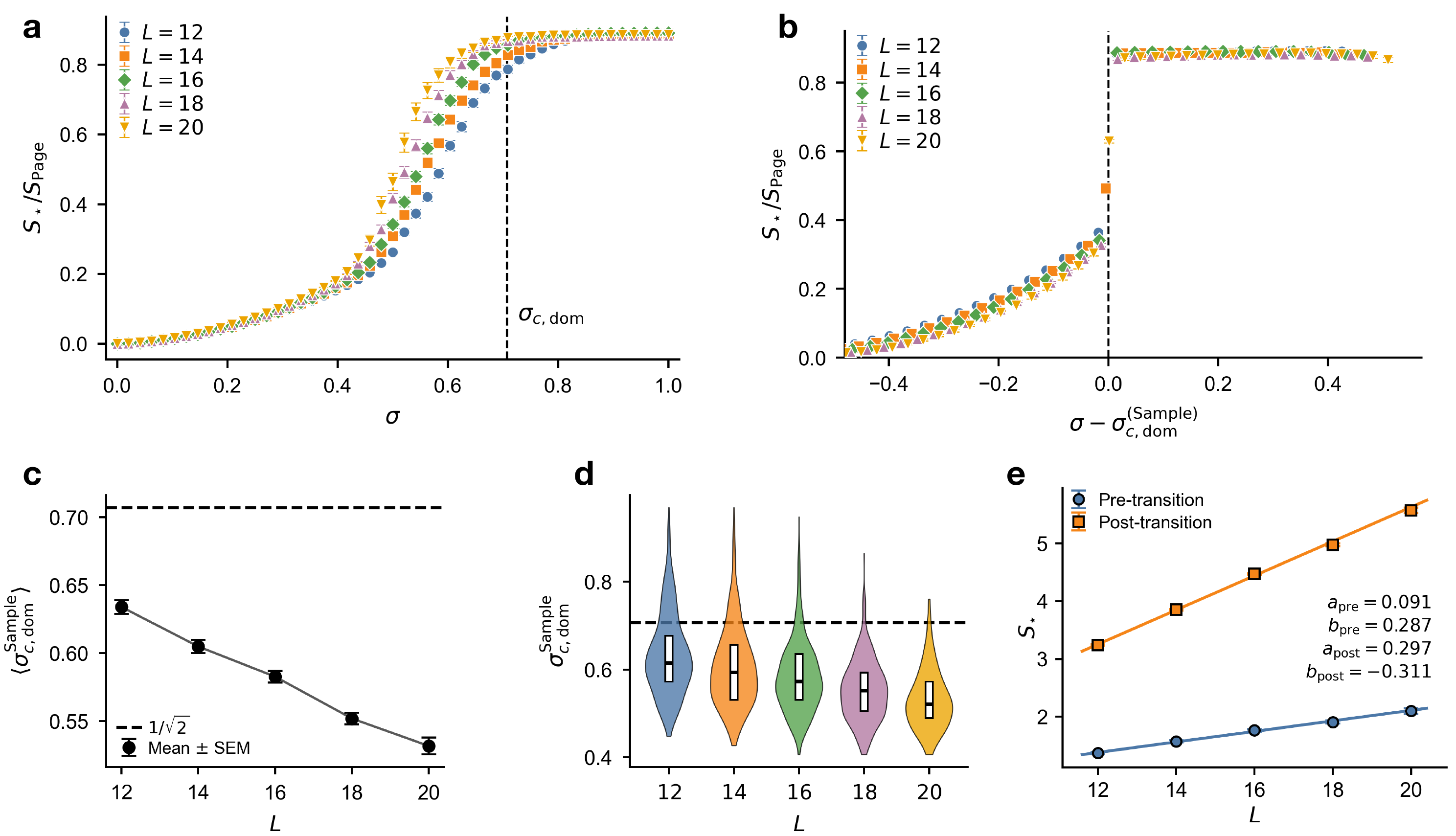}
    \caption{Entanglement transition for equal 
    mixed disorder $\sigma_{R} = \sigma_{I} = \sigma$, 
    with $k_{R} = k_{I} = 3$ and $\mu_{R} = \mu_{I} = 1$. 
    (a) Disorder-averaged normalized entanglement entropy 
    $S_{\star}/S_{\mathrm{Page}}$ as a function of $\sigma$ 
    for $L = 12, 14, 16, 18, 20$. The dashed vertical line marks 
    the dynamical dominance threshold 
    $\sigma_{c,\mathrm{dom}} = 1/\sqrt{2}$. (b) Realigned entanglement 
    entropy plotted against 
    $\sigma - \sigma_{c,\mathrm{dom}}^{(\mathrm{Sample})}$, 
    showing a sharp discontinuity within individual 
    disorder realizations. 
(c) Disorder-averaged sample-dependent critical disorder
$\langle\sigma_{I,c}^{\mathrm{Sample}}\rangle$ as a function of system
size $L$. The critical disorder decreases systematically with
increasing system size.
(d) Distribution of the sample-dependent dominance threshold
$\sigma_{c,\mathrm{dom}}^{\mathrm{Sample}}$ for different system
sizes. The violin plots show the realization-to-realization
fluctuations, while the black circles indicate the corresponding
sample means. The horizontal dotted line marks the random-matrix
prediction $\sigma_{c,\mathrm{dom}}=1/\sqrt{2}$.
(e) Pre-transition $S_{\star}(\sigma^{-}_{c, \mathrm{dom}})$ and saturated post-transition 
    entanglement entropy $S_{\star}(\sigma > \sigma_{c, \mathrm{dom}})$ as a function of 
    system size $L$, 
    fitted to 
    $S^{\rm pre}_{\star}(L) = 0.091 L + 0.287$ and $S^{\rm post}_{\star}(L) = 0.297 L - 0.311$. Error 
    bars in each panel denote the standard error of the mean over $360$ disorder realizations for $(L=12,14,16)$, $284$ realizations for $(L=18)$, and $114$ realizations for $(L=20)$.}
    \label{fig:Entanglement_sigma_R_sigma_I}
\end{figure*}

\begin{figure*}
    \centering
    \includegraphics[width=\linewidth]{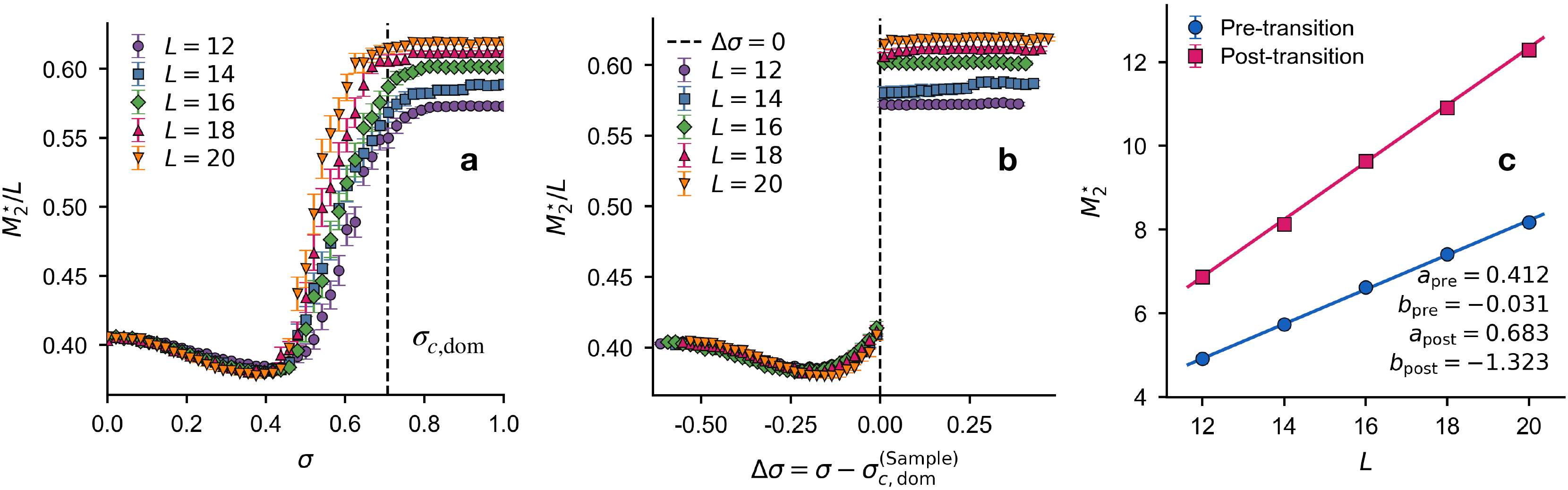}
\caption{
(a) Disorder-averaged magic density $M_2^\star/L$ of the dominant right eigenstate as a function of $\sigma$ for system sizes $L=12$--$20$. The magic increases sharply near the predicted dominance threshold $\sigma_{c,\mathrm{dom}}=1/\sqrt{2}$, indicated by the vertical dashed line. 
(b) The same data after shifting each disorder realization by its sample-dependent transition point, $\delta\sigma=\sigma-\sigma_{c,\mathrm{dom}}^{(\mathrm{Sample})}$. The collapse of the realigned curves shows that the individual disorder realizations exhibit a discontinuity in the magic. 
(c) Pre-transition and saturated post-transition magic $M_2^\star$ as a function of system size. The linear fits $M^{\star}_{2} = a L + b$ to the pre-transition and the post-transition values gives the slopes: $a_{\rm pre} = 0.412, a_{\rm post} = 0.683$ and the intercepts $b_{\rm pre} = -0.031, b_{\rm post} = -1.323$,  showing that the magic grows extensively with $L$ on both sides of the transition. Error bars denote the standard error of the mean over $60$ disorder realizations.}
    \label{fig:Magic_sigma_R_sigma_I}
\end{figure*}

In Fig.~\ref{fig:Entanglement_sigma_R_sigma_I}, we plot 
the disorder-averaged entanglement entropy $S_{\star}$ 
and in Fig.~\ref{fig:Magic_sigma_R_sigma_I} the magic $M^{\star}_{2}$ of the dominant 
right eigenstate as a function of $\sigma$ for system 
sizes $L = 12, 14, 16, 18, 20$ for locality $k_{R} = k_{I} = 3$. Both quantities remain low for 
$\sigma < \sigma_{c,\mathrm{dom}}$, then jump sharply 
near $\sigma_{c,\mathrm{dom}} = 1/\sqrt{2}$, mirroring 
the behavior observed in the purely anti-Hermitian case. 
After realigning each disorder realization relative to 
its sample-dependent critical disorder 
$\sigma_{c,\mathrm{dom}}^{(\mathrm{Sample})}$, we observe that the
discontinuity is sharp within individual realizations. The sample-averaged critical disorder decreases with system size, as shown in panels (c) and (d) of Fig.~\ref{fig:Entanglement_sigma_R_sigma_I}, but unlike the purely anti-Hermitian case, we do not have an analytical prediction for the finite size drift. We therefore do not assign it an asymptotic scaling form.

To quantify the scaling of the pre-transition and saturated post-transition entanglement and magic, we fit both quantities to linear forms,
$S_{\star}(L)=aL+b$ and $M^{\star}_{2}(L) = a L + b$, on both sides of the transition $\sigma = \sigma_{c,\mathrm{dom}}$. These fits give nonzero slopes as shown in Fig.~\ref{fig:Entanglement_sigma_R_sigma_I}(e) and Fig.~\ref{fig:Magic_sigma_R_sigma_I}(c), indicating that the entanglement and magic grow extensively with system size on both sides of the transition. In the next section, we use a random vector ansatz to explain the linear scalings we observed numerically.

\subsubsection{Post-transition entanglement and magic scaling}

In Sec.~\ref{sec:Ginibre universality of the bulk} we showed that in the equal disorder case the bulk shows the universality of Ginibre random matrices. Therefore, above the dominance threshold, the selected right eigenstate behaves as a typical vector drawn from the Ginibre ensemble. We now show that a random-vector ansatz for this state correctly predicts the linear growth of both its entanglement and its magic.

Let us first consider the entanglement. For an equal bipartition $L_A = L_B = L/2$ with Hilbert-space dimensions $N_A = N_B = 2^{L/2}$, we write the dominant right eigenstate in the product basis as
\begin{equation}
    |R_{\star}\rangle = \sum_{a=1}^{N_A}\sum_{b=1}^{N_B} X_{ab}\;|a\rangle_A|b\rangle_B ,
\end{equation}
with the entries $X_{ab}$ taken to be independent complex Gaussian variables. This ansatz is motivated by the Ginibre statistics of the bulk established numerically in Sec.~\ref{sec:Spectrum of the disordered Hamiltonian}, though we stress that Ginibre spectral statistics do not by themselves guarantee Gaussian eigenvector components; we treat this as a working ansatz and test it against the numerics.

We build the reduced density matrix from the normalized state $|R_\star\rangle$ alone,
\begin{equation}
    \rho_A = \frac{XX^{\dagger}}{\mathrm{Tr}[XX^{\dagger}]},
    \label{eq:wishart_rhoA}
\end{equation}
which is a normalized Wishart matrix~\cite{PhysRevLett.130.010401}. Our construction differs from the biorthogonal density matrix $\rho^{(i)} = |R_i\rangle\langle L_i|$ studied for Ginibre-distributed non-Hermitian Hamiltonians in Ref.~\cite{PhysRevLett.130.010401}, where they show that the entanglement entropy saturates to an $L$-independent plateau, well below the Page value, because of the biorthogonal normalization factor $\langle L_i|R_i\rangle$. Our entanglement uses only the right eigenstate, so this suppression mechanism is absent and Eq.~\eqref{eq:wishart_rhoA} reduces to the ordinary Wishart construction. Its spectrum follows the Marchenko--Pastur law, for which the average von Neumann entropy takes the Page value~\cite{PhysRevLett.130.010401, Page_1993},
\begin{equation}
    \langle S_{\star}\rangle \simeq \ln N_{A} - \frac{N_{A}}{2N_{B}}
    = \frac{\ln 2}{2}\;L - \frac{1}{2}
    \approx 0.347\;L - 0.5 .
    \label{eq:Page_law_prediction}
\end{equation}
The predicted slope $\ln 2/2 \approx 0.347$ sits close to the fitted slope $0.297$ of Fig.~\ref{fig:Entanglement_sigma_R_sigma_I}(e).

The same ansatz gives us the post-transition magic of the dominant right eigenstate. For a Haar-random state, the average second stabilizer R\'enyi entropy is~\cite{PhysRevB.111.054301}
\begin{equation}
    M_{2}^{\rm Haar} \approx L\ln 2 - \ln 4 = 0.693\;L - 1.386 .
\end{equation}
which is close to the fitted slope $0.683$ in Fig.~\ref{fig:Magic_sigma_R_sigma_I} (c).

To quantify the discontinuity in entanglement across the transition, we extract the entanglement of the dominant eigenstate on either side of the sample-aligned transition and fit its system-size dependence to the linear form $S_{\star}(L) = aL + b$. As shown in Fig.~\ref{fig:Entanglement_sigma_R_sigma_I}(c), the pre-transition data is well described by
\begin{equation}
    S^{\rm pre}_\star(L) = 0.091\;L + 0.287,
\end{equation}
while the saturated post-transition entanglement follows
\begin{equation}
    S^{\rm post}_\star(L) = 0.297\;L - 0.311 .
\end{equation}
From these fits, we obtain the discontinuity in the entanglement density
\begin{equation}
    \Delta S_{\star} \equiv \lim_{L \rightarrow \infty} \frac{S^{\rm post}_\star - S^{\rm pre}_\star}{S_{\rm Page}(L)} \approx 0.594 .
\end{equation}

We repeat the same analysis for the magic. Fitting the dominant-state magic on either side of the transition to $M_2^\star(L) = aL + b$ gives
\begin{equation}
    M_{2,\mathrm{pre}}^\star(L) \simeq 0.412\;L - 0.031,
\end{equation}
\begin{equation}
    M_{2,\mathrm{post}}^\star(L) \simeq 0.683\;L - 1.323,
\end{equation}
as shown in Fig.~\ref{fig:Magic_sigma_R_sigma_I}(c). The difference between the fitted slopes gives the jump in magic density
\begin{equation}
    \Delta m_2^\star \equiv \lim_{L \rightarrow \infty} \frac{M_{2,\mathrm{post}}^\star - M_{2,\mathrm{pre}}^\star}{L} \approx 0.271 .
\end{equation}
This value is close to the random-vector prediction
\begin{equation}
    \Delta m_2^{\rm theory} = \ln 2 - \ln(3/2) = \ln(4/3) \approx 0.288 ,
\end{equation}
obtained from the pre-transition product-state magic density $\ln(3/2)$ together with the post-transition Haar-random value.
\subsubsection{Convergence to the analytical prediction for large locality}
\label{sec:convergence-mixed-large-k}

\begin{figure}[H]
    \centering
    \includegraphics[width=\linewidth]{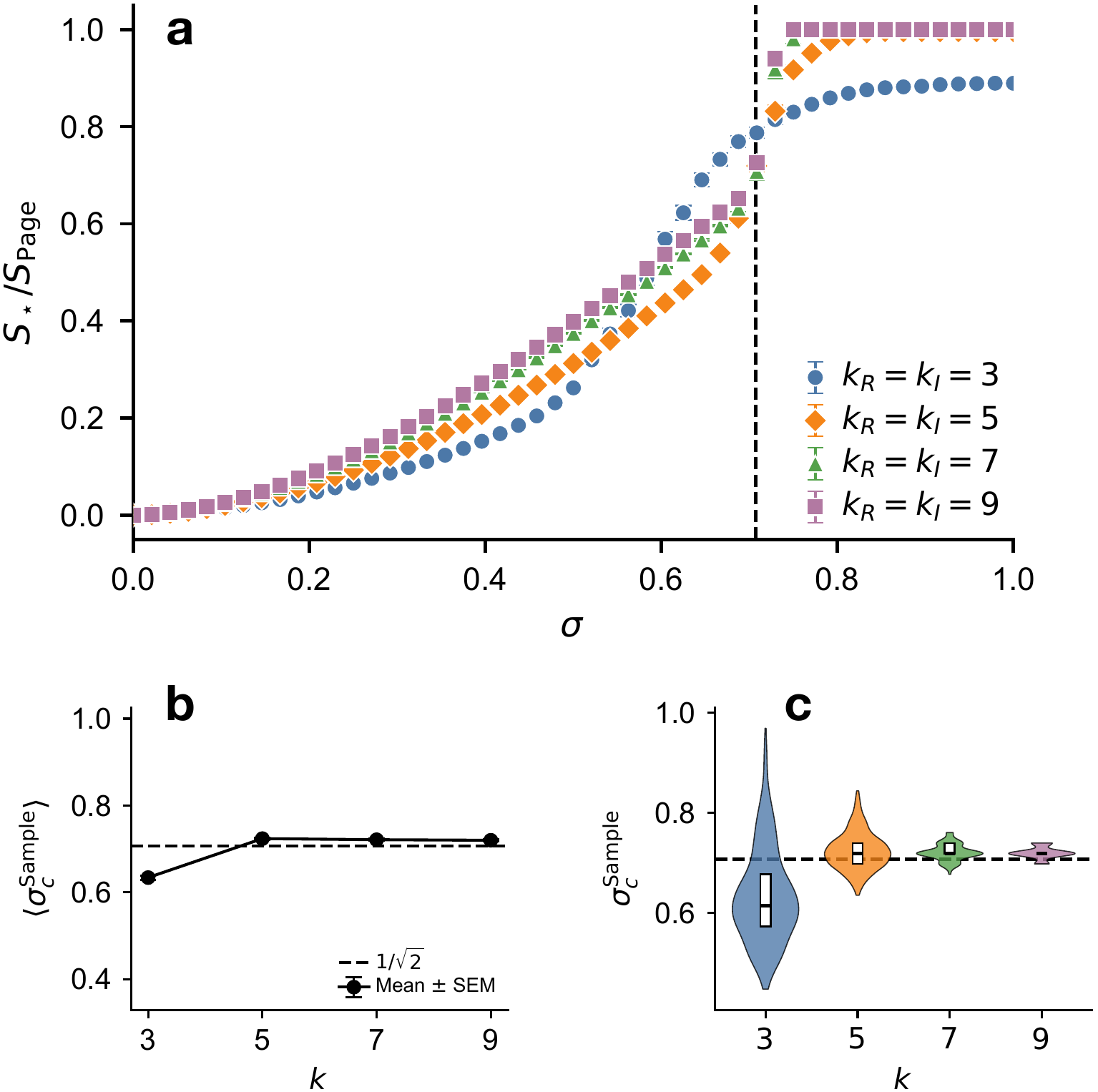}
    \caption{Entanglement entropy of the dominant eigenstate at fixed system size $L=12$, $k_{R}=k_{I}=k$, for $k=3,5,7,9$, in the mixed equal-disorder case ($\sigma_{R} = \sigma_{I} = \sigma$). The dashed line marks the analytical hard-edge prediction $\sigma^{\mathrm{dom}}_{c} = 1/\sqrt{2}$. (a) Disorder-averaged, Page-normalized entanglement entropy $S_{\star}/S_{\mathrm{Page}}$ versus $\sigma$ for each $k$. (b) Mean sample-resolved pseudocritical disorder $\langle \sigma_{c}^{\mathrm{Sample}} \rangle$ versus $k$, with error bars denoting the standard error of the mean. (c) Violin plots of $\sigma_{c}^{\mathrm{Sample}}$ versus $k$, showing the sample-to-sample spread directly. As $k$ crosses $\sqrt{L} \approx 3.46$, both the mean and the spread of $\sigma_{c}^{\mathrm{Sample}}$ converge onto $\sigma^{\mathrm{dom}}_{c}$. Statistics are based on $360$, $290$, $75$, and $49$ disorder realizations for $k=3,5,7,9$, respectively.}
    \label{fig:Equal_mixed_disorder_case_convergence_to_analytical_hard_edge_prediction}
\end{figure}

As in the anti-Hermitian disorder case, we find that the sample-resolved critical disorder converges to the analytical hard-edge prediction $\sigma^{\mathrm{dom}}_{c}$ as the locality is increased from $k \ll \sqrt{L}$ to $k \gg \sqrt{L}$. Figure~\ref{fig:Equal_mixed_disorder_case_convergence_to_analytical_hard_edge_prediction} shows the bipartite entanglement entropy of the long-time state as a function of disorder strength $\sigma$, for $k_{R} = k_{I} = 3, 5, 7, 9$ at $L = 12$ ($\sqrt{L} \approx 3.46$). The entanglement transition sharpens into a discontinuity as $k$ increases. Panels (b) and (c) quantify the distribution of sample-resolved critical disorder values and confirm convergence to the analytical prediction as $k$ crosses $\sqrt{L}$.
\section{Relation to postselected monitored dynamics}
\label{sec:postselected_monitored_dynamics}
In this section, we show that the normalized dynamics of
Eq.~\eqref{eq:normalized_time_evolved_state_overlap_section} admits an
exact reinterpretation as the postselected quantum trajectory of a
monitored system. The key observation is a gauge freedom of the
normalized evolution. If we shift the generator by an imaginary multiple
of the identity,
\begin{equation}
    G \;\to\; G - i\gamma\,\mathbf{1}, \qquad
    \mathbf{1} \equiv \mathbf{1}_{2^{L}},\;\gamma \in \mathbb{R},
    \label{eq:imaginary_shift_gauge}
\end{equation}
this multiplies $e^{-iGt}$ by the real scalar $e^{-\gamma t}$, which
cancels between the numerator and the denominator of
Eq.~\eqref{eq:normalized_time_evolved_state_overlap_section}. The
normalized state $|\psi(t)\rangle$ is therefore invariant under
Eq.~\eqref{eq:imaginary_shift_gauge}, and $\gamma$ may be chosen freely
in each disorder realization.

\emph{Spectral unraveling.}
We use this freedom to bring $G$ into the canonical form of a
conditional no-click generator. For a given realization, let
$\lambda_{\max}(H_I)$ denote the largest eigenvalue of the Hermitian
matrix $H_I$, and choose $\gamma = \lambda_{\max}(H_I)/\sqrt{2}$. We
define
\begin{equation}
    H \equiv \frac{H_R}{\sqrt{2}},
    \qquad
    \Gamma \equiv \sqrt{2}\,\big[\lambda_{\max}(H_I)\,\mathbf{1} - H_I\big],
    \label{eq:H_Gamma_definitions}
\end{equation}
so that
\begin{equation}
    G - i\,\frac{\lambda_{\max}(H_I)}{\sqrt{2}}\,\mathbf{1}
    \;=\;
    H - \frac{i}{2}\,\Gamma
    \;\equiv\; H_{\mathrm{eff}} .
    \label{eq:H_eff_no_click}
\end{equation}
By construction $\langle \psi|\Gamma|\psi\rangle \geq 0$ for every state
$|\psi\rangle$, or equivalently all eigenvalues of $\Gamma$ are
nonnegative. Therefore $H_{\mathrm{eff}} = H - i\Gamma/2$ has the
standard form of the no-click, or jump-free, effective Hamiltonian in
the quantum-trajectory description of a Lindblad evolution, with jump
operators $\{L_m\}$ satisfying
\begin{equation}
    \sum_m L_m^{\dagger}L_m = \Gamma .
    \label{eq:jump_completeness}
\end{equation}
Since $H_{I}$ is Hermitian, its eigenstates form a complete basis. Using
$\mathbf{1} = \sum_{m} |m\rangle \langle m|$ and
$H_{I} = \sum_{m} \lambda_m |m\rangle\langle m|$, we obtain an explicit
representation for the jump operators,
\begin{equation}
    L_{m} = \sqrt{\kappa_{m}}\,|m\rangle\langle m|,
    \qquad
    \kappa_{m} = \sqrt{2}\,\big[\lambda_{\max}(H_{I}) - \lambda_{m}\big] \geq 0 .
    \label{eq:jump_operators}
\end{equation}
The long-time states studied in this work can therefore be interpreted
as the conditional steady states of an interacting monitored spin
system, postselected on the absence of jumps.

\emph{A $k_{I}$-local unraveling.}
The jump operators in Eq.~\eqref{eq:jump_operators} are projections
constructed from the eigenstates of $H_I$ and are therefore generically
very nonlocal. We now present a second mapping in which the jump
operators are $k_{I}$-body. We start from $H_{I} = \sum_{\alpha} c_{\alpha}
P_{\alpha}$, where the $P_{\alpha}$ are exactly $k_{I}$-local Pauli
strings, and define the rate
\begin{equation}
    \gamma_{\mathrm{loc}} = \frac{1}{\sqrt{2}}\sum_\alpha |c_{\alpha}|
    \;\geq\; \frac{\lambda_{\max}(H_I)}{\sqrt{2}} .
    \label{eq:gamma_local}
\end{equation}
Since $P_\alpha^2 = \mathbf{1}$, the operator
$Q_{\alpha} \equiv (\mathbf{1} - \mathrm{sgn}(c_{\alpha})P_{\alpha})/2$ is a
projector, which leads to
\begin{equation}
    \Gamma_{\mathrm{loc}} \equiv
    \sqrt{2}\big[\sqrt{2}\,\gamma_{\mathrm{loc}}\,\mathbf{1} - H_{I}\big]
    = 2\sqrt{2}\sum_\alpha |c_\alpha|\, Q_\alpha \;\geq\; 0 ,
    \label{eq:gamma_loc_decomp}
\end{equation}
where the last inequality follows because a sum of positive
semidefinite operators remains positive semidefinite. The corresponding
jump operators $L_{\alpha} = \sqrt{2\sqrt{2}\,|c_{\alpha}|}\,Q_{\alpha}$ act
nontrivially only on $k_{I}$ sites.

Both unravelings generate the same normalized no-click state, since
their effective Hamiltonians differ only by the imaginary identity
shift in Eq.~\eqref{eq:imaginary_shift_gauge}. They correspond,
however, to different jump-resolved Lindblad dynamics and assign
different probabilities to the selected no-click record, as we discuss
next.

\emph{Postselection cost.}
The probability of postselection in general scales exponentially in
time~\cite{Molmer:93, 10.1093/acprof:oso/9780199213900.001.0001}. Since
$H_{I}$ is Hermitian, we have
$\lambda_{\max}(H_{I}) \leq \|H_{I}\|$~\cite{Horn_Johnson_1985}. Using
$\|P_{\alpha}\| = 1$ together with the triangle inequality gives
$\|H_{I}\| = \|\sum_{\alpha} c_{\alpha} P_{\alpha}\| \leq \sum_{\alpha}
|c_{\alpha}|$, and combining the two bounds yields
$\lambda_{\max}(H_{I}) \leq \sum_{\alpha} |c_{\alpha}|$. Inserting this
into Eq.~\eqref{eq:gamma_local} gives
$\gamma_{\mathrm{loc}} \geq \lambda_{\max}(H_I)/\sqrt{2}$.

The nonnegativity requirement
$\Gamma = 2\gamma\,\mathbf{1} - \sqrt{2}H_{I} \geq 0$ gives
$2\gamma - \sqrt{2}\lambda_{m} \geq 0$ for every eigenvalue $\lambda_{m}$
of $H_{I}$, so the smallest admissible gauge shift is
$\gamma_{\min} = \lambda_{\max}(H_{I})/\sqrt{2}$. Since the no-click
probability is
\begin{equation}
    p(\gamma, t) = e^{-2\gamma t}\,\big\| e^{-iGt}|\psi(0)\rangle \big\|^{2} ,
    \label{eq:no_click_probability}
\end{equation}
and $\gamma_{\mathrm{loc}} \geq \gamma_{\min}$, the no-click probability
is smaller for the exactly $k_{I}$-local unraveling. Locality in this
case therefore comes at the cost of a lower postselection probability.
This tradeoff between locality and postselection efficiency is
consistent with the general complexity-theoretic framework of
Ref.~\cite{lhv6-6t1r}.

\emph{Weak-measurement form and connection to measurement-induced transitions.}
The local unraveling can be recast as an explicit sequence of weak
measurements, which makes direct contact with the circuit models
studied in the measurement-induced phase transition (MIPT) literature.
We Trotterize the short-time evolution generated by
$G = (H_R + iH_I)/\sqrt{2}$ with $H_I = \sum_\alpha c_{\alpha}
P_{\alpha}$. To first order,
\begin{equation}
    e^{-iG\delta t} = e^{-iH_{R}\delta t/\sqrt{2}}\;
    e^{H_{I}\delta t/\sqrt{2}} + O(\delta t^{2}) .
    \label{eq:trotter_first_split}
\end{equation}
Writing $H_{I}\delta t/\sqrt{2} = \sum_{\alpha} \theta_{\alpha}
P_{\alpha}$ with $\theta_{\alpha} \equiv c_{\alpha}\delta t/\sqrt{2}$ and
applying the first-order Trotter decomposition again,
\begin{equation}
    e^{\sum_{\alpha} \theta_{\alpha} P_{\alpha}}
    = \prod_{\alpha} e^{\theta_{\alpha} P_{\alpha}} + O(\delta t^{2}) ,
    \label{eq:trotter_second_split}
\end{equation}
which yields
\begin{equation}
    e^{-iG\delta t} = e^{-iH_{R}\delta t/\sqrt{2}}
    \prod_{\alpha} e^{\theta_{\alpha} P_{\alpha}} + O(\delta t^{2}) .
    \label{eq:trotter_weak_measurement}
\end{equation}
Even when different $P_{\alpha}$ fail to commute, their commutators
first contribute at order $\delta t^{2}$. Over a fixed total evolution
time the per-step error in Eq.~\eqref{eq:trotter_weak_measurement}
accumulates to $O(\delta t)$.

Each factor $e^{\theta P}$ behaves like a measurement. Since every Pauli
string satisfies $P^{2} = \mathbf{1}$, we define the projections
\begin{equation}
    \Pi_{\pm} = \frac{\mathbf{1} \pm P}{2} ,
    \label{eq:pauli_projectors}
\end{equation}
so that $P\Pi_{\pm} = \pm\Pi_{\pm}$ and
$e^{\theta P} = e^{\theta}\Pi_{+} + e^{-\theta}\Pi_{-}$. For
$\theta > 0$ the operator $e^{\theta P}$ amplifies the $P = +1$ component
by $e^{\theta}$ and suppresses the $P = -1$ component by $e^{-\theta}$.
For $\theta < 0$ it favors the $P = -1$ eigenspace instead.

Accordingly, we define the Kraus operators
\begin{equation}
    K_\alpha^{\pm} = \frac{e^{\pm \theta_\alpha P_\alpha}}
    {\sqrt{2\cosh(2\theta_\alpha)}} ,
    \label{eq:kraus_branches}
\end{equation}
which satisfy
\begin{equation}
    K^{+\dagger}_{\alpha} K^{+}_{\alpha}
    + K^{-\dagger}_{\alpha} K^{-}_{\alpha} = \mathbf{1}
    \label{eq:kraus_completeness}
\end{equation}
and therefore represent a valid two-outcome measurement of $P_\alpha$.
The corresponding positive operator-valued measure (POVM) elements are
\begin{equation}
    E_{\alpha}^{\pm} = K^{\pm\dagger}_{\alpha} K^{\pm}_{\alpha}
    = \frac{1}{2}\big[\mathbf{1} \pm \tanh(2\theta_{\alpha}) P_{\alpha}\big] .
    \label{eq:povm_elements}
\end{equation}
As $\theta_{\alpha} \to 0$ we obtain
$E_{\alpha}^{\pm} \simeq (\mathbf{1} \pm 2\theta_{\alpha}P_{\alpha})/2$,
in which case the measurement extracts very little information about the
system. As $\theta_{\alpha} \to \infty$ we have
$E_{\alpha}^{\pm} \to \Pi_{\pm}$, and the measurement approaches an
ordinary projective measurement of $P_{\alpha}$. For a normalized state
$|\psi\rangle$, the probability of the $+$ branch is
\begin{equation}
    p_{+} = \langle \psi| K^{+\dagger}_{\alpha} K^{+}_{\alpha} |\psi\rangle
    = \frac{1}{2}\Big[1 + \tanh(2\theta_\alpha)\,\langle P_{\alpha}\rangle\Big] .
    \label{eq:born_weight}
\end{equation}

The no-click trajectory corresponds to selecting the $K^{+}_{\alpha}$
branch at every step. This identifies our model with the
unitary-plus-weak-measurement circuits studied in the MIPT
literature~\cite{PhysRevX.9.031009, PhysRevB.100.134306}, 
and separates the postselection cost of
Eq.~\eqref{eq:no_click_probability} into a state-independent piece
arising from the binary outcome count and a state-dependent piece
$\sum_{\alpha} \tanh(2\theta_{\alpha})\langle P_{\alpha}\rangle$.
\section{Discussion and outlook}
\label{sec:discussion_and_outlook}
We introduced a structured non-Hermitian many-body ensemble in which the Hermitian and anti-Hermitian parts are drawn independently from exactly $k$--local spin Hamiltonians with tunable means and variances. The aim was to construct a
model that interpolates, in a controlled way, between two
limits that are usually difficult to connect---the universal
predictions of random-matrix theory and the microscopic
structure of physical many-body Hamiltonians. Fully
random non-Hermitian matrices capture universal spectral correlations but contain none of the locality structure
present in physical systems. Generic many-body non-Hermitian Hamiltonians, on the other hand, are
too complex to treat analytically. Our model
occupies a useful middle ground. It retains an explicit $k$--local Pauli-string structure while admitting a controlled
random-matrix description of the bulk spectrum and
exact closed-form expressions for the eigenvalues of the deterministic part
through Krawtchouk polynomials.

Taken together, the three disorder cases studied in Sec.~\ref{sec:Outlier-to-bulk transitions of dominant eigenvalue} show that the outlier-to-bulk switching transition is not a generic consequence of disorder, but requires the random-matrix bulk to outgrow the deterministic outlier specifically along the imaginary direction. This competition is present for purely anti-Hermitian and for mixed disorder, where it produces a sharp switch in the dominant eigenvalue and simultaneous jumps in entanglement and magic, and it is absent for purely Hermitian disorder, where the bulk instead grows along the real axis and no switching event occurs.

In the mixed disorder case, the post-transition state admits a compact statistical description. A random-vector ansatz predicts a Haar-random magic density and a Page-law entanglement slope. The magic prediction matches the fitted value to within $1.4\%$ with no free parameters beyond the ansatz itself, while the entanglement prediction overshoots the fitted slope by $14\%$. We attribute this difference to the fact that the selected right eigenstate lies at the edge of the bulk spectrum rather than being a typical Ginibre eigenvector. Extremal eigenstates of chaotic local Hamiltonians are known to violate the eigenstate thermalization hypothesis and carry sub-Page entanglement even when mid-spectrum eigenstates of the same Hamiltonian thermalize fully~\cite{PhysRevE.105.014109}. The same edge-state picture explains the anti-Hermitian disorder case at small $k$, where the post-transition plateau sits below $S_{\rm Page}$ and rises toward it only as $k$ crosses $\sqrt L$, the same crossover that governs whether the bulk itself has a hard spectral edge.

The reinterpretation of the dynamics under $G$ as a postselected quantum trajectory connects our model to measurement-induced transitions. For purely anti-Hermitian disorder, the disorder strength $\sigma_{I}$ plays the role of a monitoring strength, and the dynamical-dominance transition is a sharp change in the conditional steady state of that monitored system.

Our results are consistent with, and give a microscopic route into, the broader picture emerging from non-Hermitian many-body quantum chaos. In the non-Hermitian SYK model, universal local spectral correlations were shown to be fixed by symmetry data such as the fermion parity and interaction range~\cite{PhysRevX.12.021040}. Here the analogous microscopic data are the system size $L$ and the localities $k_{R},k_{I}$, whose parities fix the antiunitary symmetry class of the bulk directly, without positing an unstructured Ginibre ensemble at the outset.

Several extensions of this work are natural. The most immediate is a fully rigorous treatment of the outlier trajectory. We used the finite-rank deformed-random-matrix result of Ref.~\cite{orourke2014lowrankperturbationslarge} as an effective description for our deterministic perturbation which has full rank. A systematic treatment of the full rank case could determine the finite size corrections.

The double-scaling limit in which $k^{2}/L$ is held fixed, in which the density of states of the zero mean disordered ensemble is known in closed form~\cite{Erd_s_2014}, is a natural setting in which to solve the anti-Hermitian ray exactly rather than through the two limiting regimes $k\ll\sqrt L$ and $k\gg\sqrt L$ treated separately here. This would unify the sharp transition and the finite-size drift into a single formula.

The role of spatial locality also remains open. The present ensemble is all-to-all interacting and carries no spatial structure. A spatially local version in one or two dimensions
would allow one to ask sharper questions about
area-law, logarithmic, and volume-law entanglement in
the selected non-Hermitian eigenstate. One possible direction
is to couple a spatially local chaotic Hermitian
Hamiltonian to an anti-Hermitian component designed
to produce an isolated outlier. Increasing the strength
of the chaotic component could then drive a transition
between an outlier-selected low-entanglement state and a
bulk-selected highly entangled state. Such a construction
would connect the present random-matrix mechanism to
more conventional notions of area and volume law entanglement phases in extended
many-body systems.

Finally, the anti-Hermitian disorder strength $\sigma_{I}$ has a direct experimental meaning as an engineered dissipation rate, and small-$k$ realizations of this model are within reach of existing platforms with programmable multi-body interactions, including trapped-ion and Rydberg-atom systems with engineered loss channels.
\section*{Acknowledgments}
\textcolor{red}{S.D. thanks Jamir Marino for discussions on magic and for suggesting the study of magic dynamics in spin chains, which motivated the investigation of magic transitions in this work.}
This research was carried out with the support from the National Science Foundation (NSF)
under Grant No. OSI-2228725. The authors thank the High
Performance Computing facility at The University of Texas at
Dallas (HPC@UTD) for providing computational resources. S.D. acknowledges the use of ChatGPT$-5.6$ (OpenAI) and Claude Sonnet $5$ (Anthropic) as coding assistants during the implementation of the GPU-based matrix-free Arnoldi solver described in the Supplemental Material, and the sampling algorithm for magic. The authors subsequently reviewed, modified, and validated all code, and takes full responsibility for the results reported here.
\section*{Data Availability}
All computer programs used to generate the data are available in the Zenodo repository at \href{https://doi.org/10.5281/zenodo.21862099}{10.5281/zenodo.21862099}.

\appendix
\section{Antiunitary spectral constraints}\label{sec:Antiunitary spectral constraints}
In this section, we prove the spectral constraints on $G$ due to the antiunitary symmetry imposed by $T = U \mathcal{K}$, listed in Table~\ref{tab:nonHermitian_symmetry_classification}.
\subsubsection{Even $k_{R}$, odd $k_{I}$}
Here $TGT^{-1}= G$ and $TG^{\dagger}T^{-1} = G^{\dagger}$. Let us start with the eigenvalue equation for the right eigenstate $G|R_{n}\rangle = z_{n} |R_{n}\rangle$. Now apply $T$ on both sides of this equation, $TG|R_{n}\rangle = z^{*}_{n}T|R_{n}\rangle$. Using the commutation relation $TG = GT$, we get $G(T|R_{n}\rangle) = z^{*}_{n} (T|R_{n}\rangle)$. Thus, for every eigenvalue $z_{n}$, the spectrum has its complex conjugate $z^{*}_{n}$. For odd $L$, we have $T^{2}=-1$. Thus every real eigenvalue $z_{n} \in R$ is doubly degenerate.
\subsubsection{Odd $k_{R}$, even $k_{I}$}
Here $TGT^{-1}=-G$ and $TG^{\dagger}T^{-1}=-G^{\dagger}$, applying $T$ to the right eigenvalue equation, $-GT|R_{n}\rangle = z^{*}_{n}T|R_{n}\rangle$ which gives $GT|R_{n}\rangle = - z^{*}_{n}T|R_{n}\rangle$. So for $z_{n}$, $-z^{*}_{n}$ is also an eigenvalue of $G$. For odd $L$, we get $T^{2} = -1$, so every purely imaginary eigenvalue, if present, is doubly degenerate.
\subsubsection{Odd $k_{R}$, odd $k_{I}$}
In this case, $TGT^{-1} = -G^{\dagger}$. Now we use $T = U\mathcal{K}$, which gives $U G^{*} U^{-1} = -G^{\dagger} = -(G^{*})^{T}$. Taking complex conjugate of both sides $G^{T} = -UGU^{-1}$. Now consider the characteristic polynomial for the matrix $G$: $\mathrm{det}(zI - G) = \mathrm{det}(zI - G^{T}) = \mathrm{det}(zI + UGU^{-1}) = \mathrm{det}(zI+G)$, where to get the last equality we have used the fact that similar matrices have the same determinant. From the equality $\mathrm{det}(zI-G) = \mathrm{det}(zI+G)$, we conclude that for every eigenvalue $z_{n}$, the spectrum also contains $-z_{n}$. For odd $L$, every zero eigenvalue is doubly degenerate.
\subsubsection{Even $L$, even $k_{R}$, even $k_{I}$}
For this case, one can show that $(U|L\rangle)^{*} \propto |R\rangle$~\cite{PhysRevResearch.2.023286}.
\subsubsection{Odd $L$, even $k_{R}$, even $k_{I}$}

For odd $L$, we have $U^{T}=-U$, where $T=U\mathcal{K}$ is the
antiunitary symmetry operator and $\mathcal{K}$ denotes complex
conjugation. The symmetry relation $TGT^{-1}=G^{\dagger}$ can be
written as
\begin{equation}
UG^{*}U^{-1}=G^{\dagger}=(G^{*})^{T}.
\end{equation}
Taking the complex conjugate of this relation and using $U^{*}=U$,
we obtain $G^{T}=UGU^{-1}$, or equivalently,
\begin{equation}
UG=G^{T}U.
\label{eq:self-dual-relation}
\end{equation}

Consider a right eigenstate of $G$,
\begin{equation}
G|R_{n}\rangle=z_{n}|R_{n}\rangle.
\label{eq:right-eigenstate-AII-dagger}
\end{equation}
Taking the ordinary transpose, rather than the Hermitian conjugate,
gives
\begin{equation}
\left(|R_{n}\rangle\right)^{T}G^{T}
=
z_{n}\left(|R_{n}\rangle\right)^{T}.
\end{equation}
Multiplying from the right by $U$ and using
Eq.~\eqref{eq:self-dual-relation}, we find
\begin{align}
\left(|R_{n}\rangle\right)^{T}G^{T}U
&=
\left(|R_{n}\rangle\right)^{T}UG
\nonumber \\
&=
z_{n}\left(|R_{n}\rangle\right)^{T}U.
\end{align}
Therefore,
\begin{equation}
\langle L_{n}|
\equiv
\left(|R_{n}\rangle\right)^{T}U
\label{eq:left-from-right-AII-dagger}
\end{equation}
is a left eigenstate of $G$ with the same eigenvalue,
$\langle L_{n}|G=z_{n}\langle L_{n}|$.

We now show that $z_{n}$ cannot be a nondegenerate eigenvalue.
Using Eq.~\eqref{eq:left-from-right-AII-dagger}, the biorthogonal
overlap between the corresponding left and right eigenstates is
\begin{equation}
\langle L_{n}|R_{n}\rangle
=
\left(|R_{n}\rangle\right)^{T}U|R_{n}\rangle.
\end{equation}
This quantity is a scalar and therefore equal to its own transpose,
\begin{equation}
\left(|R_{n}\rangle\right)^{T}U|R_{n}\rangle
=
\left[\left(|R_{n}\rangle\right)^{T}U|R_{n}\rangle\right]^{T}
=
\left(|R_{n}\rangle\right)^{T}U^{T}|R_{n}\rangle.
\end{equation}
Since $U^{T}=-U$, this gives
\begin{equation}
\left(|R_{n}\rangle\right)^{T}U|R_{n}\rangle
=
-\left(|R_{n}\rangle\right)^{T}U|R_{n}\rangle,
\end{equation}
and hence
\begin{equation}
\langle L_{n}|R_{n}\rangle=0.
\label{eq:self-biorthogonal-AII-dagger}
\end{equation}

Suppose, for contradiction, that $z_{n}$ is nondegenerate. For a
diagonalizable non-Hermitian matrix, the left and right eigenstates
associated with a nondegenerate eigenvalue can always be normalized
such that
\begin{equation}
\langle L_{n}|R_{n}\rangle=1.
\end{equation}
This contradicts Eq.~\eqref{eq:self-biorthogonal-AII-dagger}.
Therefore, $z_{n}$ cannot be nondegenerate, and every eigenvalue of
$G$ is at least twofold degenerate.
\section{Relation to existing Non-Hermitian classification}\label{sec:Relation to existing Non-Hermitian classification}
We now discuss the relation between our classification scheme and the existing classification for non-Hermitian random matrices, by showing that each possible conjugation $T G T^{-1}$ maps to one of the non-Hermitian random-matrix ensembles in~\cite{PhysRevResearch.2.023286, PhysRevX.9.041015}.

The condition $TGT^{-1} = G$ translates to $UG^{*}U^{-1} = G$, and for even $L$, $U U^{*} = +1$ (class AI), and for odd $L$, $U U^{*} = -1$ (class AII).

For $T G T^{-1} = - G$ gives $U G ^{*} U^{-1} = - G$ and for even $L$, $U U^{*} = 1$ (class D$^{\dagger}$), and for odd $L$, $U U^{*} = -1$ (class C$^{\dagger}$).

The condition $T G T^{-1} = G^{\dagger}$ gives $U G^{*} U ^{-1} = G^{\dagger}$. Taking the Hermitian conjugate, we obtain $U G^{T} U = G$. For even $L$, $U U^{*} = +1$ (class AI$^{\dagger}$), and for odd $L$, $U U^{*} = -1$ (class AII$^{\dagger}$). Further, one can show that if there are no degeneracies in the spectrum, then the right and the left eigenvectors are related by $U(|L\rangle)^{*} \propto |R\rangle$~\cite{PhysRevResearch.2.023286}.

The condition $T G T^{-1} = - G^{\dagger}$ gives $U G^{T} U^{-1} = - G$, and $U U^{*} = +1$ (class D) for even $L$, and $U U^{*} = -1$ (class C) for odd $L$.
\section{Derivation of the disorder-averaged trace moments}\label{app:trace_moments}

In this section, we derive Eqs.~\eqref{eq:m11_main} and \eqref{eq:m20_main} for the disorder-averaged trace moments
\begin{equation}
    m_{11} = \frac{1}{N}\left\langle \mathrm{Tr}\left(GG^{\dagger}\right)\right\rangle, \qquad
    m_{20} = \frac{1}{N}\left\langle \mathrm{Tr}\left(G^{2}\right)\right\rangle ,
\end{equation}
used in Sec.~\ref{sec:Spectrum of the disordered Hamiltonian} to fix the effective scale and ellipticity of the disordered bulk.

For simplicity we set $k_{R}=k_{I}=k$. Denote an exactly $k$--local Pauli string by $P_{a}$, with corresponding coupling constants $J_{a}$ in $H_{R}$ and $M_{a}$ in $H_{I}$, so that
\begin{equation}
    H_{R} = \sum_{a} J_{a}P_{a}, \qquad H_{I} = \sum_{a} M_{a}P_{a} .
\end{equation}
Since $G^{\dagger} = (H_{R}-iH_{I})/\sqrt{2}$, we obtain
\begin{multline}
    GG^{\dagger} = \frac{1}{2}\left(H_{R}+iH_{I}\right)\left(H_{R}-iH_{I}\right) \\
    = \frac{1}{2}\left(H_{R}^{2}+H_{I}^{2}+iH_{I}H_{R}-iH_{R}H_{I}\right).
\end{multline}
Taking the trace removes the commutator term, $\mathrm{Tr}(H_{I}H_{R}-H_{R}H_{I})=0$, so
\begin{equation}
    m_{11} = \frac{1}{2N}\left\langle \mathrm{Tr}\;H_{R}^{2} + \mathrm{Tr}\;H_{I}^{2}\right\rangle .
\end{equation}
Expanding in the Pauli-string basis,
\begin{equation}
    m_{11} = \frac{1}{2N}\left\langle \sum_{a,b}J_{a}J_{b}\;\mathrm{Tr}(P_{a}P_{b}) + \sum_{a,b}M_{a}M_{b}\;\mathrm{Tr}(P_{a}P_{b})\right\rangle ,
\end{equation}
and using the Pauli-string orthogonality relation $\mathrm{Tr}(P_{a}P_{b})=N\delta_{ab}$,
\begin{equation}
    m_{11} = \frac{1}{2}\sum_{a}\left(\left\langle J_{a}^{2}\right\rangle + \left\langle M_{a}^{2}\right\rangle\right) .
\end{equation}
Writing $\langle J_{a}^{2}\rangle = (\mathbb{E}[J_{a}])^{2}+\mathrm{Var}(J_{a})$ and similarly for $M_{a}$, and inserting the coupling normalizations of Eqs.~\eqref{eq:Mean_J_M_couplings}, we find
\begin{equation}
    \sum_{a}\left\langle J_{a}^{2}\right\rangle = L^{2}\left[\sigma_{R}^{2} + \frac{\mu_{R}^{2}}{\binom{L}{k}}\right], \;
    \sum_{a}\left\langle M_{a}^{2}\right\rangle = L^{2}\left[\sigma_{I}^{2} + \frac{\mu_{I}^{2}}{\binom{L}{k}}\right] ,
\end{equation}
which gives Eq.~\eqref{eq:m11_main},
\begin{equation}
    m_{11} = \frac{L^{2}}{2}\left[\sigma_{R}^{2}+\sigma_{I}^{2} + \frac{\mu_{R}^{2}+\mu_{I}^{2}}{\binom{L}{k}}\right] .
\end{equation}
An identical calculation, keeping track of the relative sign between the $H_{R}^{2}$ and $H_{I}^{2}$ contributions arising from $G^{2}=(H_{R}^{2}-H_{I}^{2}+i\{H_{R},H_{I}\})/2$ rather than $GG^{\dagger}$, yields Eq.~\eqref{eq:m20_main},
\begin{equation}
    m_{20} = \frac{L^{2}}{2}\left[\sigma_{R}^{2}-\sigma_{I}^{2} + \frac{(\mu_{R}+i\mu_{I})^{2}}{\binom{L}{k}}\right] .
\end{equation}

It is also useful to record the component-wise form of these moments, which follows directly from the definition of the matrix trace. Writing
\begin{align}
    (GG^{\dagger})_{nn} &= \sum_{m=1}^{N} G_{nm}(G^{\dagger})_{mn} \\ &= \sum_{m=1}^{N} G_{nm}G_{nm}^{*} = \sum_{m=1}^{N}|G_{nm}|^{2},
\end{align}
we obtain
\begin{equation}
    \mathrm{Tr}[GG^{\dagger}] = \sum_{n=1}^{N}(GG^{\dagger})_{nn} = \sum_{n,m=1}^{N}|G_{nm}|^{2} ,
\end{equation}
so that
\begin{equation}
    m_{11} = \frac{1}{N}\sum_{n,m=1}^{N}\left\langle |G_{nm}|^{2}\right\rangle .
    \label{eq:m11_component_appendix}
\end{equation}
Similarly,
\begin{equation}
    m_{20} = \frac{1}{N}\left\langle \mathrm{Tr}[G^{2}]\right\rangle = \frac{1}{N}\sum_{n,m=1}^{N}\left\langle G_{nm}G_{mn}\right\rangle .
    \label{eq:m20_component_appendix}
\end{equation}
Equations~\eqref{eq:m11_component_appendix} and \eqref{eq:m20_component_appendix} make explicit that $m_{11}$ and $m_{20}$ are, respectively, the disorder-averaged sum of squared matrix-element magnitudes and the disorder-averaged sum of products of symmetric matrix-element pairs; together they fix the elliptic-ensemble parameters $\langle X_{ij}^{2}\rangle$ and $\langle X_{ij}X_{ji}\rangle$ quoted in Sec.~\ref{sec:Spectrum of the disordered Hamiltonian}.
\section{Invariance of the random-matrix ensembles under unitary conjugation}\label{sec:Invariance of the random-matrix ensembles under unitary conjugation}
In this section, we show that the random-matrix ensemble associated with the completely disordered Hamiltonians remains invariant under the unitary conjugations defined in Eq.~\eqref{eq:A_B_under_unitary_conjugations}.
To begin with, we define the transformed antiunitary operator
\begin{equation}
    \widetilde{T} = U^\dagger T U,
\end{equation}
which remains antiunitary since $T$ is antiunitary and $U$ is unitary.
Using Eq.~\eqref{eq:Action-of-sigma_y-K}, $T M T^{-1} = (-1)^k M$ for any exactly $k$-local
Hermitian matrix $M$, a direct calculation gives
\begin{eqnarray}
    \widetilde{T}\; A_{k_{R}}\; \widetilde{T}^{-1}
    &=& U^\dagger \bigl[T H_R(0,1) T^{-1}\bigr] U
        \cdot \bigl(3^{k_{R}}\tbinom{L}{k_{R}}\bigr)^{-1/2}\nonumber\\
     &=& (-1)^{k_{R}}\; A_{k_{R}}, \label{eq:Asymm} \\
    \widetilde{T}\; B_{k_{I}}\; \widetilde{T}^{-1}
    &=& U^\dagger \bigl[T H_I(0,1) T^{-1}\bigr] U
        \cdot \bigl(3^{k_{I}}\tbinom{L}{k_{I}}\bigr)^{-1/2}\nonumber \\
     &=& (-1)^{k_{I}}\; B_{k_{I}}. \label{eq:Bsymm}
\end{eqnarray}
Second, since $T^2 = (-1)^L \mathbf{1}$ [Eq.~\eqref{eq:Sigma_y_K^2}] is proportional to
the identity and therefore commutes with any unitary,
\begin{equation}
    \widetilde{T}^2 = (U^\dagger T U)^2 = U^\dagger T^2 U 
    = (-1)^L\; \mathbf{1}.  \label{eq:Ttildesq}
\end{equation}
The pair $\bigl((-1)^{k_{R}}, (-1)^L\bigr)$ in
Eqs.~(\ref{eq:Asymm}) and~(\ref{eq:Ttildesq}) is identical to the pair
that classifies $H_R(0,1)$ in the original basis,
and the same holds for $B_{k_{I}}$ via Eq.~(\ref{eq:Bsymm}).
The three Dyson classes correspond to the following cases we list below.
\begin{itemize}
\item \textit{GUE} ($k_{R/I}$ odd, any $L$): 
  $\widetilde{T} M \widetilde{T}^{-1} = -M$, so there is no antiunitary
  symmetry mapping the matrix to itself. $A_{k_{R}}$ and $B_{k_{I}}$ have no
  effective time-reversal symmetry and belong to the GUE. Moreover,
  since the GUE distribution is invariant under conjugation by any fixed
  unitary, we conclude
  \begin{equation}
      A_{k_{R}} \stackrel{d}{=}
      \frac{H_R(0,1)}{\sqrt{3^{k_{R}}\binom{L}{k_{R}}}}, \qquad
      B_{k_{I}} \stackrel{d}{=}
      \frac{H_I(0,1)}{\sqrt{3^{k_{I}}\binom{L}{k_{I}}}},
  \end{equation}
  so the transformed matrices have exactly the same distribution as the
  originals.

\item \textit{GOE} ($k_{R/I}$ even, $L$ even): 
  $\widetilde{T} M \widetilde{T}^{-1} = M$ and $\widetilde{T}^2 =
  +\mathbf{1}$. This is the GOE antiunitary condition. Note that
  $A_{k_{R}}$ and $B_{k_{I}}$ are generically \emph{not} entrywise
  real-symmetric in the eigenbasis of $G_0$, because $U$ has complex
  entries [see Eq.~\eqref{eq:Psi_s}].

\item \textit{GSE} ($k_{R/I}$ even, $L$ odd): 
  $\widetilde{T} M \widetilde{T}^{-1} = M$ and $\widetilde{T}^2 =
  -\mathbf{1}$. By the standard Kramers argument, for any eigenvector
  $|\psi\rangle$ of $A_{k_{R}}$ (or $B_{k_{I}}$) with eigenvalue $\lambda$,
  the state $\widetilde{T}|\psi\rangle$ is also an eigenvector with the
  same eigenvalue. Computing the inner product using antilinearity of
  $\widetilde{T}$ and $\widetilde{T}^2 = -\mathbf{1}$,
  \begin{equation}
      \langle\psi|\widetilde{T}\psi\rangle
      = \langle\widetilde{T}^2\psi|\widetilde{T}\psi\rangle^*
      = -\langle\psi|\widetilde{T}\psi\rangle^*,
  \end{equation}
  which forces $\langle\psi|\widetilde{T}\psi\rangle = 0$. Thus $|\psi\rangle$
  and $\widetilde{T}|\psi\rangle$ are orthogonal and every eigenvalue of
  $A_{k_{R}}$ and $B_{k_{I}}$ is at least doubly degenerate. The local level
  statistics within each degenerate sector follow the GSE Wigner-Dyson
  distribution ($\beta = 4$), consistent with Table~\ref{tab:nonHermitian_symmetry_classification}.
\end{itemize}
\section{Secondary outliers never become dynamically dominant}
\label{sec:Secondary outliers never become dynamically dominant}

In this section, we establish that the dominant eigenvalue originates from only two possible sources: the outlier branch generated by the extremal deterministic eigenvalue $d_{0}$ (or its symmetry-related degenerate partner), or the upper edge of the bulk. No other deterministic eigenvalue can produce a dominant eigenvalue at any disorder strength.

For simplicity, we take $k_{R} = k_{I} = k$ and $\mu_{R} = \mu_I = 1$. The deterministic eigenvalues in Eq.~\eqref{eq:Deterministic_eigenvalues_G0_matrix} then take the form
\begin{equation}
    d_n
    =
    \frac{L}{\sqrt{2}}\;a_n(1+i),
    \qquad
    a_n
    \equiv
    \binom{L}{k}^{-1}K_k^{(L,2)}(n),
    \label{eq:secondary_an_def}
\end{equation}
with $a_0=1$, so that
\begin{equation}
    d_{0}=\frac{L}{\sqrt{2}}(1+i).
\end{equation}
The first secondary deterministic eigenvalue is given by
\begin{equation}
    d_{1}=\frac{La}{\sqrt{2}}(1+i),
    \qquad
    a\equiv a_1
    =
    1-\frac{2k}{L},
    \label{eq:secondary_d1}
\end{equation}
and therefore $0<a<1$ when $k<L/2$.

We restrict our attention to the purely anti-Hermitian disorder case, $\sigma_{R}=0$, for which the elliptic parameters entering the outlier equation are
\begin{equation}
    s=\frac{L\sigma_{I}}{\sqrt{2}},
    \qquad
    \tau=-1.
\end{equation}
Applying the outlier formula of Eq.~\eqref{eq:General_outlier_formula} to a deterministic eigenvalue $d_n$ yields
\begin{align}
    z_n^{\rm out}(\sigma_{I})
    &=
    d_n-\frac{s^2}{d_n}
    =
    \frac{L}{\sqrt{2}}
    \left[
        a_n-\frac{\sigma_{I}^2}{2a_n}
        +i\left(
            a_n+\frac{\sigma_{I}^2}{2a_n}
        \right)
    \right].
    \label{eq:secondary_outlier_trajectory}
\end{align}
This outlier exists provided $|d_n|>s$, equivalently
\begin{equation}
    \sigma_{I}<\sqrt{2}\;|a_n|.
    \label{eq:secondary_outlier_existence}
\end{equation}

Branches with $a_n<0$ may be excluded immediately since they produce outliers with negative imaginary parts
\begin{equation}
    \operatorname{Im}z_n^{\rm out}
    =
    \frac{L}{\sqrt{2}}
    \left(
        a_n+\frac{\sigma_{I}^2}{2a_n}
    \right)
    <0
\end{equation}

We now bound the coefficients $a_n$. Writing
\begin{equation}
    a_n
    =
    \sum_j p_j(-1)^j,
    \qquad
    p_j
    \equiv
    \frac{\binom{n}{j}\binom{L-n}{k-j}}
    {\binom{L}{k}},
\end{equation}
Vandermonde's identity gives $\sum_jp_j=1$. Since each $p_j>0$, it follows that $|a_n|\leq1$. For $0<n<L$ and $0<k<L$, the sum contains both even and odd values of $j$, so this bound holds strictly: $|a_n|<1$. The edge values are $a_0=1$ and $a_L=(-1)^k$. For even $k$, $a_L=1$ corresponds only to the symmetry-related degeneracy $d_L=d_{0}$ and therefore does not constitute an independent branch. Consequently, any distinct upper-half-plane branch satisfies
\begin{equation}
    0<a_n<1.
\end{equation}

The imaginary part of the outlier generated by $d_{0}$ is
\begin{equation}
    \operatorname{Im}z_0^{\rm out}
    =
    \frac{L}{\sqrt{2}}
    \left(
        1+\frac{\sigma_{I}^2}{2}
    \right).
\end{equation}
Comparing this to any secondary upper-half-plane outlier, and subtracting Eq.~\eqref{eq:secondary_outlier_trajectory}, we obtain
\begin{align}
    \operatorname{Im}z_0^{\rm out}
    -
    \operatorname{Im}z_n^{\rm out}
    &=
    \frac{L}{\sqrt{2}}
    \left[
        1+\frac{\sigma_{I}^2}{2}
        -
        a_n-\frac{\sigma_{I}^2}{2a_n}
    \right]
    \nonumber\\
    &=
    \frac{L}{\sqrt{2}}
    (1-a_n)
    \left(
        1-\frac{\sigma_{I}^2}{2a_n}
    \right).
    \label{eq:secondary_outlier_difference}
\end{align}
Whenever the secondary outlier exists, Eq.~\eqref{eq:secondary_outlier_existence} implies
\begin{equation}
    \sigma_{I}^2<2a_n^2<2a_n,
\end{equation}
where the second inequality follows from $0<a_n<1$. Hence
\begin{equation}
    1-\frac{\sigma_{I}^2}{2a_n}>0.
\end{equation}
Together with $1-a_n>0$, Eq.~\eqref{eq:secondary_outlier_difference} therefore gives
\begin{equation}
    \boxed{
    \operatorname{Im}z_0^{\rm out}
    >
    \operatorname{Im}z_n^{\rm out}
    }
\end{equation}
for every existing secondary outlier in the upper half-plane. This inequality is the central result from which the remaining conclusions follow directly.

Consider first the case in which the $d_{0}$ outlier lies above the bulk,
\begin{equation}
    \operatorname{Im}z_0^{\rm out}
    >
    \operatorname{Im}z_{\rm bulk}^{\max}.
\end{equation}
It then also lies above every secondary outlier and is therefore dynamically dominant. Consider instead the case in which the bulk overtakes the $d_{0}$ branch,
\begin{equation}
    \operatorname{Im}z_{\rm bulk}^{\max}
    >
    \operatorname{Im}z_0^{\rm out}.
\end{equation}
The inequality established above guarantees
\begin{equation}
    \operatorname{Im}z_{\rm bulk}^{\max}
    >
    \operatorname{Im}z_0^{\rm out}
    >
    \operatorname{Im}z_n^{\rm out}
\end{equation}
for every secondary outlier that still exists. A further observation reinforces this conclusion. Since $a_n<1$, the existence threshold of each secondary outlier,
\begin{equation}
    \sigma_{I,c}^{(n),{\rm outlier}}
    =
    \sqrt{2}\;a_n,
\end{equation}
is always smaller than the corresponding threshold for the $d_{0}$ outlier,
\begin{equation}
    \sigma_{I,c}^{\rm outlier}
    =
    \sqrt{2}.
\end{equation}
Secondary outliers therefore cease to exist before the $d_{0}$ outlier does.

In summary, a secondary outlier can never become dynamically dominant. The dominant eigenvalue $z_{\star}$ is determined either by the $d_{0}$ outlier branch or by the eigenvalue at the upper edge of the bulk. This conclusion does not require an explicit expression for $\operatorname{Im}z_{\rm bulk}^{\max}$, and it therefore applies both in the hard-edge regime $k_{I}\gg\sqrt{L}$ and in the soft-edge regime $k_{I}\ll\sqrt{L}$.
\section{Eigenvalue bound for long-tail regime}
\label{sec:max_eigenvalue_bound}
In this section, we derive a probabilistic upper bound on the largest imaginary part 
of the bulk eigenvalues of $G(\sigma_{I})$ in the purely 
anti-Hermitian disorder case. For $\mu_{R} = \mu_{I} = 0$ and 
$\sigma_{R} = 0$, the matrix reduces to
\begin{equation}
    G(\sigma_{I}) = \frac{i}{\sqrt{2}} H_{I}(0, \sigma^{2}_{I}), \label{eq:G_sigma_I}
\end{equation}
whose eigenvalues are purely imaginary, $z_{n} = i\lambda_{n}/\sqrt{2}$, 
where $\lambda_{n}$ are the real eigenvalues of the Hermitian 
matrix $H_{I}$. Bounding the maximum imaginary part of $G(\sigma_{I})$ in Eq.~\eqref{eq:G_sigma_I}
therefore reduces to bounding $\lambda_{\max}(H_{I})$, which is the maximum eigenvalue of $H_{I}(0, \sigma^{2}_{I})$.

We start by computing the disorder-averaged second operator moment of $H_{I}$.
Writing $H_{I} = \sum_{\boldsymbol{\alpha}} M_{\boldsymbol{\alpha}} 
P_{\boldsymbol{\alpha}}$, where the sum runs over all $3^{k_{I}}\binom{L}{k_{I}}$ 
exactly $k_{I}$-local Pauli strings, and using the Pauli 
orthogonality $\mathrm{Tr}[P_{\boldsymbol{\alpha}}P_{\boldsymbol{\beta}}] 
= 2^{L}\delta_{\boldsymbol{\alpha},\boldsymbol{\beta}}$ together with 
$P^{2}_{\boldsymbol{\alpha}} = I$, we obtain
\begin{align}
    \mathbb{E}[H^{2}_{I}] 
    &= \sum_{\boldsymbol{\alpha}} 
    \mathrm{Var}[M_{\boldsymbol{\alpha}}]\; P^{2}_{\boldsymbol{\alpha}} 
    \nonumber \\
    &= \underbrace{3^{k_{I}}\binom{L}{k_{I}}}_{\text{number of Pauli strings}} 
    \times \underbrace{\frac{\sigma^{2}_{I} L^{2}}{3^{k_{I}}\binom{L}{k_{I}}}}_{\text{variance per coupling}} 
    \times I \nonumber \\
    &= \sigma^{2}_{I} L^{2}\; I.
\end{align}

Next, we apply the matrix Gaussian series inequality 
from Ref.~\cite{Tropp_2011}. Writing $H_{I} 
= \sum_{\boldsymbol{\alpha}} g_{\boldsymbol{\alpha}} 
\bigl(\sigma_{\mathrm{op}}/\sqrt{3^{k_{I}}\binom{L}{k_{I}}}\bigr) 
P_{\boldsymbol{\alpha}}$ with $g_{\boldsymbol{\alpha}} \sim 
\mathcal{N}(0, 1)$ independent, and using 
$\sigma^{2}_{\mathrm{op}} = \sigma^{2}_{I}L^{2}$, 
leads to
\begin{equation}
    \mathrm{Pr}\bigl[\lambda_{\max}(H_{I}) \geq t\bigr] 
    \leq 2^{L} \exp\!\left(-\frac{t^{2}}{2\sigma^{2}_{I}L^{2}}\right).
\end{equation}
In words, the probability of finding an eigenvalue of $H_{I}$ above $t$ is exponentially suppressed in $t^{2}$. 
Setting the failure probability equal to  $\delta$ and 
solving for $t$ gives the bound on the imaginary part of the eigenvalues $z_{n}$ of $G(\sigma_{I})$ in Eq.~\eqref{eq:G_sigma_I}
\begin{equation}
    \mathrm{Im}\; z^{\rm max}_{\rm bulk} 
    = \frac{\lambda_{\max}(H_{I})}{\sqrt{2}} 
    \leq \sigma_{I} L \sqrt{L\ln 2 + \ln(1/\delta)} \label{eq:Imaginary_general_delta},
\end{equation}
with probability at least $1 - \delta$. Choosing an 
exponentially small failure probability $\delta = 2^{-L}$ 
gives the bound
\begin{equation}
    \mathrm{Im}\; z^{\rm max}_{\rm bulk} 
    \leq \sigma_{I} L \sqrt{2L\ln 2}.
    \label{eq:max_Im_bound}
\end{equation}
\clearpage
\onecolumngrid

\setcounter{section}{0}
\setcounter{subsection}{0}
\setcounter{equation}{0}
\setcounter{figure}{0}
\setcounter{table}{0}

\renewcommand{\thesection}{S\arabic{section}}
\renewcommand{\thesubsection}{S\arabic{section}.\arabic{subsection}}
\renewcommand{\theequation}{S\arabic{equation}}
\renewcommand{\thefigure}{S\arabic{figure}}
\renewcommand{\thetable}{S\arabic{table}}

\makeatletter
\renewcommand{\theHsection}{supp.\arabic{section}}
\renewcommand{\theHsubsection}{supp.\arabic{section}.\arabic{subsection}}
\renewcommand{\theHequation}{supp.eq.\arabic{equation}}
\renewcommand{\theHfigure}{supp.fig.\arabic{figure}}
\renewcommand{\theHtable}{supp.tab.\arabic{table}}
\makeatother

\begin{center}
{\large\bfseries Supplemental Material}\\[0.5em]
{\large\bfseries Spectral-Outlier Control of Entanglement and Magic in
Non-Hermitian \(k\)-Local Spin Ensembles}\\[1em]

Sasanka Dowarah and Michael Kolodrubetz\\
Department of Physics, The University of Texas at Dallas,
Richardson, Texas 75080, USA
\end{center}

\vspace{1em}
\section*{S2: Residual plots of outlier}
\begin{figure}[H]
    \centering
    \includegraphics[width=0.39\linewidth]{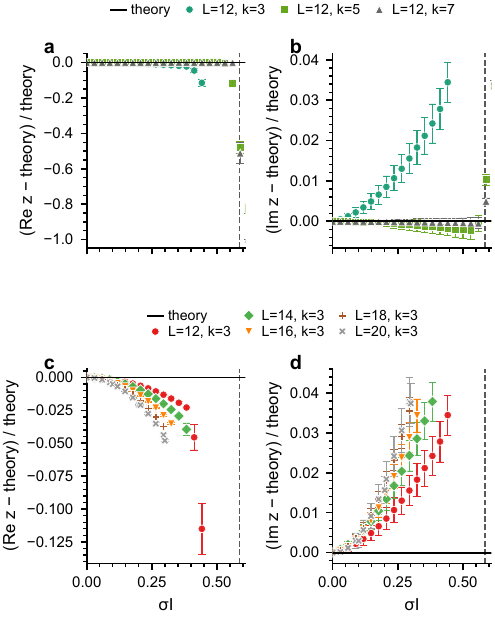}
    \caption{
Deviation of the numerically computed dominant eigenvalue $z_\star$ from the analytical outlier prediction $z_{\rm out}^{\rm th}$ for purely anti-Hermitian disorder, $\sigma_R=0$. 
(a,b) Relative deviations of the real and imaginary parts for $L=12$ and different $k$. 
(c,d) Corresponding results for $k=3$ and different system sizes $L$. 
Markers show disorder averages and error bars denote the standard error of the mean. The solid black line indicates exact agreement with theory, while the vertical dashed line marks the dominance threshold $\sigma_{I,c}^{\rm dom}=2-\sqrt{2}$. 
Agreement improves with increasing $k$, whereas deviations grow with $L$ at fixed $k=3$ due to the soft spectral tails in the $k\ll\sqrt{L}$ regime.
}
    \label{fig:placeholder}
\end{figure}

\section*{S1: Numerical method for computing the dominant eigenvalue}
\label{sec:SI_GPU_eigenvalue_method}
\addcontentsline{toc}{section}{S1. Numerical method for computing the dominant eigenvalue}\label{sec:SI_GPU_eigenvalue_method}

We compute the eigenvalue of $G$ with the largest imaginary part using a matrix-free implementation, in which the action of $G$ on a state vector is evaluated directly from its Pauli-string representation, without ever constructing the matrix explicitly.

\subsection*{Reduction to a real-part problem}
Define $A \equiv -iG$. If $G|R\rangle = z|R\rangle$, then $A|R\rangle = \lambda|R\rangle$ with $\lambda = -iz$. Writing $z = x+iy$ gives $\operatorname{Re}(\lambda) = \operatorname{Im}(z)$, so the eigenvalue of $G$ with the largest imaginary part is obtained from the eigenvalue of $A$ with the largest real part. The right eigenvectors of $A$ and $G$ coincide, and after the Arnoldi calculation returns $\lambda_\star$, the corresponding eigenvalue of $G$ is recovered as $z_\star = i\lambda_\star$.

\subsection*{Matrix-free Pauli-string representation}
Each Pauli string $P_t = \sigma_{\alpha_0}^{(0)} \otimes \cdots \otimes \sigma_{\alpha_{L-1}}^{(L-1)}$, with $\alpha_j \in \{0,x,y,z\}$, is exactly $k$-local if $k$ of the $\alpha_j$ are nonzero. Computational basis states $|b_0 b_1 \cdots b_{L-1}\rangle$ are identified with integers $x = \sum_j b_j 2^{L-1-j}$.

Each Pauli string is encoded by two bitmasks: a flip mask $f_t = \sum_{j:\alpha_j = x\,{\rm or}\,y} 2^{L-1-j}$, marking sites where the bit is flipped, and a phase mask $g_t = \sum_{j:\alpha_j = y\,{\rm or}\,z} 2^{L-1-j}$, marking sites contributing a sign. Together with $m_t = n_y(t) \bmod 4$, where $n_y(t)$ is the number of $\sigma_y$ factors, these masks fully specify the operator. One can show that
\begin{equation}
    (P_t\psi)_x
    =
    i^{m_t}(-1)^{\operatorname{pop}(s\& g_t)}\psi_s,
    \qquad
    s \equiv x\oplus f_t,
    \label{eq:SI_pauli_mask_action}
\end{equation}
where $\operatorname{pop}(\cdot)$ denotes the population count (number of set bits) of a bitwise AND. This expression is evaluated directly on the GPU. For a linear combination of $k$-local strings, $A = \sum_{t\in\mathcal{P}_k} c_t P_t$, linearity gives the matrix-free matrix-vector product
\begin{equation}
    (A\psi)_x
    =
    \sum_{t\in{\cal P}_k}
    c_t\,i^{m_t}(-1)^{\operatorname{pop}((x\oplus f_t)\& g_t)}\psi_{x\oplus f_t}.
    \label{eq:SI_matrix_free_matvec}
\end{equation}
\subsection*{Restarted Arnoldi iteration}
We extract the eigenvalue of $A$ with the largest real part using a restarted Arnoldi iteration~\cite{Sorensen1997} with Krylov dimension $n_{\rm cv}=50$ and up to $80$ restarts, using two-pass modified Gram--Schmidt orthogonalization for numerical stability. At each restart we examine the four Ritz values with the largest real parts and retain the one with the largest real part, breaking near-degeneracies by smaller residual.

For a candidate Ritz pair $(\lambda,q)$, we require the relative residual
\begin{equation}
    r_{\rm rel}
    =
    \frac{\|Aq-\lambda q\|_2}
    {\|Aq\|_2+|\lambda|\|q\|_2}
    <10^{-10}.
    \label{eq:SI_relative_residual}
\end{equation}
Each converged eigenvalue is then mapped back via $z = i\lambda$.
\subsubsection*{Convergence data}
In this section, we show the convergence data for different system sizes. In each table, $N$ is the total number of eigenpairs computed
(realizations $\times$ sampled $\sigma$ values); converged fraction is
the fraction with $r_{\rm rel} < 10^{-10}$; restarts are averaged over
all $\sigma$ and, separately, over points within $|\sigma-\sigma_c|<0.1$.
\begin{table}[H]
\centering
\caption{Convergence statistics of the Arnoldi eigensolver across
system sizes for anti-Hermitian disorder used in Fig.~\ref{fig:sigma-c-sample-distribution}.}
\label{tab:convergence}
\begin{tabular}{ccccccc}
\hline\hline
$L$ & $N$ & Converged & Median $r_{\rm rel}$ & Max $r_{\rm rel}$ & Restarts (all) & Restarts (near $\sigma_c$) \\
\hline
14 & 9800  & 100.00\% & $1.199\times10^{-12}$ & $9.992\times10^{-11}$ & 2.55 & 3.68 \\
16 & 9849  & 100.00\% & $1.816\times10^{-12}$ & $9.996\times10^{-11}$ & 2.85 & 4.52 \\
18 & 12740 & 99.98\%  & $3.606\times10^{-12}$ & $1.151\times10^{-3}$  & 3.19 & 5.21 \\
20 & 12789 & 99.95\%  & $3.802\times10^{-12}$ & $1.270\times10^{-3}$  & 3.28 & 5.08 \\
\hline\hline
\end{tabular}
\end{table}

\begin{table}[H]
\centering
\caption{Convergence statistics of the Arnoldi eigensolver across
locality $k$ for anti-Hermitian disorder ($\sigma_R = 0$), fixed $L = 12$ used in Fig.~\ref{fig:k_scan_threshold_recovery}.}
\label{tab:convergence_antiHermitian_k_scan}
\begin{tabular}{ccccccc}
\hline\hline
$k$ & $N$ & Converged & Median $r_{\rm rel}$ & Max $r_{\rm rel}$ & Restarts (all) & Restarts (near $\sigma_c$) \\
\hline
3 & 9800 & 100.00\% & $3.669\times10^{-13}$ & $9.990\times10^{-11}$ & 2.14 & 2.80 \\
5 & 9800 & 100.00\% & $3.849\times10^{-12}$ & $9.997\times10^{-11}$ & 3.01 & 3.08 \\
7 & 3612 & 100.00\% & $3.265\times10^{-12}$ & $9.977\times10^{-11}$ & 2.94 & 3.01 \\
9 & 3207 & 99.97\% & $3.654\times10^{-12}$ & $2.633\times10^{-3}$ & 3.37 & 3.57 \\
\hline\hline
\end{tabular}
\end{table}

\begin{table}[H]
\centering
\caption{Convergence statistics of the Arnoldi eigensolver across
system sizes for mixed disorder ($\sigma_R = \sigma_I$), $k_R = k_I = 3$ used in Fig.~\ref{fig:Entanglement_sigma_R_sigma_I}.}
\label{tab:convergence_mixed}
\begin{tabular}{ccccccc}
\hline\hline
$L$ & $N$ & Converged & Median $r_{\rm rel}$ & Max $r_{\rm rel}$ & Restarts (all) & Restarts (near $\sigma_c$) \\
\hline
12 & 17640 & 98.92\% & $8.120\times10^{-15}$ & $9.106\times10^{-2}$ & 6.59 & 12.98 \\
14 & 17640 & 98.46\% & $1.046\times10^{-13}$ & $7.410\times10^{-2}$ & 9.42 & 19.00 \\
16 & 17640 & 95.56\% & $8.017\times10^{-13}$ & $6.177\times10^{-2}$ & 12.54 & 26.32 \\
18 & 13903 & 89.20\% & $4.998\times10^{-12}$ & $7.210\times10^{-2}$ & 11.93 & 23.99 \\
20 & 5582 & 84.40\% & $1.322\times10^{-11}$ & $5.251\times10^{-2}$ & 11.73 & 23.70 \\
\hline\hline
\end{tabular}
\end{table}

\begin{table}[H]
\centering
\caption{Convergence statistics of the Arnoldi eigensolver across
locality $k$ for mixed disorder ($\sigma_R = \sigma_I$), fixed $L = 12$ used in Fig.~\ref{fig:Equal_mixed_disorder_case_convergence_to_analytical_hard_edge_prediction}.}
\label{tab:convergence_mixed_k_scan}
\begin{tabular}{ccccccc}
\hline\hline
$k$ & $N$ & Converged & Median $r_{\rm rel}$ & Max $r_{\rm rel}$ & Restarts (all) & Restarts (near $\sigma_c$) \\
\hline
3 & 17640 & 98.92\% & $8.120\times10^{-15}$ & $9.106\times10^{-2}$ & 6.59 & 12.98 \\
5 & 14200 & 98.85\% & $9.455\times10^{-15}$ & $7.469\times10^{-2}$ & 7.59 & 12.04 \\
7 & 1948 & 98.41\% & $2.759\times10^{-14}$ & $6.434\times10^{-2}$ & 6.93 & 11.73 \\
9 & 1510 & 97.95\% & $4.367\times10^{-14}$ & $6.278\times10^{-2}$ & 8.56 & 13.43 \\
\hline\hline
\end{tabular}
\end{table}
\clearpage
\bibliography{apssamp}
\end{document}